\documentclass[a4paper,12pt]{article}

\usepackage{amsmath,amsfonts,amssymb} 
\usepackage{authblk}                  
\usepackage{geometry}                 
\usepackage[numbers]{natbib}          
\usepackage{hyperref}                 
\usepackage{times}                    
\usepackage{parskip}                  
\usepackage{physics}
\usepackage[nocomma]{optidef}
\usepackage{mathbbol}
\usepackage[bb=libus]{mathalpha}
\usepackage{subcaption}
\usepackage{float}
\usepackage[toc,page]{appendix}
\usepackage{graphicx}      
\usepackage{subcaption}    
\usepackage{caption}       
\usepackage{rotating}     
\usepackage{pdflscape}
\usepackage{booktabs}
\usepackage{multirow}
\usepackage{siunitx} 
\usepackage{caption}
\usepackage[table]{xcolor}
\usepackage{pifont}
\usepackage{longtable} 
\usepackage{setspace}
\usepackage{pdfpages}

\usepackage{booktabs,siunitx,threeparttable,tabularx,array}
\newcolumntype{L}{>{\raggedright\arraybackslash}X}
\renewcommand{\arraystretch}{1.2}
\DeclareFontFamily{U}{mathc}{}
\DeclareFontShape{U}{mathc}{m}{it}%
{<->s*[1.03] mathc10}{}

\DeclareMathAlphabet{\mathscr}{U}{mathc}{m}{it}

\title{Modeling Bilateral Lymphatic Head and Neck Tumour Progression for Personalized Elective Target Volume Definition}

\author[1,2]{Kristoffer Moos}
\author[3]{Anne Ivalu Sander Holm}
\author[4]{Yoel Perez Haas}
\author[4]{Roman Ludwig}
\author[2,3]{Jesper Grau Eriksen}
\author[1,2]{Stine Sofia Korreman\thanks{Corresponding author. Email: stine.korreman@clin.au.dk}}

\affil[1]{\small Department of Clinical Medicine, Aarhus University, Aarhus, Denmark}
\affil[2]{\small Danish Centre for Particle Therapy, Aarhus University Hospital, Aarhus, Denmark}
\affil[3]{\small Department of Experimental Clinical Oncology, Aarhus University Hospital, Aarhus, Denmark}
\affil[4]{\small Department of Radiation Oncology, University Hospital of Zurich, Zurich, Switzerland}

\date{}

\begin{document}

\maketitle

\begin{abstract}
\textbf{Objective:} Large irradiated volumes are a major contributor to severe side-effects in patients with head and neck cancer undergoing curatively intended radiotherapy. We propose a data-driven approach for defining the elective clinical target volume (CTV-E) on a patient-specific basis, with the potential to reduce irradiated volumes compared to standard guidelines. \textbf{Approach:} We introduce a bilateral Bayesian Network (BN), trained on a large cohort, to estimate the patient-specific risk of undetected nodal involvement for both ipsilateral and contralateral lymph node levels (LNLs) I, II, III, and IV, using clinical features, such as patterns of nodal involvement, T-stage, tumour location. By applying risk thresholds, we generated individualized, risk-dependent CTV-E's for representative patient scenarios and compared the resulting treatment volumes and residual risk to those recommended by standard clinical guidelines. \textbf{Main results:} We computed the risks for a set of representative patient scenarios including 1) N0 (T1 and T2 tumour stage), 2) N+ in ipsilateral LNL II (T1 and T2 tumour stage), 3) N+ in ipsilateral LNL II and III (T1 and T2 tumour stage), and 4) N+ of both ipsilateral and contralateral LNL II (T3 and T4 tumour stage). Depending on the chosen risk threshold, the bilateral BN allowed for reductions in irradiated volumes relative to standard clinical protocols. For every patient scenario considered, the CTV-E's defined by the applied risk thresholds were associated with a low estimated probability of undetected nodal involvement in any excluded LNL. \textbf{Significance:} We present a data-driven framework for personalized CTV-E definition, encouraging the discussion of more patient-specific elective nodal target volumes, with potential for de-escalation of irradiated elective volumes.
\end{abstract}

\section{Introduction}

Radiotherapy (RT) is an essential part of the standard treatment for head and neck cancer (HNC) and is routinely administered to the majority of patients, either as definitive therapy or in combination with surgery and systemic treatment. Patients are treated with high doses of irradiation to the primary tumour and involved lymph nodes, while attempting to preserve as low dose as possible to surrounding healthy tissue. During the past decades, technological advancements in radiotherapy have led to the development of highly personalized treatments on an individual patient basis, including significantly better treatment outcomes with higher cure rates, and reduced acute and long-term side effects \citep{thompson_practice-changing_2018}. Key enhancements consist of multi-modality imaging to identify target volume, the development of optimization methods to accurately conform the radiation beams to tumours, and introduction of high precision beam delivery capable of targeting with sub-millimeter precision \citep{baumann_radiation_2016}.

Despite the efforts to minimize exposure of healthy tissue, irradiation of such tissue remains inevitable and often leads to undesirable complications. A substantial part of patients with HNC whose primary treatment is radiotherapy, suffer from acute and/or long-term side effects, including xerostomia, oral discomfort, difficulties chewing, speaking, and swallowing, as well as loss of taste and reduced periodontal hygiene. Common for these side effects is their significant impact on the patient's quality of life, affecting daily activities and overall well-being. 

In HNC, a low-dose elective clinical target volume (CTV-E) is defined when there is suspicion of malignant lymph nodes (LNs). Despite the relatively low dose prescribed to the CTV-E, it includes a considerable portion of the cervical volume, resulting in substantial irradiation of adjacent organs-at-risk (OAR) and surrounding healthy tissue. Several studies have shown that large irradiated volumes are one of the leading causes of acute- and long-term side effects \citep{Mestdagh_2019, Mestdagh_2020, navran_impact_2019, Langendijk_2009, Mamgani_2015}.

Current studies suggest that it may be relevant to adjust the clinical guidelines, particularly with respect to reconsidering conventional approaches for establishing the CTV-E. This is primarily due to instances of elective neck failures being exceedingly rare with the majority of regional recurrences appearing directly in the high-dose target volume \cite{dandekar_patterns_failure_2014, sher_treatment_2012, garden_patterns_2013, duprez_regional_2011, van_den_bosch_patterns_2016, kristensen_high_dose_2024}. Dose and volume de-escalation, in particular, have received a lot of attention as promising treatment options, all while maintaining regional control, resulting in favorable reduction in toxicity profiles, and improving patient-reported outcomes \cite{sher_prospective_2021, deschuymer_randomized_2020}. 

A strategy that could potentially allow for systematic volume de-escalation is defining the CTV-E using a probabilistic approach. By allowing for statistical evaluations of malignant LNs, we can selectively omit any LNs associated with sufficiently low risk of being malignant, all while preserving the therapeutic effect. Two previous studies have investigated the viability of using risk-based approaches to assess distinction of malignant LNs. \citet{pouymayou_bayesian_2019} have previously demonstrated that by using a statistical framework to estimate lymphatic nodal involvement, it is possible to assess the risk of subclinical disease in different lymph node levels (LNLs). Similarly, \citet{ludwig_hidden_2021} proposed a probabilistic approach for estimating the risk of lymphatic tumour progression. 

However, the initial approaches outlined in \citet{pouymayou_bayesian_2019} and \citet{ludwig_hidden_2021} are limited to ipsilateral tumour progression. Only addressing ipsilateral tumour progression may overlook essential aspects of the disease dynamics and underestimate the full extent of subclinical disease. This could lead to incomplete and/or inaccurate assessment of the overall pattern of spread, increasing the risk of missing metastatic developments that may affect the patient outcome. While Ludwig and colleagues have since expanded their probabilistic framework to incorporate bilateral lymphatic tumour progression \cite{ludwig_hidden_2025}, further investigations are needed to validate the clinical utility of such models in informing target volume definition. In particular, it is of interest to investigate whether such probabilistic frameworks are generalizable and clinically applicable across broader populations. 

The primary objective of this study is to investigate the potential of a probabilistic framework to support patient-tailored CTV-E definitions, and to evaluate whether such individualization can lead to clinically meaningful volume de-escalation. To this end, section 2 provides a detailed description of the patient cohort utilized for model development, highlighting key clinical characteristics relevant to the analysis. Section 3 describes the extension of the models to account for bilateral lymphatic spread. This extension explicitly accounts for the different patterns of ipsilateral and contralateral LN involvement. Section 4 evaluates the CTV-E definitions derived from the probabilistic model through direct comparison with clinically established CTV-E recommendations. This is completed by carrying out a systematic sensitivity analysis to investigate how variations in the choice of risk threshold influence the extent of CTV-E volume inclusion and the associated residual risks.

\section{Patients}

Our analysis is based on a retrospective cohort of 428 patients treated for oropharyngeal squamous cell carcinoma with definitive radiotherapy at Aarhus University Hospital between 2013 and 2020. Detailed clinical and pathological data were compiled for each patient, including information on nodal involvement for each individual LNL, as well as primary tumor characteristics. In this study, the model development was based on three key clinical variables: 1) the anatomical origin of the primary tumour, 2) the tumour stage, and 3) the detailed pattern of LN involvement. A summary of these key characteristics for the cohort are provided in table \ref{tab:baseline_char}.

\begin{table}[H]
\centering
\caption{Patient characteristics of retrospective included patients (N = 428)}
\label{tab:baseline_char}
\begin{tabular}{llcc}
\toprule
\textbf{Characteristic} & \textbf{Category} & \textbf{N} & \textbf{\%} \\
\midrule
\textbf{Sex} \\
\hspace{3mm} & Male   & 309 & 72.2 \\
\hspace{3mm} & Female & 119 & 27.8 \\
\addlinespace
\textbf{Primary Tumor Origin} \\
\hspace{3mm} & Base of tongue            & 141 & 33.0 \\
\hspace{3mm} & Vallecula                 & 23  & 5.4  \\
\hspace{3mm} & Tonsil                    & 170 & 39.7 \\
\hspace{3mm} & Pharyngeal arches/Tonsillar fossa & 57 & 13.3 \\
\hspace{3mm} & Glossotonsillar sulcus    & 7   & 1.6  \\
\hspace{3mm} & Posterior pharyngeal wall & 12  & 2.8  \\
\hspace{3mm} & Inferior soft palate      & 13  & 3.0  \\
\hspace{3mm} & Uvula                     & 5   & 1.2  \\
\addlinespace
\textbf{T97 Classification} \\
\hspace{3mm} & T1    & 104 & 24.3 \\
\hspace{3mm} & T2    & 171 & 40.0 \\
\hspace{3mm} & T3    & 89  & 20.8 \\
\hspace{3mm} & T4    & 64  & 14.9  \\
\addlinespace
\textbf{T-Stage Stratification} \\
\hspace{3mm} & Early (T1/T2)          & 275 & 64.7 \\
\hspace{3mm} & Advanced (T3/T4)   & 153 & 35.3 \\
\bottomrule
\end{tabular}
\end{table}

\subsection{Primary Tumour Origin}

For model development, the primary tumor's origin was categorized according to the anatomical sub-site within the oropharynx, including the tonsil, base of tongue, soft palate (including tonsillar fossa), vallecula, glossotonsillar sulcus, posterior pharyngeal wall, inferior surface of the soft palate, and uvula. In this cohort, the most common primary sites were the tonsil (39.7\%) and base of tongue (33.0\%). The soft palate and tonsillar fossa together accounted for 13.3\% of cases, followed by the vallecula (5.4\%), inferior surface of the soft palate (3.0\%), posterior pharyngeal wall (2.8\%), glossotonsillar sulcus (1.6\%), and uvula (1.2\%). 

Because tumor lateralization has been shown to significantly influence the risk of contralateral nodal involvement, particular consideration was given to this factor in cohort selection. Tumors located closer to the midline are more likely to exhibit bilateral lymphatic spread compared to those situated more laterally. Accordingly, the final cohort consisted of 428 patients with centrally located primary tumors, defined as those situated $\leq$ 1 cm of the midsagittal line.

\subsection{Tumour Stage}

Tumor stage was classified according to the UICC7 criteria, providing a detailed assessment of the size and local extent of the primary tumor. The most common T-category was T2, comprising 171 patients (40.0\%). This was followed by T1 (24.3\%), T3 (20.8\%), and T4 (14.9\%). The T-category, which reflects the size and local extent of the primary tumor, was used as an approximation for greater risk of undetected nodal involvement since larger or more locally advanced tumors, have a higher probability of harbouring occult metastasis. Among the 428 patients included in the cohort, the distribution of T-category highlights the predominance of early-stage tumors. Specifically, 275 patients (64.2\%) presented with early-stage disease (T1 or T2), while 153 patients (35.8\%) had advanced-stage tumors (T3 or T4).

\subsection{Patterns of Nodal Involvement}

For each patient, the anatomical locations of nodal metastases were systematically documented by physicians as part of routine clinical practice, distinguishing between ipsilateral and contralateral involvement relative to the primary tumor site. Nodal status was determined primarily by radiological assessment, based on MRI combined with PET/CT imaging. In situations where imaging findings were uncertain, additional diagnostic workup was performed with fine needle aspiration biopsy (FNA) to establish nodal status as definitively as possible.

\begin{figure}[H]
    \centering
    \begin{subfigure}{0.9\textwidth} 
        \caption{}
        \label{fig:ipsilateral_upset}
        \includegraphics[width=\textwidth]{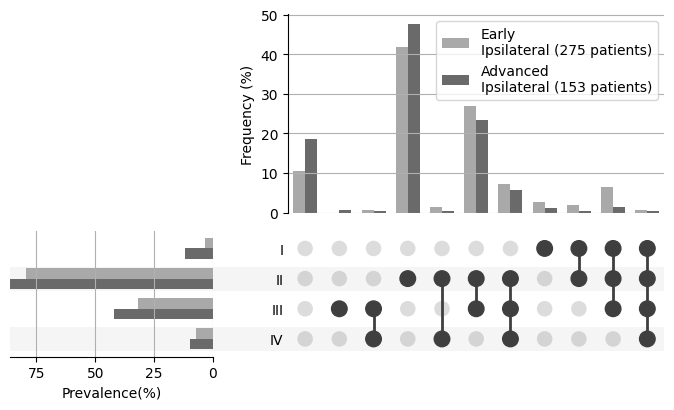}
    \end{subfigure}
    
    \vspace{0.3cm} 

    \begin{subfigure}{0.9\textwidth} 
        \caption{}
        \label{fig:contralateral_upset}
        \includegraphics[width=\textwidth]{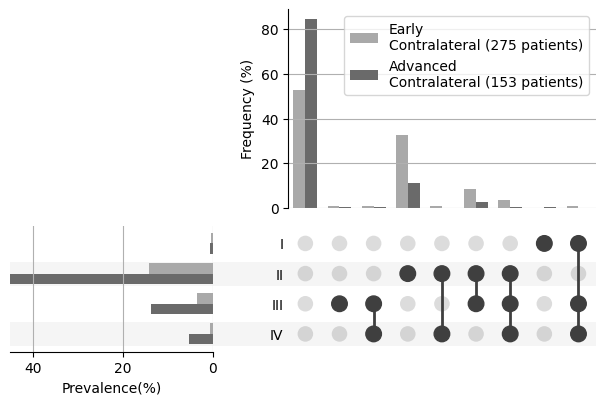}
    \end{subfigure}        

    \caption{Upset plots illustrating the patterns of nodal involvement. (\textbf{a}) UpSet plot of lymphatic nodal spread patterns for different involvement patterns in early- and advanced-stage ipsilateral LNLs. (\textbf{b}) UpSet plot of lymphatic nodal spread patterns for different involvement patterns in early- and advanced-stage contralateral LNLs. The left panels in each figure showcase the prevalence of involvement for each LNL individually, while the main panels highlight the most frequent patterns of simultaneous nodal involvement. }
    \label{fig:UpSet_plot_data_characteristics}
\end{figure}

The patterns of LN involvement for the population are summarized in figure \ref{fig:UpSet_plot_data_characteristics}, which presents upset plots stratified by tumor stage (early vs. advanced) and by laterality (ipsilateral vs. contralateral). For the ipsilateral neck, advanced-stage tumors exhibited broader and more frequent nodal involvement compared to early-stage disease. As shown in figure \ref{fig:ipsilateral_upset}, simultaneous involvement of LNL II and III and more extensive multi-level combinations (including LNL IV) were most common in advanced tumors. Among early-stage tumors, isolated involvement of LNL II was the predominant pattern. For the contralateral neck, nodal involvement was overall less common. As shown in figure \ref{fig:contralateral_upset}, the majority of both early- and advanced-stage tumors exhibited no detectable contralateral nodal involvement. Among contralateral node-positive cases, isolated LNL II involvement was most common, with multi-level (e.g., II+III or II+III+IV) combinations seen more frequently as tumor stage increased.

\section{Bayesian Network for Bilateral Lymph Node Involvement}

We propose a probabilistic model based on a graph representation of the lymphatic drainage pathways that describes the bilateral spread of lymphatic tumor progression in HNC. Our formulation is influenced by recent data-driven models, such as that of \citet{ludwig_hidden_2025}. For this purpose, we utilize the Python package "lymph" \footnote{\url{https://github.com/lycosystem/lymph}}.

One should keep in mind, that the proposed framework applies exclusively to the previously untreated neck. Once a patient has undergone prior surgery or radiotherapy, the lymphatic system may be altered, creating new drainage routes and thereby diminish the reliability of predictions regarding tumor spread patterns \cite{flach_sentinel_2012}.

\subsection{Bilateral Extension}

We model lymphatic spread using a Bayesian Network (BN) that encodes the probabilistic dependencies between nodal involvement across multiple LNLs on both the ipsilateral ($s = i$) and contralateral ($s = c$) sides of the neck. For each LNL $\ell \in \mathcal{L} = \{\mathrm{I}, \mathrm{II}, \mathrm{III}, \mathrm{IV}\}$ and each side $s \in \{i, c\}$, we introduce a binary latent variable $x_\ell^s \in \{0,1\}$ representing the true state of the LNL: $x_\ell^s = 1$ indicates metastatic involvement (microscopic or macroscopic), and $x_\ell^s = 0$ indicates absence of involvement. The corresponding observed variable $z_\ell^s \in \{0,1\}$ captures whether involvement is radiologically detectable, with $z_\ell^s = 1$ denoting positive imaging findings (e.g., CT, MRI, or PET) and $z_\ell^s = 0$ denoting absence of detectable involvement. Under this formulation, $x_\ell^s = 1, z_\ell^s = 1$ corresponds to macroscopic involvement (detected), while $x_\ell^s = 1, z_\ell^s = 0$ corresponds to microscopic or occult involvement (undetected). This distinction also permits modeling of imperfect imaging sensitivities and specificities, thereby accounting for both false positives and false negatives.

\begin{figure}[H]
\centering
\includegraphics[width=0.8\textwidth]{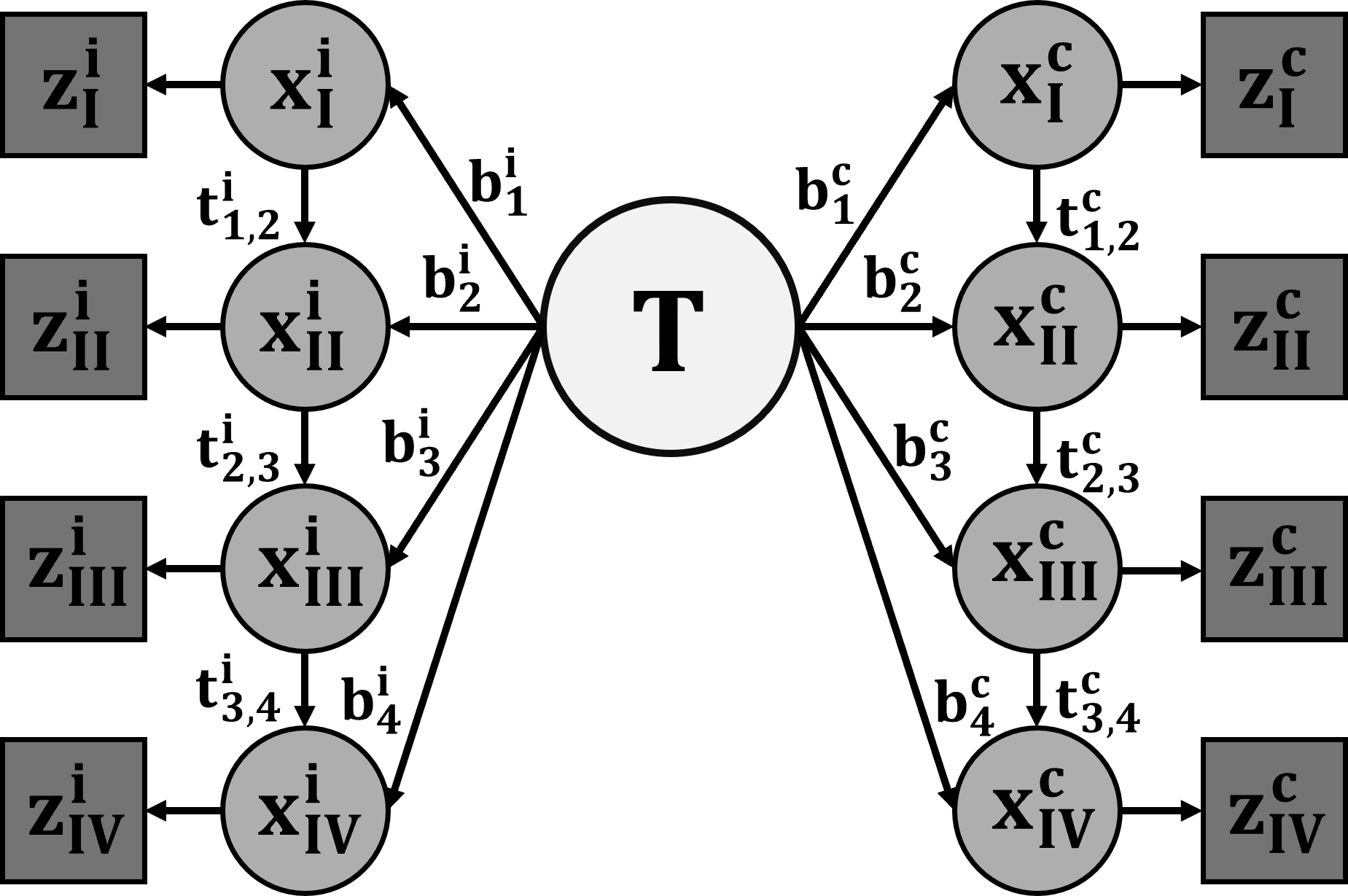}
\caption{Structure of the Bilateral Bayesian Network modeling lymphatic drainage. The unobserved variables $x^s_k$ represent the true involvement status of the $\mathrm{k}^{th}$ LNL on side $s \in {i,c}$, while $z^s_k$ denotes the observed status as defined by diagnostic imaging. $T$ represents the primary tumour. Each directed arc encodes either a base probability ($b^s_k$) or a transition probability ($t^s_k$).}
\label{BN_Graph}
\end{figure}

The BN is structured such that each $x_\ell^s$ node is connected to its anatomical parent(s) from which it receives efferent lymphatic flow, most commonly the preceding LNL on the same side ($x_{\ell-1}^s$ for $\ell > 1$) and the primary tumor. This structure reflects the anatomical progression pathways for tumor spread through the lymphatic system and allows for explicit modeling of dependencies between neighboring LNLs as well as the influence of the primary tumor, as illustrated in figure \ref{BN_Graph}.

For notational clarity, we define the vectors $\mathbf{x}^s = (x^s_\ell)_{\ell \in \mathcal{L}}$ and $\mathbf{z}^s = (z^s_\ell)_{\ell \in \mathcal{L}}$ to represent the full unobserved and observed states across all LNLs on side $s$, respectively. The joint prior distribution over all unobserved involvement states on both sides is given by:

\begin{equation}
P(\mathbf{x}^i, \mathbf{x}^c) = \sum_{T} P(T) \, P(\mathbf{x}^i \mid T) \, P(\mathbf{x}^c \mid T)
\end{equation}

Here, $T$ is a binary variable indicating the presence ($T=1$) or absence ($T=0$) of a primary tumor at the site of interest, while $P(T)$ denotes the prior probability that the tumor is present at that site. The conditional probability of a full involvement pattern on side $s$ given $T$ is factorized according to the BN structure:

\begin{equation}
P(\mathbf{x}^s \mid T) = \prod_{\ell \in \mathcal{L}} P(x^s_\ell \mid \mathrm{Pa}(x^s_\ell), T)
\end{equation}

Here, $\mathrm{Pa}(x^s_\ell)$ denotes the parent nodes of $x^s_\ell$, including $x^s_{\ell-1}$ for $\ell > 1$ and the tumor variable $T$. This structure ensures that the model represents biologically plausible pathways of microscopic tumour spread, reflecting both local anatomical spread and tumor-specific risk. We model the radiological findings under the assumption of conditional independence, such that the probability of a radiological finding in any LNL depends only on its true microscopic involvement and is unaffected by the state of other LNLs:

\begin{equation}
P(\mathbf{z}^s \mid \mathbf{x}^s) = \prod_{\ell \in \mathcal{L}} P(z^s_\ell \mid x^s_\ell)
\end{equation}

where $P(z^s_\ell \mid x^s_\ell)$ specifies the probability of a positive or negative radiological finding in LNL $\ell$ conditional on its true microscopic involvement, reflecting the sensitivity and specificity of the diagnostic imaging modality. Given observed imaging findings $\mathbf{z}^i$ and $\mathbf{z}^c$, we compute the posterior distribution over all microscopic involvement configurations using Bayes' theorem:

\begin{equation}
P(\mathbf{x}^i, \mathbf{x}^c \mid \mathbf{z}^i, \mathbf{z}^c) = 
\frac{P(\mathbf{x}^i, \mathbf{x}^c) \, P(\mathbf{z}^i \mid \mathbf{x}^i) \, P(\mathbf{z}^c \mid \mathbf{x}^c)}{\sum_{\mathbf{x}^i,\, \mathbf{x}^c} P(\mathbf{x}^i, \mathbf{x}^c) \, P(\mathbf{z}^i \mid \mathbf{x}^i) \, P(\mathbf{z}^c \mid \mathbf{x}^c)}
\end{equation}

where the denominator ensures normalization across all possible combinations of microscopic involvement on both sides. This formulation enables the explicit computation of marginal and conditional probabilities for any subset of LNLs or clinically relevant pattern. 

\subsection{Parameter learning}

First, we formally define the likelihood function, quantifying how the observed nodal involvement relates to the underlying microscopic disease process and model parameters. The BN is parameterized by $\boldsymbol{\theta} = \{t_{p \to v}, b_v\}$, where $t_{p \to v}$ denotes the probability of microscopic tumor spread from parent node $p$ to LNL $v$ (transition probability), and $b_v$ is the probability of direct spread from the primary tumor to LNL $v$ (base probability). 

Given observed radiological findings for $N$ patients, $\mathcal{D} = \{\mathbf{z}_1, \dots, \mathbf{z}_N\}$, where each $\mathbf{z}_n = (\mathbf{z}_n^i, \mathbf{z}_n^c)$ encodes the imaging results across LNLs for patient $n$, the per-patient likelihood marginalized over microscopic states is given by:

\begin{equation}
\mathcal{L}_n = P(\mathbf{z}_n^i, \mathbf{z}_n^c \mid T_n; \boldsymbol{\theta}) = \sum_{\mathbf{x}^i, \mathbf{x}^c}
    P(\mathbf{x}^i, \mathbf{x}^c \mid T_n; \boldsymbol{\theta})\,
    P(\mathbf{z}_n^i \mid \mathbf{x}^i)\,
    P(\mathbf{z}_n^c \mid \mathbf{x}^c).
\end{equation}

The total log-likelihood for the dataset is then:

\begin{equation}
\log \mathcal{L}(\mathcal{D} \mid \boldsymbol{\theta}) = \sum_{n=1}^N \log \mathcal{L}_n.
\end{equation}

In practice, this marginalization and likelihood computation is performed using efficient matrix-based methods.

\subsubsection{Sampling Using Markov Chain Monte Carlo}

Posterior inference for the BN parameters $\boldsymbol{\theta}$ was performed using Markov Chain Monte Carlo (MCMC) sampling. We employed the affine-invariant ensemble sampler implemented in the Python package \texttt{emcee} \footnote{\url{https://github.com/dfm/emcee}} to explore the posterior distribution defined by the likelihood from the previous section and a uniform prior over all parameters \citep{foreman_mackey_emcee_2013}.

We sampled a 14-dimensional parameter space corresponding to the transition and base probabilities in the bilateral network (four base probabilities for direct spread to each LNL, and three transition probabilities for stepwise progression between LNLs on each side). More details on the sampling configurations can be found in table \ref{tab:sampling_config} in appendix \ref{app:Sampling}. For parameter learning, we assume that imaging observations reflect the ground truth of nodal involvement, effectively treating observed variables as perfect indicators of the underlying microscopic states. This corresponds to sensitivity of 1 and specificity of 1. Under this assumption, the observation model simplifies to a deterministic mapping \(z_\ell^s = x_\ell^s\), which eliminates uncertainty from the diagnostic process.

The learned spread parameters are summarized in table \ref{tab:BN_Params} found in appendix \ref{app:params}, which illustrates the estimated probabilities for both direct and stepwise progression between LNLs in the bilateral network. 

\subsection{Risk Assessment and Derivation of Model-Based Elective Target Volumes}

Within any diagnostic scenario, risk assessment seeks to evaluate the probability of undetected nodal involvement in each LNL, using the patient’s observed imaging findings as the basis for inference. While the marginal risk per-LNL (the probability of undetected nodal involvement for each LNL) provides valuable information, it does not capture the combinatorial, correlated risk of missing nodal involvement when multiple LNLs are simultaneously spared from the CTV-E. For this reason, the true (unobserved) distribution of microscopic disease may include multiple involved LNLs in a manner that is not captured by considering each LNL in isolation. Relying solely on per-LNL marginal risks may therefore overestimate the actual probability of leaving microscopic disease untreated, particularly in cases where certain constellations of involvement are more likely.

Instead, the goal is to ensure that the total probability of missing undetected nodal involvement remains within an acceptable risk limit, accounting for all possible microscopic patterns compatible with the observed diagnosis and model uncertainty. Therefore, in the following, we introduce a global, joint risk minimization strategy for constructing risk-adaptive, individualized CTV-E definitions.

\subsubsection{Joint Posterior Distribution of Nodal Involvement}

We compute the risk of nodal involvement for any pattern or specific LNL, conditional on an individual patient’s observed diagnosis. For a patient with observed imaging findings $(\mathbf{z}^i, \mathbf{z}^c)$, the posterior probability of any pair of nodal involvement states $(\mathbf{x}^i, \mathbf{x}^c)$ is given by Bayes’ theorem:

\begin{equation} \label{eq:computing_risk}
P(\mathbf{x}^i, \mathbf{x}^c \mid \mathbf{z}^i, \mathbf{z}^c, T; \boldsymbol{\theta}) =
\frac{
    P(\mathbf{x}^i, \mathbf{x}^c \mid T; \boldsymbol{\theta})\,
    P(\mathbf{z}^i \mid \mathbf{x}^i)\,
    P(\mathbf{z}^c \mid \mathbf{x}^c)
}{
    \displaystyle\sum_{\mathbf{x}^{i'} \in \mathcal{X}^i}\sum_{\mathbf{x}^{c'} \in \mathcal{X}^c}
    P(\mathbf{x}^{i'}, \mathbf{x}^{c'} \mid T; \boldsymbol{\theta})\,
    P(\mathbf{z}^i \mid \mathbf{x}^{i'})\,
    P(\mathbf{z}^c \mid \mathbf{x}^{c'})
}
\end{equation}

Here, the numerator represents the product of the joint prior probability for the given unobserved state and the likelihood of the observed imaging findings for that state. The denominator is a normalization constant ensuring the posterior sums to one. Here, the prime notation ($\mathbf{x}^{i'}$, $\mathbf{x}^{c'}$) in the denominator indicates summation over all possible configurations of the unobserved states, different from the specific $(\mathbf{x}^i, \mathbf{x}^c)$ considered in the numerator.

Unlike during parameter estimation, where we treated the observed imaging findings as ground truth (i.e., assumed perfect sensitivity and specificity), risk assessment for individual patients instead uses the sensitivity and specificity of the diagnostic modality as estimated from literature. In Denmark, diagnosis is most often established by PET/CT combined MRI. However, to our knowledge, published data and literature on the sensitivity and specificity of PET/CT combined with MRI remain sparse. Therefore, we parameterize the observation model $P(\mathbf{z}^s \mid \mathbf{x}^s)$ using a sensitivity of 0.71 and a specificity of 0.90 for PET/CT alone, as reported by \citet{guedj_fdg_2024}.

\subsubsection{Computation of Marginal Involvement Risks}

While equation \ref{eq:computing_risk} provides the posterior probability for any specific unobserved state, clinical decisions require the marginal probability that a particular LNL $\ell$ on side $s$ is truly involved. This marginal probability is obtained by summing the joint posterior over all unobserved state configurations in which $x^s_\ell = 1$, while summing over all possible configurations of the remaining unobserved variables. Specifically, the marginal risk for LNL $\ell$ on side $s$ is given by:

\begin{equation}
P(x^s_\ell = 1 \mid \mathbf{z}^i, \mathbf{z}^c, T; \boldsymbol{\theta}) = 
\begin{cases}
\displaystyle
\sum_{\mathbf{x}^i : x^i_\ell = 1} \sum_{\mathbf{x}^c}
P(\mathbf{x}^i, \mathbf{x}^c \mid \mathbf{z}^i, \mathbf{z}^c, T; \boldsymbol{\theta}) & \text{if } s = i \\[3ex]
\displaystyle
\sum_{\mathbf{x}^i} \sum_{\mathbf{x}^c : x^c_\ell = 1}
P(\mathbf{x}^i, \mathbf{x}^c \mid \mathbf{z}^i, \mathbf{z}^c, T; \boldsymbol{\theta}) & \text{if } s = c
\end{cases}
\end{equation}

That is, when computing the marginal risk for an ipsilateral LNL ($s = i$), the summation over $\mathbf{x}^i$ is restricted to those configurations where $x^i_\ell = 1$, and the summation over $\mathbf{x}^c$ is over all possible contralateral configurations. Conversely, for a contralateral LNL ($s = c$), the restriction is applied to $\mathbf{x}^c$, while the sum over $\mathbf{x}^i$ is unrestricted. This procedure is repeated for each LNL and side as required, yielding a specific risk for every diagnostic scenario.

\subsubsection{Propagation of Parameter Uncertainty in Risk Estimates}

To account for uncertainty in model parameters, we propagate this uncertainty into the estimated marginal risks by repeating the marginalization procedure for each posterior sample $\boldsymbol{\theta}_k$ obtained from the MCMC algorithm. Specifically, for each diagnostic scenario and each posterior sample $k = 1, \ldots, N_s$, we compute the marginal risk for every LNL $\ell$ on side $s$ as:

\begin{equation}
\mathcal{R}^{(k)}_{\ell,s} =
\begin{cases}
\displaystyle
\sum_{\mathbf{x}^i : x^i_\ell = 1} \sum_{\mathbf{x}^c}
P(\mathbf{x}^i, \mathbf{x}^c \mid \mathbf{z}^i, \mathbf{z}^c, T; \boldsymbol{\theta}_k) & \text{if } s = i\\[3ex]
\displaystyle
\sum_{\mathbf{x}^i} \sum_{\mathbf{x}^c : x^c_\ell = 1}
P(\mathbf{x}^i, \mathbf{x}^c \mid \mathbf{z}^i, \mathbf{z}^c, T; \boldsymbol{\theta}_k) & \text{if } s = c
\end{cases}
\end{equation}

Here, $\mathcal{R}^{(k)}{\ell,s}$ denotes the marginal risk for LNL $\ell$ on side $s$ under the $k$-th posterior draw of the parameters. Repeating this calculation for each $k$ yields a set of risk estimates ${\mathcal{R}^{(k)}_{\ell,s}}_{k=1}^{N_s}$, which can then be summarized using the posterior mean and a credible interval for each LNL and side:

\begin{align}
    \bar{\mathcal{R}}_{\ell,s} &= \frac{1}{N_s} \sum_{k=1}^{N_s} \mathcal{R}^{(k)}_{\ell,s} \\[2ex]
    \mathrm{CI}_{\ell,s} &= \left[\mathcal{R}^{(k_\mathrm{low})}_{\ell,s},\, \mathcal{R}^{(k_\mathrm{high})}_{\ell,s}\right]
\end{align}

Where $k_\mathrm{low}$ and $k_\mathrm{high}$ correspond to the desired percentiles of the empirical distribution of $\mathcal{R}^{(k)}_{\ell,s}$ (e.g., 2.5th and 97.5th percentiles for a 95\% credible interval). This approach provides a full quantification of the parameter-driven uncertainty in the estimated risk for each lymph node level and side.

\subsubsection{Joint Risk-Based Optimization of CTV-E Protocols}

Based on the computed risks, LNLs are iteratively included in the CTV-E in order of descending marginal risk. At each step, the set of excluded LNLs is identified, and the total probability of missing disease in these spared levels is recomputed. Let $\mathcal{E}^i$ and $\mathcal{E}^c$ denote the sets of excluded ipsilateral and contralateral LNLs, respectively. The total risk of missed microscopic disease is defined as the posterior probability that \emph{at least one} excluded LNL harbors occult tumor involvement. This quantity can be computed by summing the posterior probability over all configurations $(\mathbf{x}^i, \mathbf{x}^c)$ where at least one of the excluded lymph node levels is involved:

\begin{equation}
\label{eq:total_missed_risk}
\mathcal{R}_{\mathrm{miss}} =
\sum_{\substack{(\mathbf{x}^i, \mathbf{x}^c) \text{ s.t. }\\
\exists \ell \in \mathcal{E}^i : x^i_\ell = 1 \;\;\text{or}\;\; \exists \ell \in \mathcal{E}^c : x^c_\ell = 1}}
P(\mathbf{x}^i, \mathbf{x}^c \mid \mathbf{z}^i, \mathbf{z}^c, T; \boldsymbol{\theta})
\end{equation}

To explicitly account for parameter uncertainty, the upper bound of the credible interval for the missed risk, $\mathcal{R}_\mathrm{missed}^{(\mathrm{high})}$, is compared to a user-specified risk threshold $\tau$ (e.g., 5\%) after each step:

\begin{equation}
\mathcal{R}_\mathrm{miss}^{(\mathrm{high})} < \tau.
\end{equation}

This process is repeated until the stopping criterion is satisfied, ensuring that the probability of missing occult disease in any excluded LNL remains within the acceptable risk limit for all plausible parameter values captured by the credible interval. This approach guarantees that the defined CTV-E protocol is robust to parameter uncertainty and that the prescribed risk limit is not exceeded due to model variability.

For every clinically relevant diagnostic scenario, that is, for each possible combination of observed LN involvement, the joint risk minimization algorithm is systematically applied. For each scenario, the algorithm outputs a protocol specifying which LNLs should be included in the CTV-E, as well as the resulting joint residual risk and its credible interval. This protocol can be translated into a clinical lookup table, that allows for individualized, risk-adaptive CTV-E definitions that transparently reflect the modeled posterior uncertainty for any patient presentation.

\section{Application on Cases of Oropharyngeal Cancer}

In this section, we present four representative patient cases to illustrate how the model-based CTV-E definition compares with conventional guidelines. Given the $2^{8}$ possible nodal involvement scenarios, we focus on the most clinically relevant and commonly encountered patterns in oropharyngeal cancer. Importantly, no single risk threshold is universally optimal; its selection may depend on institutional policy, clinical philosophy, or patient preference. For transparency and comparability, we report model-based CTV-E recommendations at multiple risk thresholds (2\%, 5\%, 8\%, 10\%, 12\%, 15\%, and 20\%), providing side-by-side comparisons for each patient example to show how varying the risk tolerance affects the extent of irradiation and the residual risk of occult nodal involvement. Extended listings of model-based treatment recommendations for all possible nodal constellations and thresholds are provided in appendix \ref{app:proto_early_2}-\ref{app:proto_advanced_20}. 

\subsection{Patient 1: N0, Early-stage (T1-T2)}

\begin{figure}[H]
\centering
\scriptsize

\begin{subfigure}{0.48\linewidth}
    \centering
    \caption{Current Clinical Practice (DAHANCA)}
    \label{fig:Treatment_Protocols_N0_Clinical}
    \begin{minipage}{0.48\linewidth}
        \includegraphics[width=\linewidth]{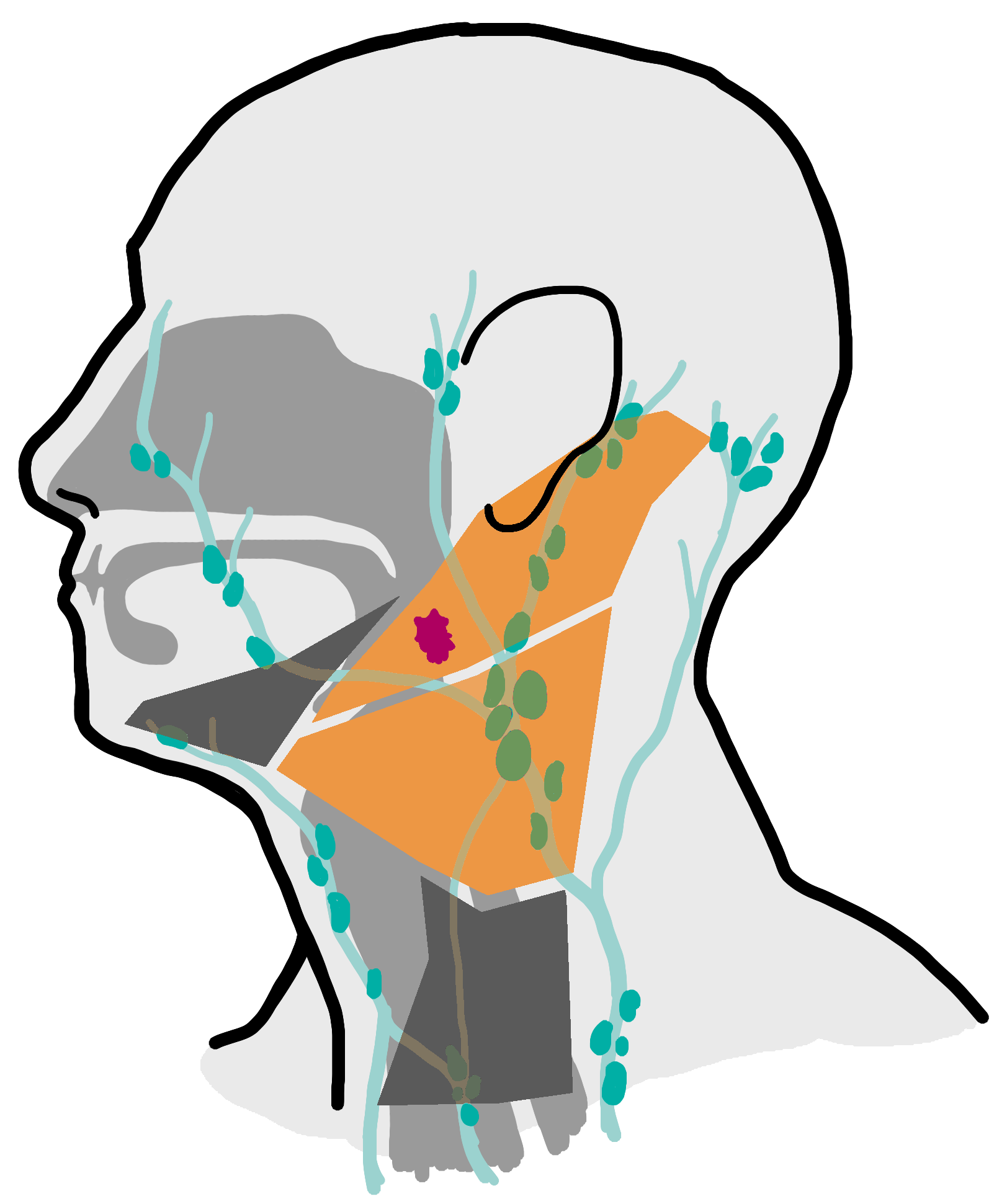}
    \end{minipage}%
    \begin{minipage}{0.48\linewidth}
        \includegraphics[width=\linewidth]{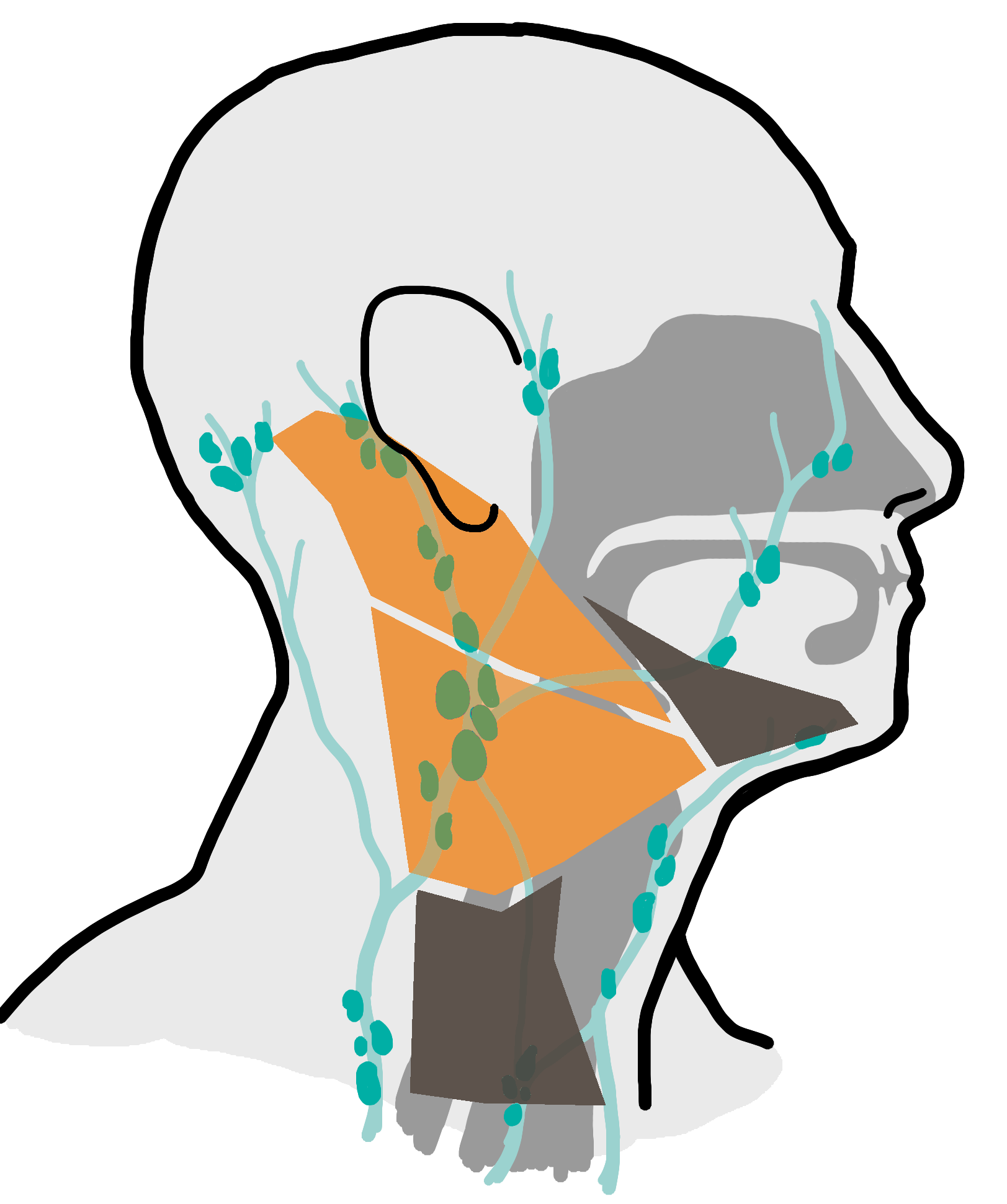}
    \end{minipage}
\end{subfigure}
\hfill
\begin{subfigure}{0.48\linewidth}
    \centering
    \caption{Model-based for 2\% Risk Threshold}
    \label{fig:Treatment_Protocols_N0_2}
    \begin{minipage}{0.48\linewidth}
        \includegraphics[width=\linewidth]{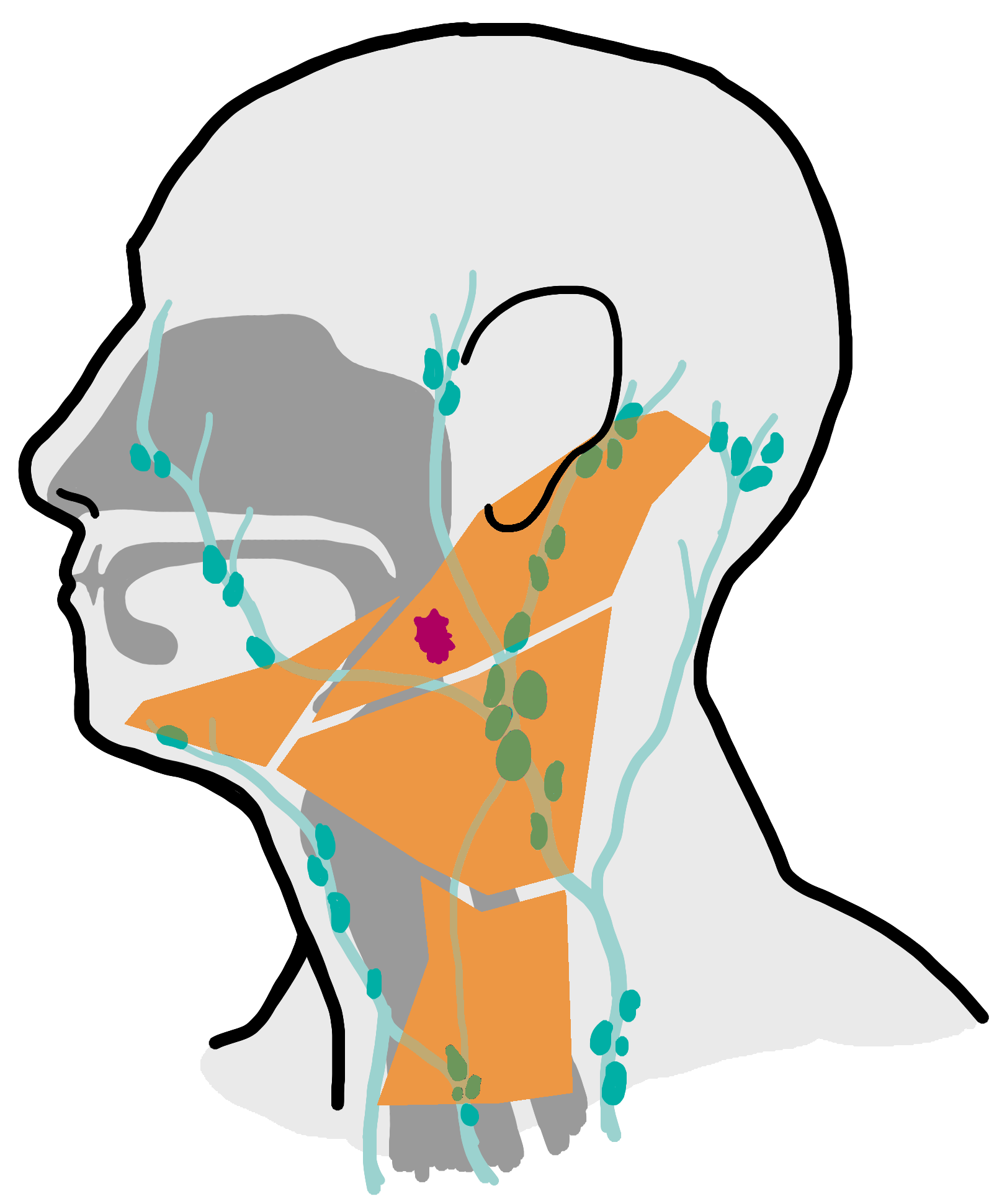}
    \end{minipage}%
    \begin{minipage}{0.48\linewidth}
        \includegraphics[width=\linewidth]{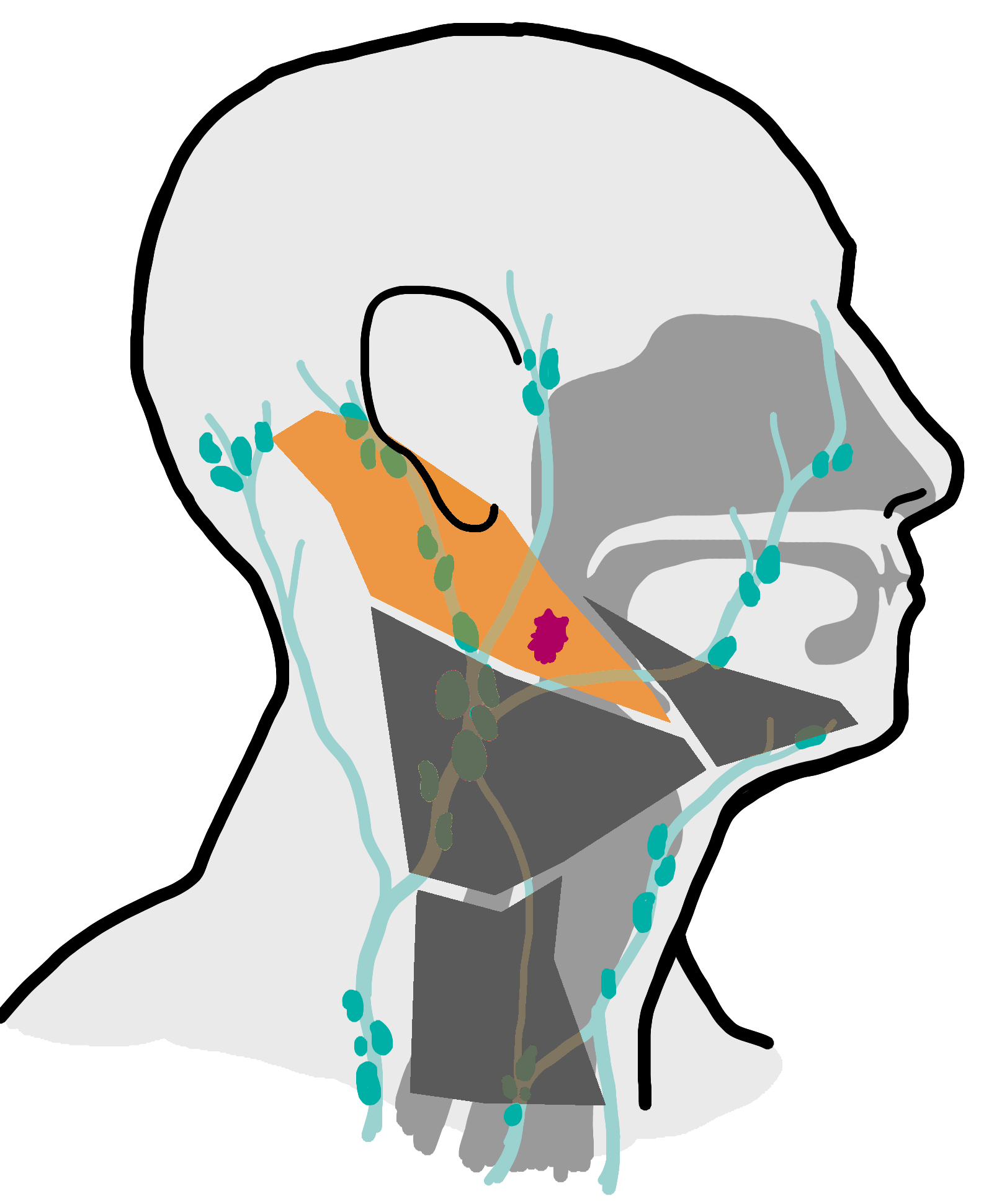}
    \end{minipage}

\end{subfigure}

\vspace{1mm}

\begin{subfigure}{0.48\linewidth}
    \centering
    \caption{Model-based for 5\% Risk Threshold}
    \label{fig:Treatment_Protocols_N0_5}
    \begin{minipage}{0.48\linewidth}
        \includegraphics[width=\linewidth]{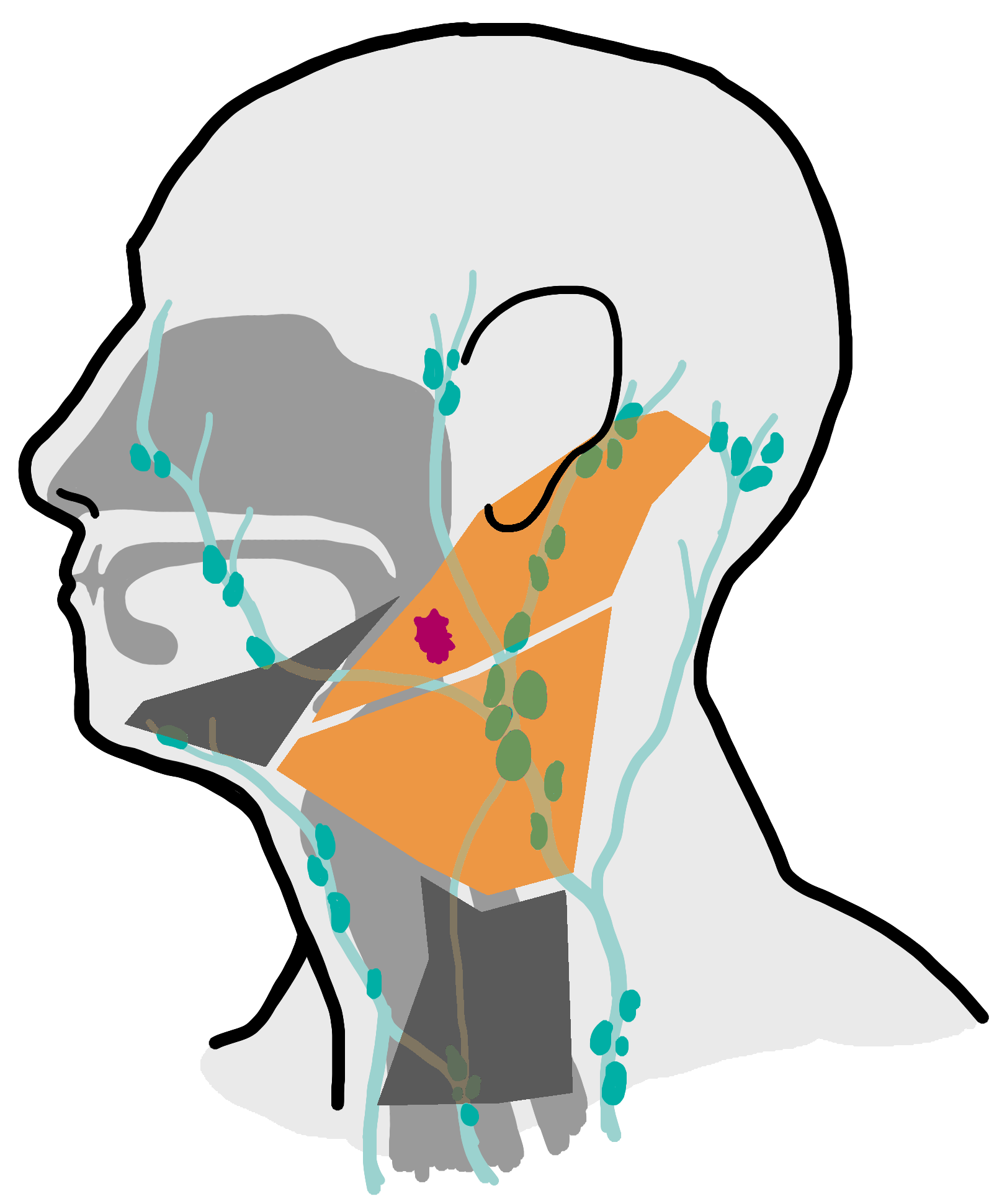}
    \end{minipage}%
    \begin{minipage}{0.48\linewidth}
        \includegraphics[width=\linewidth]{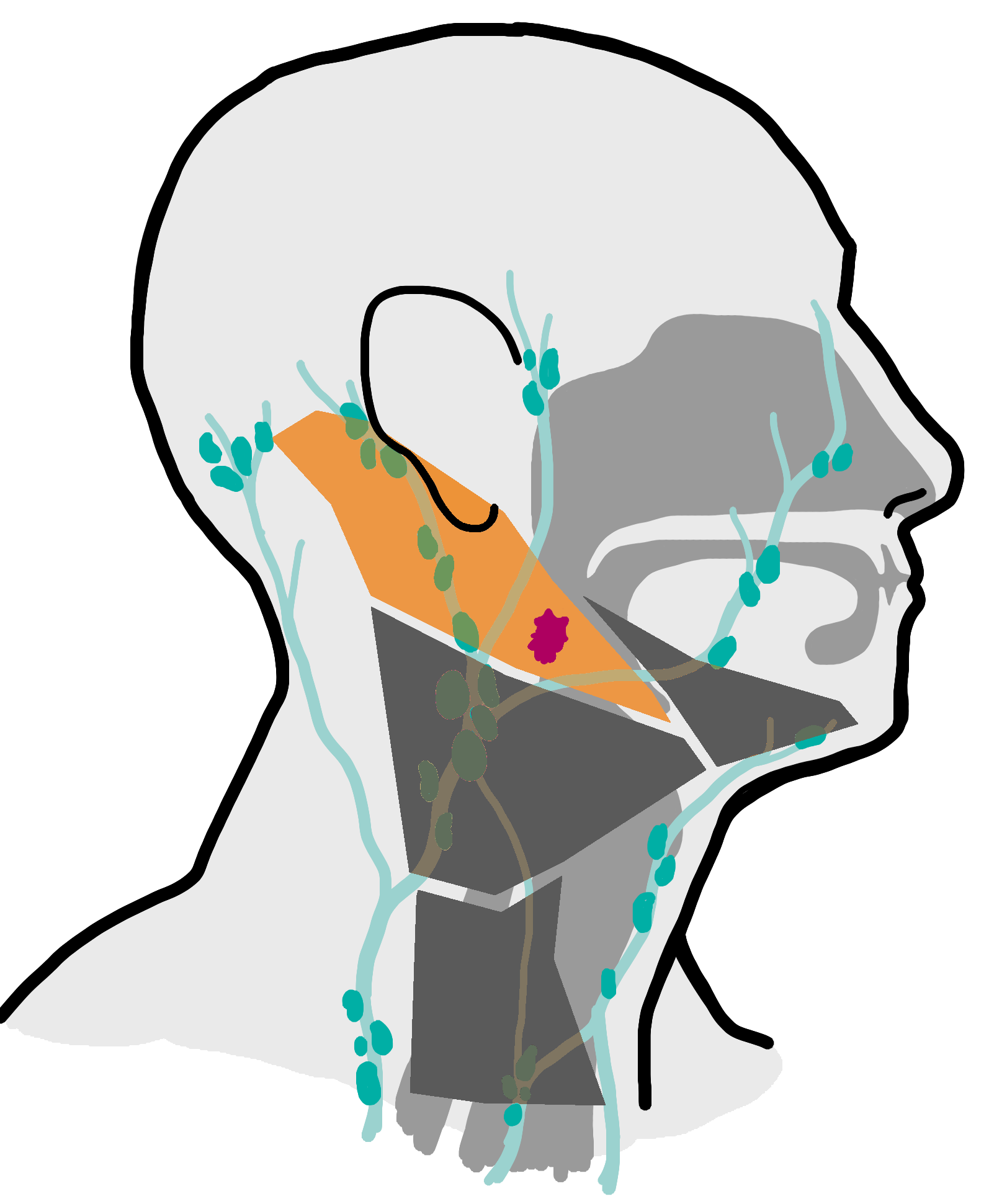}
    \end{minipage}
\end{subfigure}
\hfill
\begin{subfigure}{0.48\linewidth}
    \centering
    \caption{Model-based for 8\% Risk Threshold}
    \label{fig:Treatment_Protocols_N0_8}
    \begin{minipage}{0.48\linewidth}
        \includegraphics[width=\linewidth]{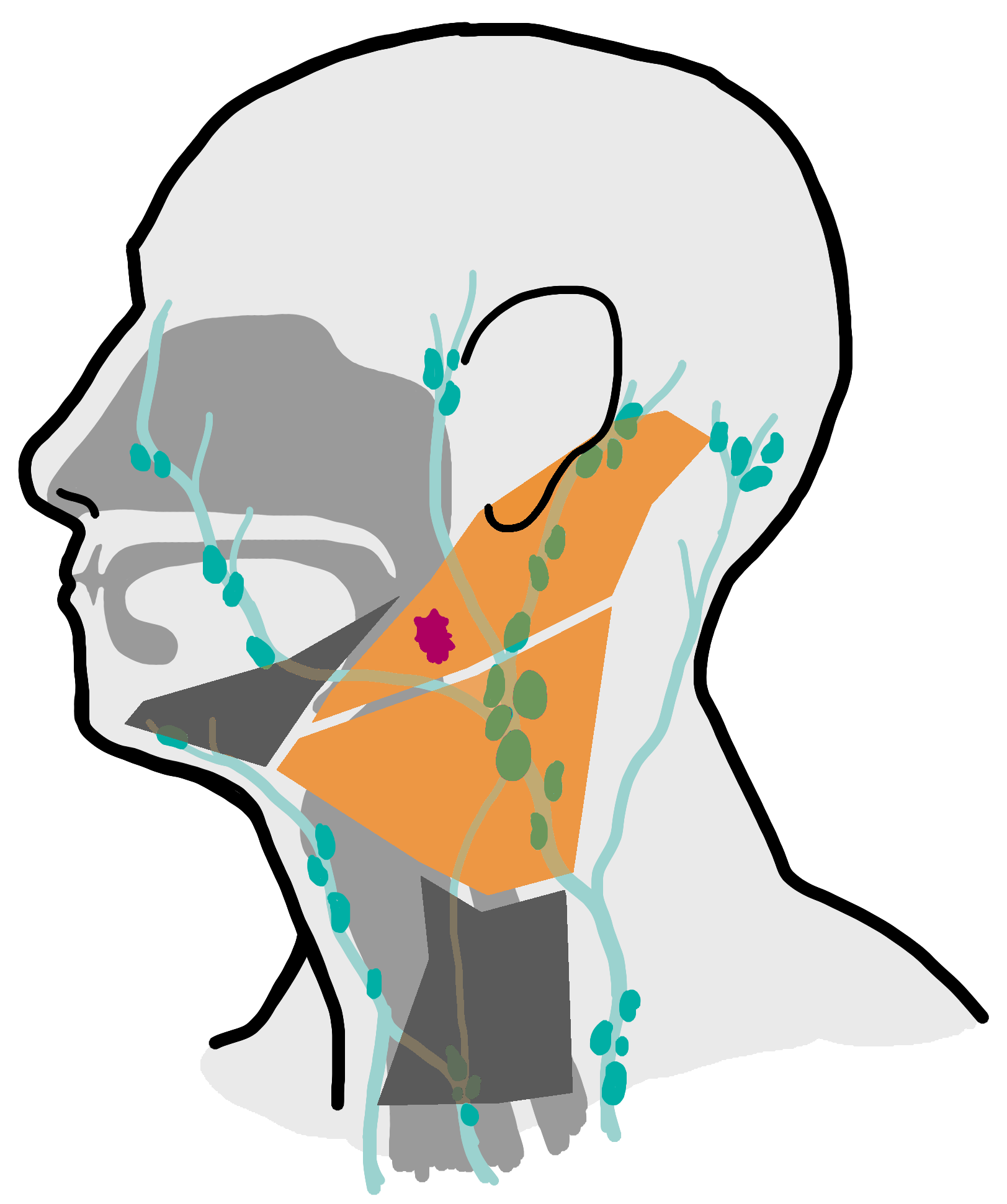}
    \end{minipage}%
    \begin{minipage}{0.48\linewidth}
        \includegraphics[width=\linewidth]{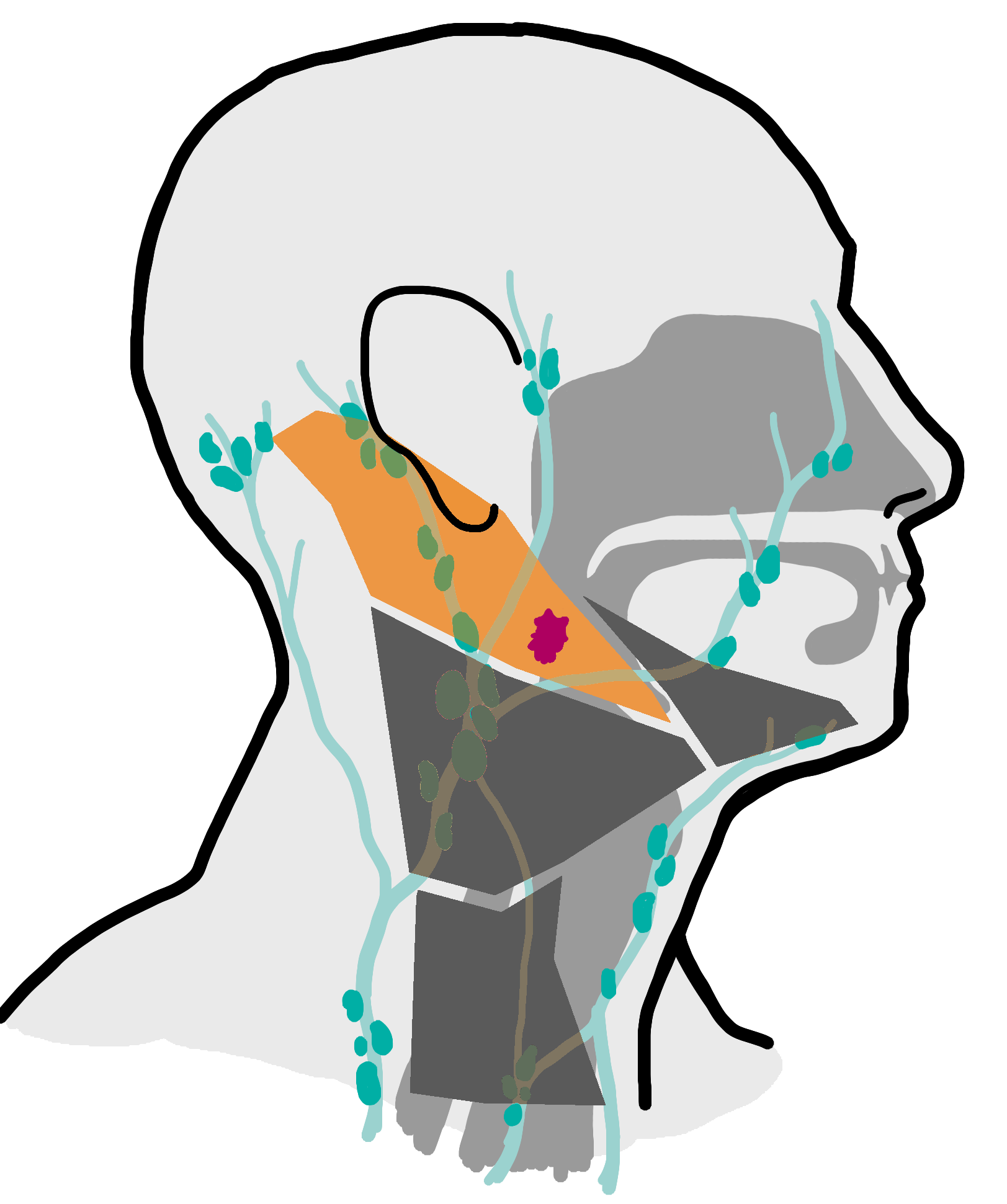}
    \end{minipage}
    
\end{subfigure}

\vspace{1mm}

\begin{subfigure}{0.48\linewidth}
    \centering
    \caption{Model-based for 10\% Risk Threshold}
    \label{fig:Treatment_Protocols_N0_10}
    \begin{minipage}{0.48\linewidth}
        \includegraphics[width=\linewidth]{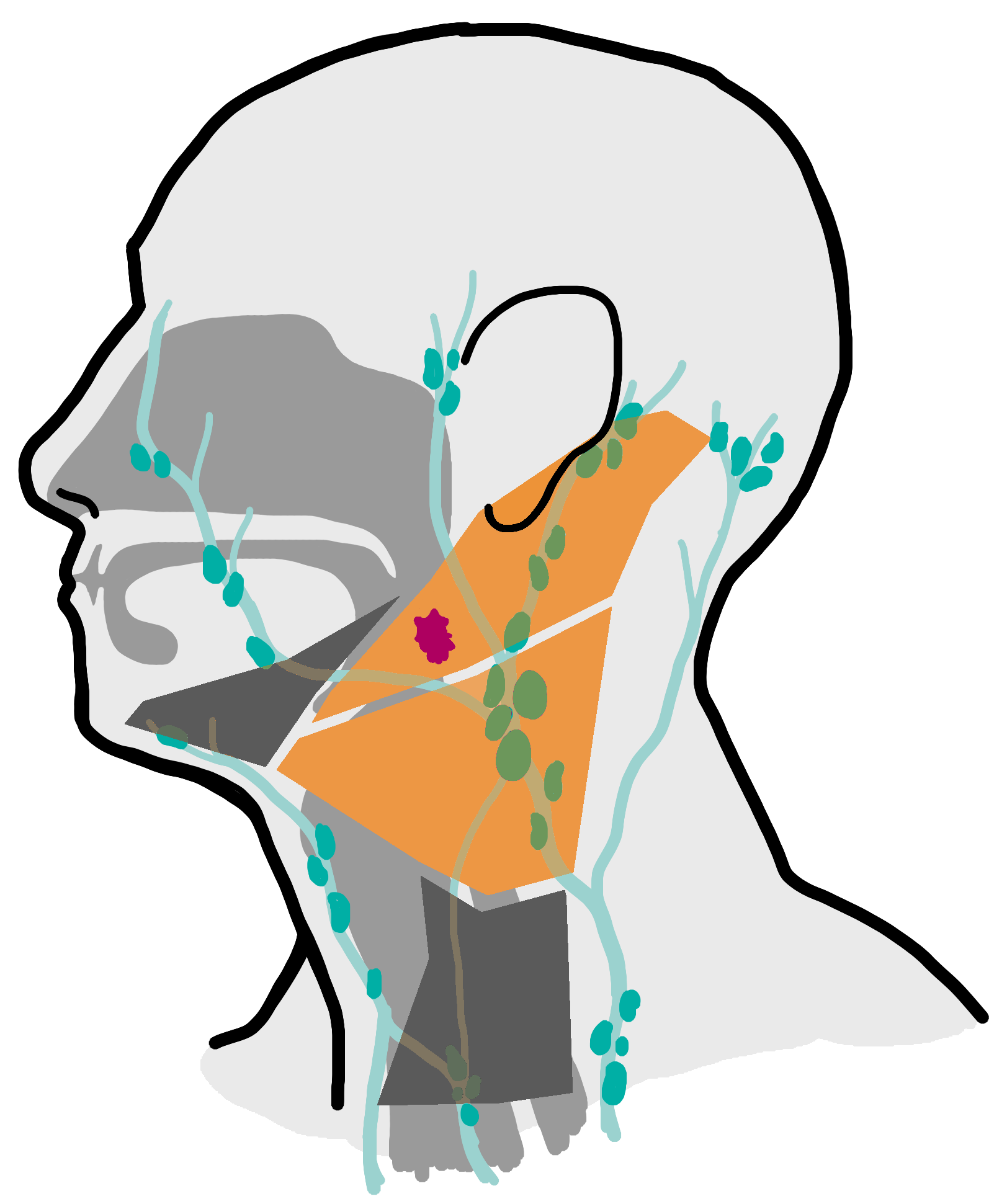}
    \end{minipage}%
    \begin{minipage}{0.48\linewidth}
        \includegraphics[width=\linewidth]{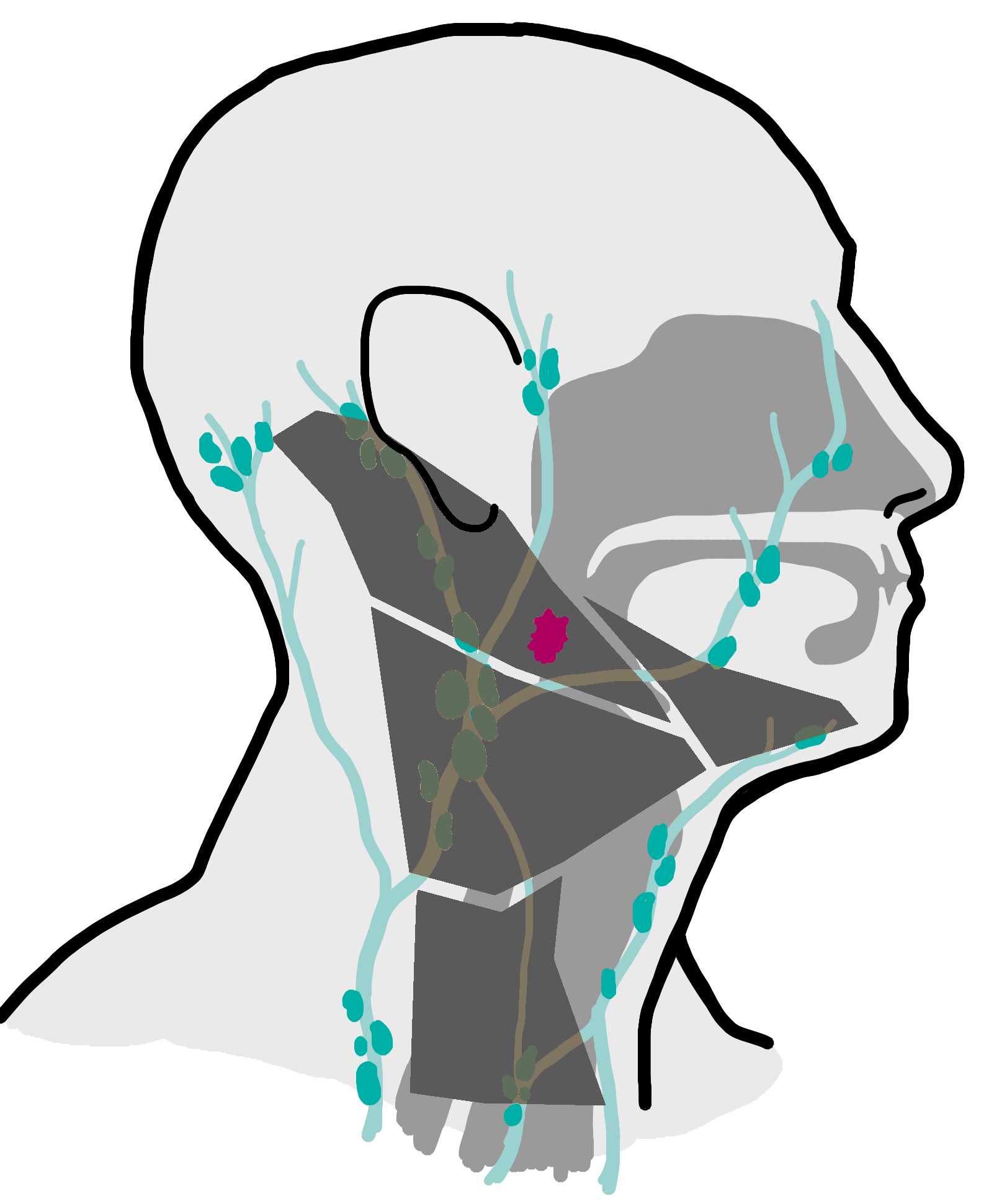}
    \end{minipage}
    
\end{subfigure}
\hfill
\begin{subfigure}{0.48\linewidth}
    \centering
    \caption{Model-based for 12\% Risk Threshold}
    \label{fig:Treatment_Protocols_N0_12}
    \begin{minipage}{0.48\linewidth}
        \includegraphics[width=\linewidth]{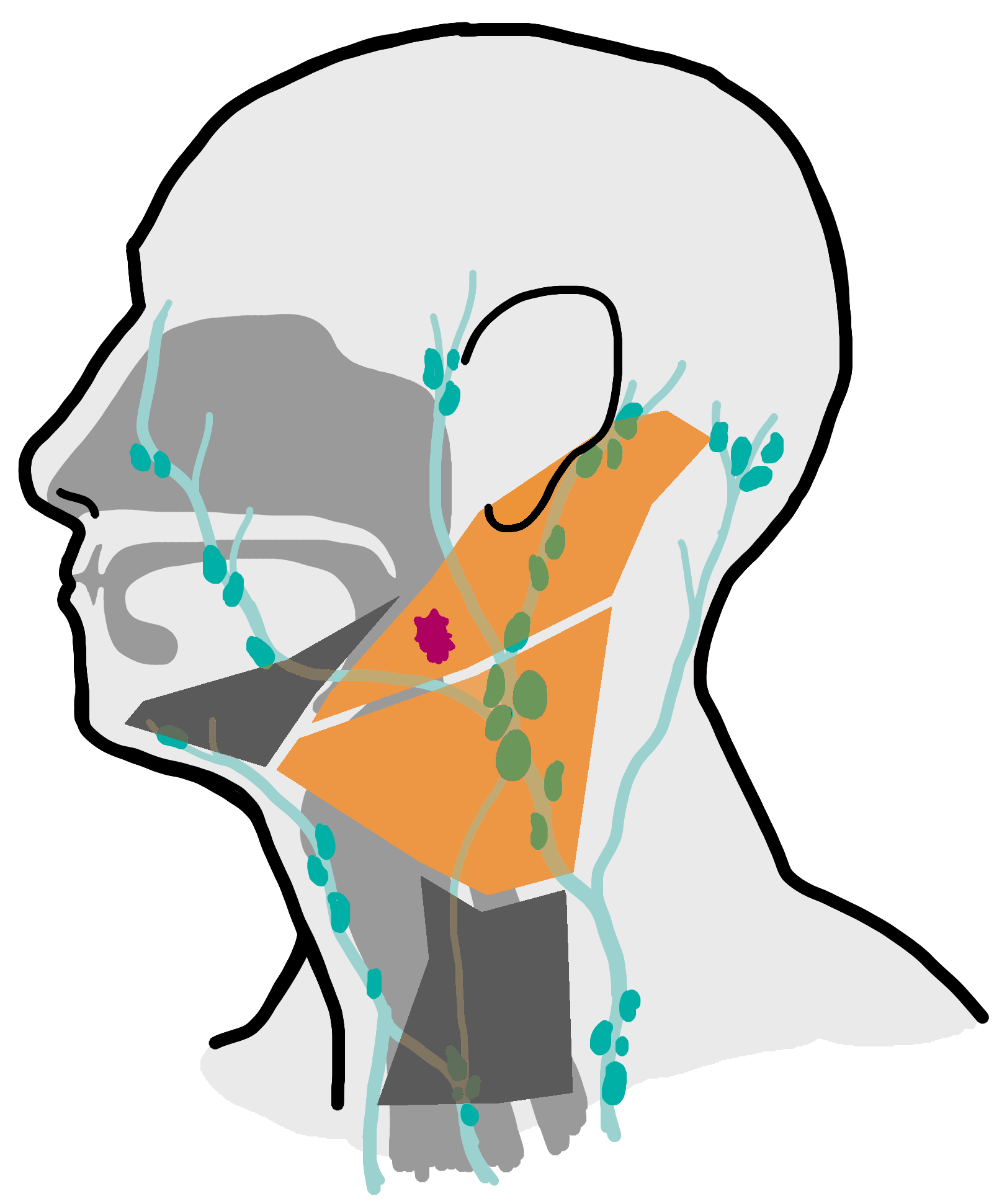}
    \end{minipage}%
    \begin{minipage}{0.48\linewidth}
        \includegraphics[width=\linewidth]{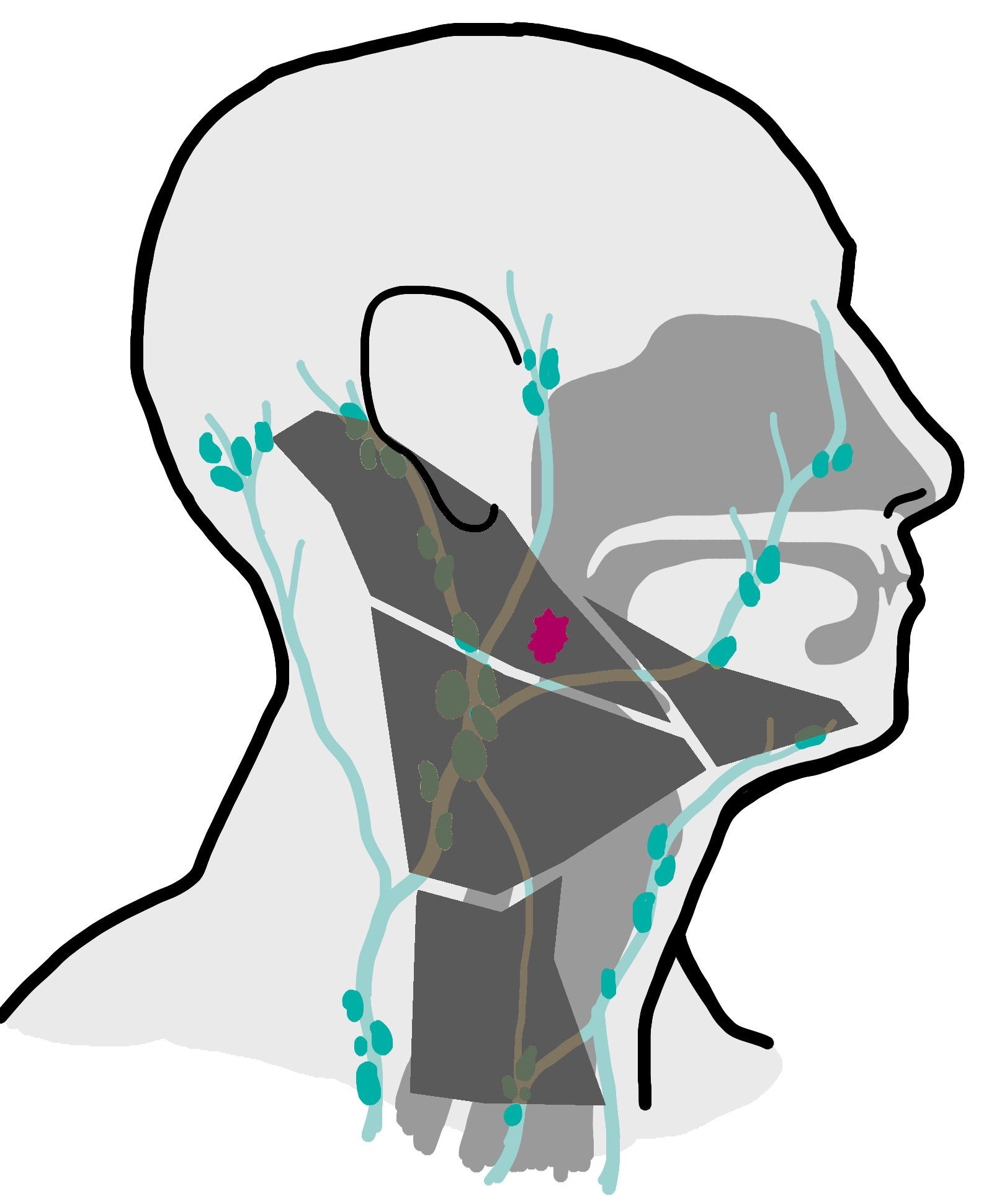}
    \end{minipage}
    
\end{subfigure}

\vspace{1mm}

\begin{subfigure}{0.48\linewidth}
    \centering
    \caption{Model-based for 15\% Risk Threshold}
    \label{fig:Treatment_Protocols_N0_15}
    \begin{minipage}{0.48\linewidth}
        \includegraphics[width=\linewidth]{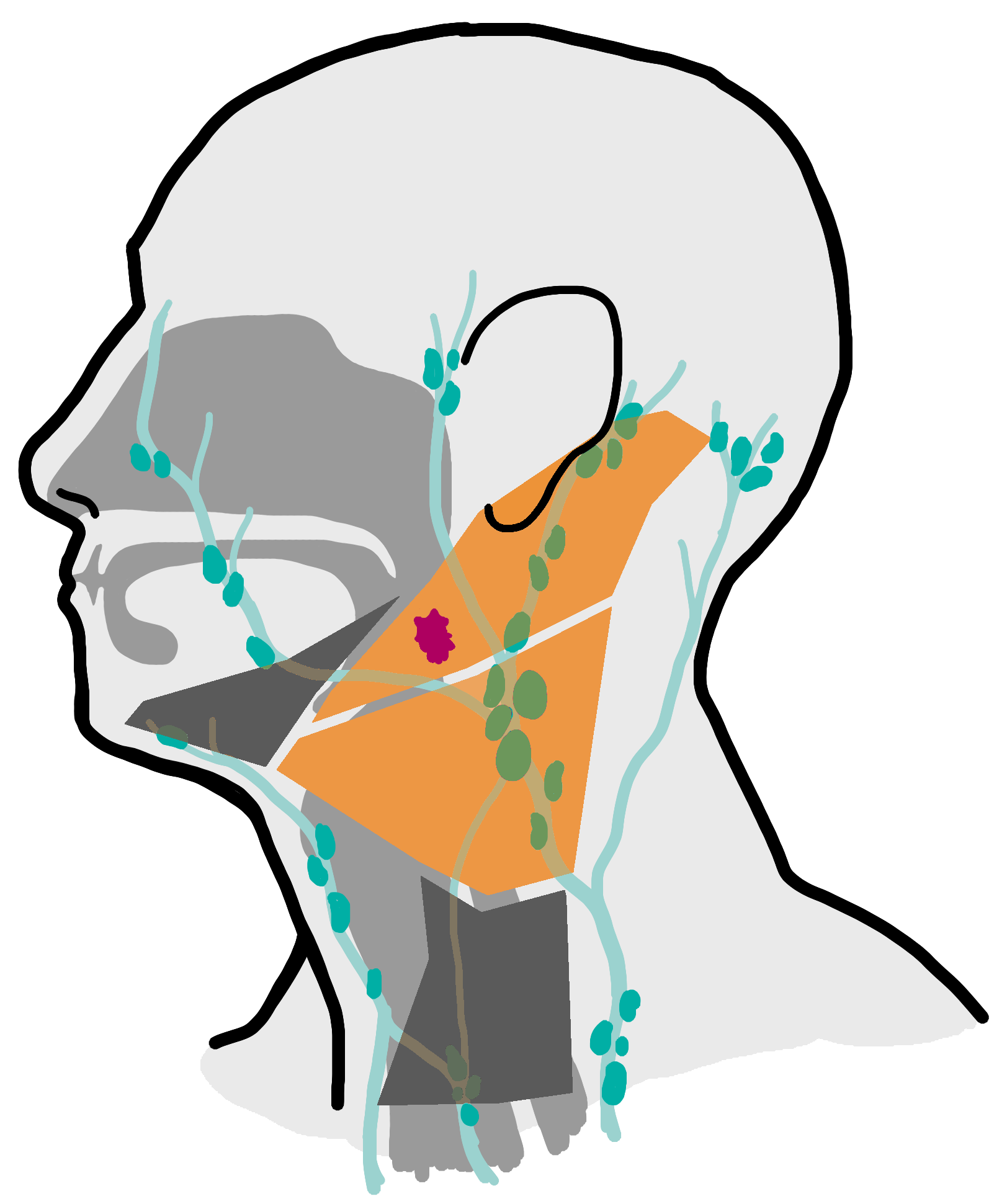}
    \end{minipage}%
    \begin{minipage}{0.48\linewidth}
        \includegraphics[width=\linewidth]{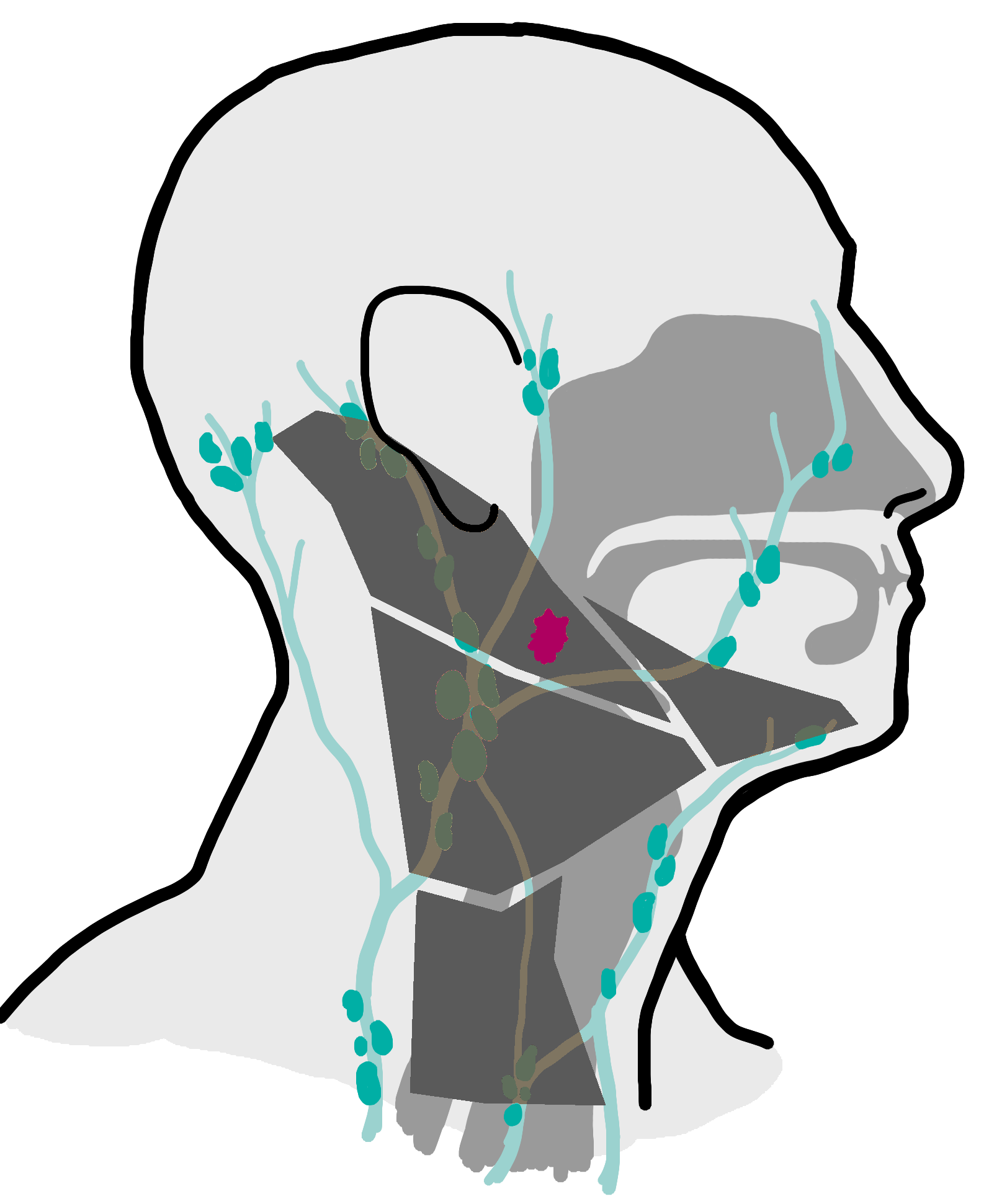}
    \end{minipage}
    
\end{subfigure}
\hfill
\begin{subfigure}{0.48\linewidth}
    \centering
    \caption{Model-based for 20\% Risk Threshold}
    \label{fig:Treatment_Protocols_N0_20}
    \begin{minipage}{0.48\linewidth}
        \includegraphics[width=\linewidth]{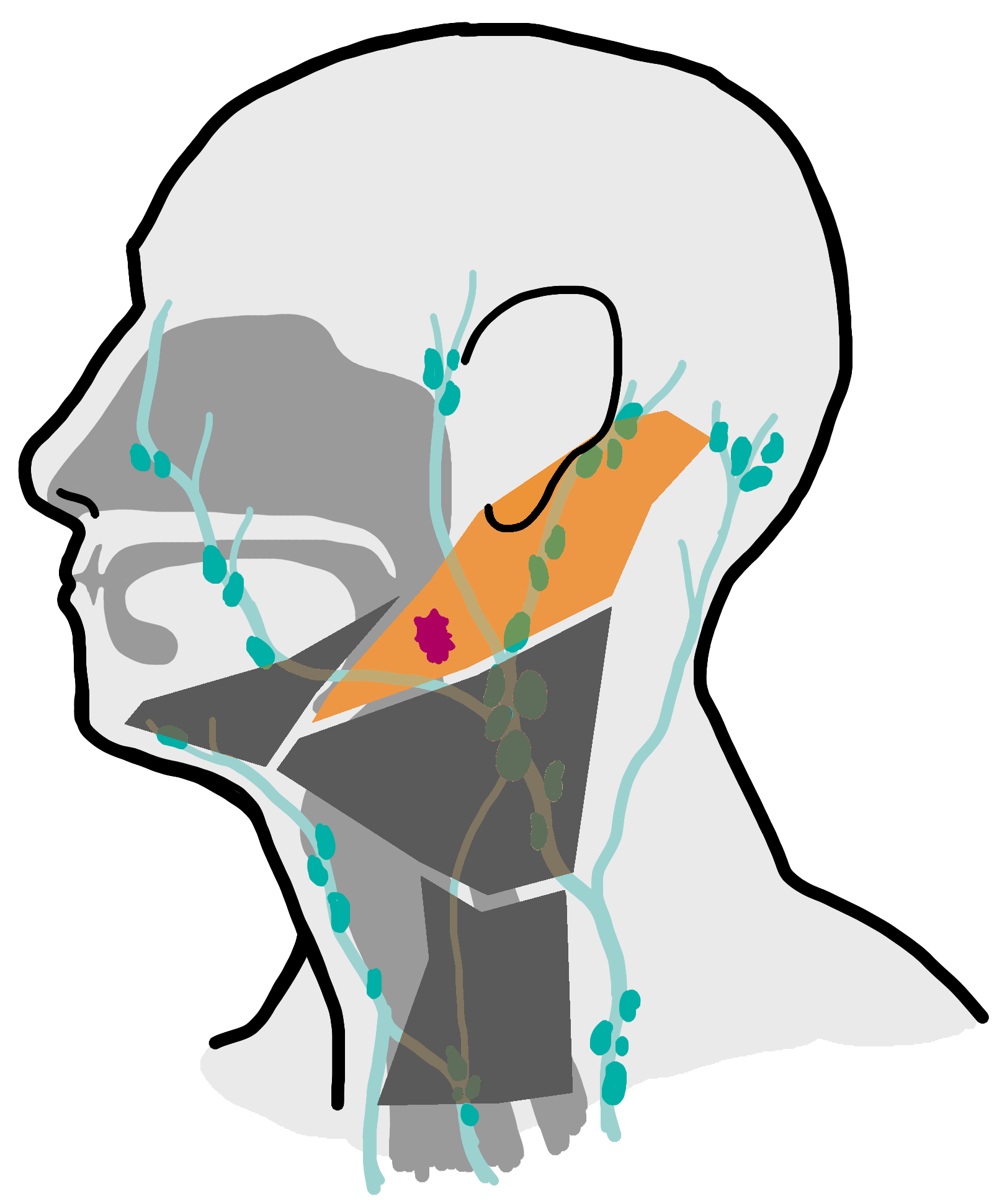}
    \end{minipage}%
    \begin{minipage}{0.48\linewidth}
        \includegraphics[width=\linewidth]{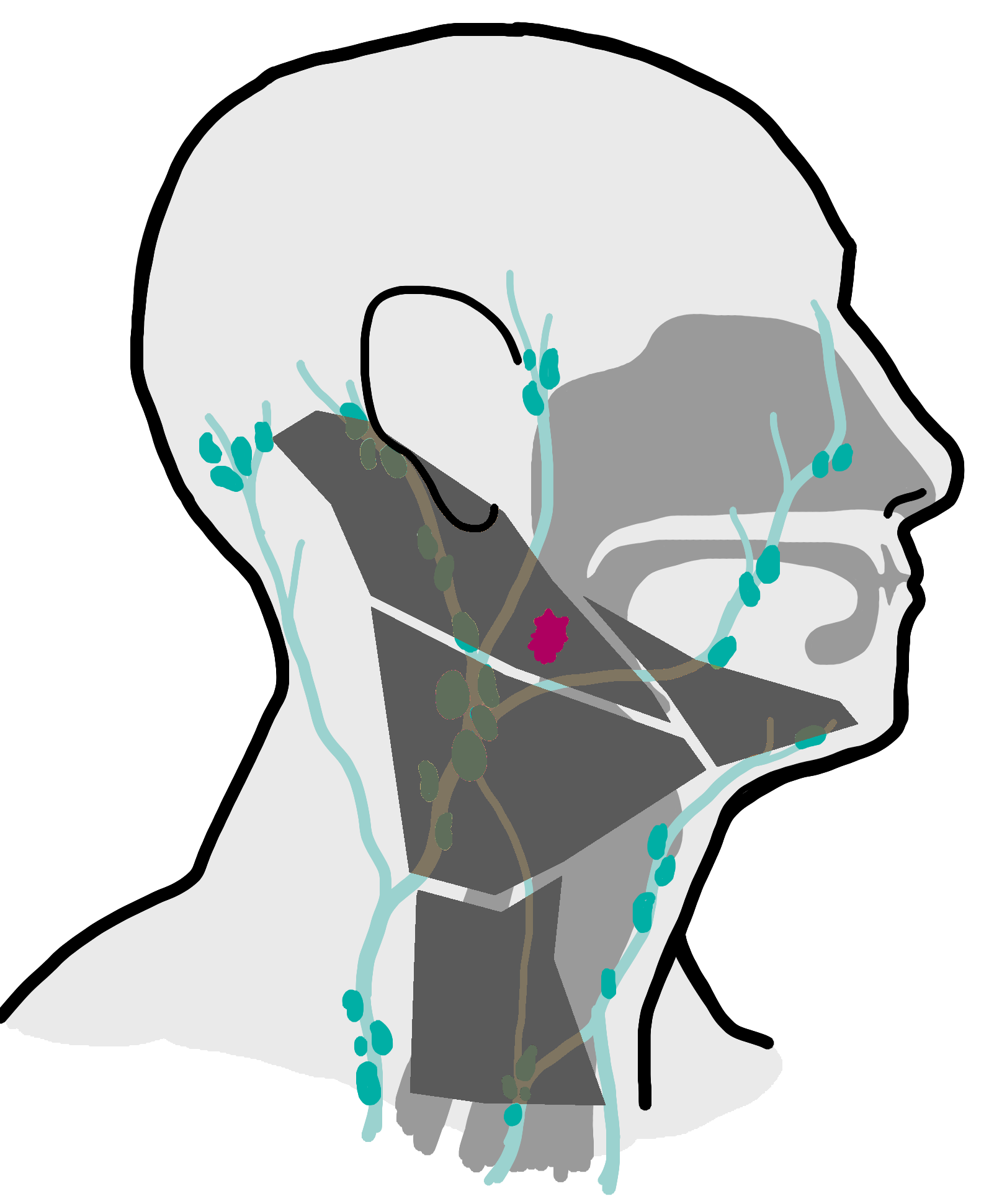}
    \end{minipage}
    
\end{subfigure}

\caption{Estimated elective nodal irradiation strategies for a patient with an early-stage central oropharyngeal tumour and N0 disease under varying risk thresholds. LNLs above the inclusion threshold are shown in orange (included into the CTV-E), while LNLs below the threshold are shown in gray (excluded from treatment)}
\label{fig:Treatment_Protocols_N0}
\end{figure}

Figure \ref{fig:Treatment_Protocols_N0} visualizes the CTV-E for the standard guideline-based approach as well as for a range of thresholds for the model-based, risk-adaptive strategy. In line with current clinical practice, elective irradiation for the guideline-based approach typically (for a Danish patient) includes bilateral LNLs II and III (figure \ref{fig:Treatment_Protocols_N0_Clinical}).

For a conservative risk threshold of 2\% (figure \ref{fig:Treatment_Protocols_N0_2}), the model recommends including ipsilateral LNLs I–IV and contralateral LNL II to the CTV-E. This, in turn, yields an estimated probability of undetected nodal involvement in any of the unirradiated LNLs of only 0.8\%. However, this comes at the cost of expanding the irradiated volume beyond what is recommended by standard clinical guidelines. With a moderate risk threshold of 5\% (figure \ref{fig:Treatment_Protocols_N0_5}), the model suggests limiting elective irradiation to ipsilateral LNLs II–III and contralateral LNL II, closely matching standard clinical protocols. This results in a residual risk of 2.9\% for undetected involvement in the omitted LNLs. If the risk threshold is increased to 10\% (figure \ref{fig:Treatment_Protocols_N0_10}), elective irradiation can be limited to the ipsilateral side, including only ipsilateral LNLs II and III while omitting all contralateral LNLs. This results in a residual risk of 6.8\% in the omitted LNLs.

As an extreme comparison, omitting all elective nodal levels and restricting irradiation to only high- and intermediate-risk volumes increases the residual risk to 51.6\%, a finding that cannot be generalized beyond this case (as it reflects a scenario without any clinically involved lymph nodes) but highlights the necessity of some degree of elective coverage. The corresponding treatment strategies for the remaining risk thresholds are visualized in figures \ref{fig:Treatment_Protocols_N0_2}-\ref{fig:Treatment_Protocols_N0_20}.


\subsection{Patient 2: N+ Ipsilateral LNL II, Early-stage (T1-T2)}

\begin{figure}[H]
\centering
\scriptsize

\begin{subfigure}{0.45\linewidth}
    \centering
    \caption{Current Clinical Practice (DAHANCA)}
    \label{fig:Treatment_Protocols_N+_II_Clinical}
    \begin{minipage}{0.48\linewidth}
        \includegraphics[width=\linewidth]{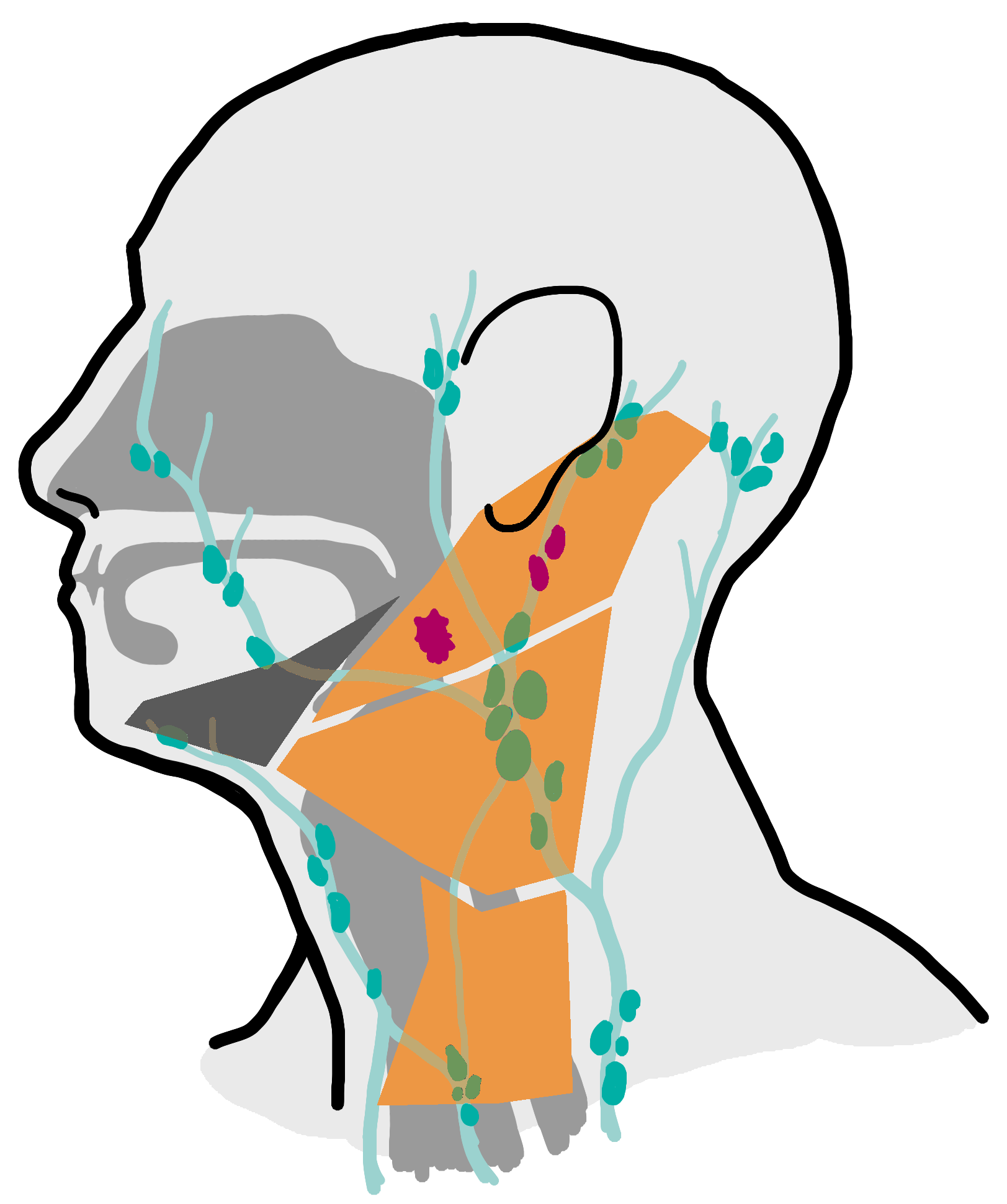}
    \end{minipage}%
    \begin{minipage}{0.48\linewidth}
        \includegraphics[width=\linewidth]{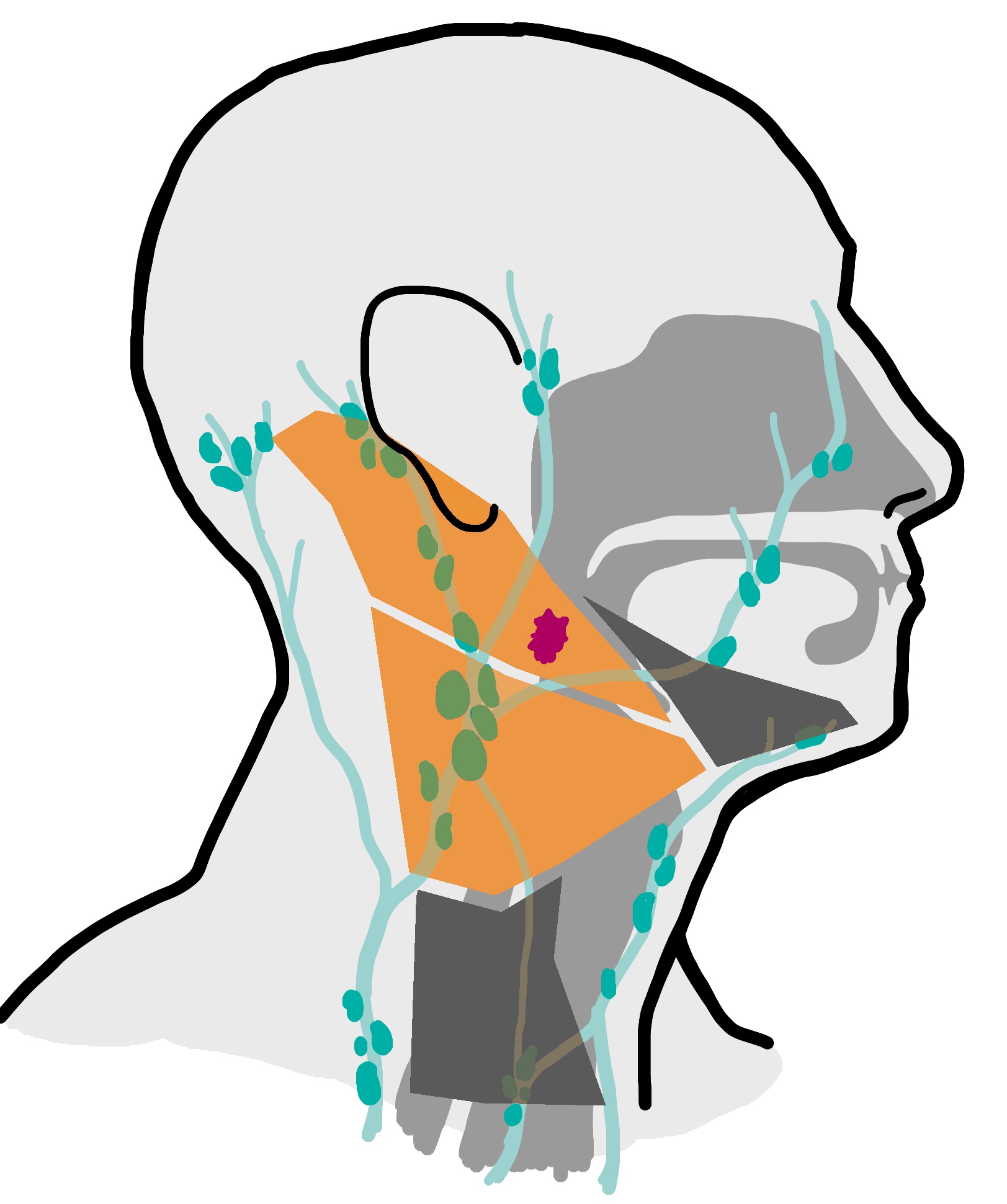}
    \end{minipage}
    
\end{subfigure}
\hfill
\begin{subfigure}{0.45\linewidth}
    \centering
    \caption{Model-based for 2\% Risk Threshold}
    \label{fig:Treatment_Protocols_N+_II_2}
    \begin{minipage}{0.48\linewidth}
        \includegraphics[width=\linewidth]{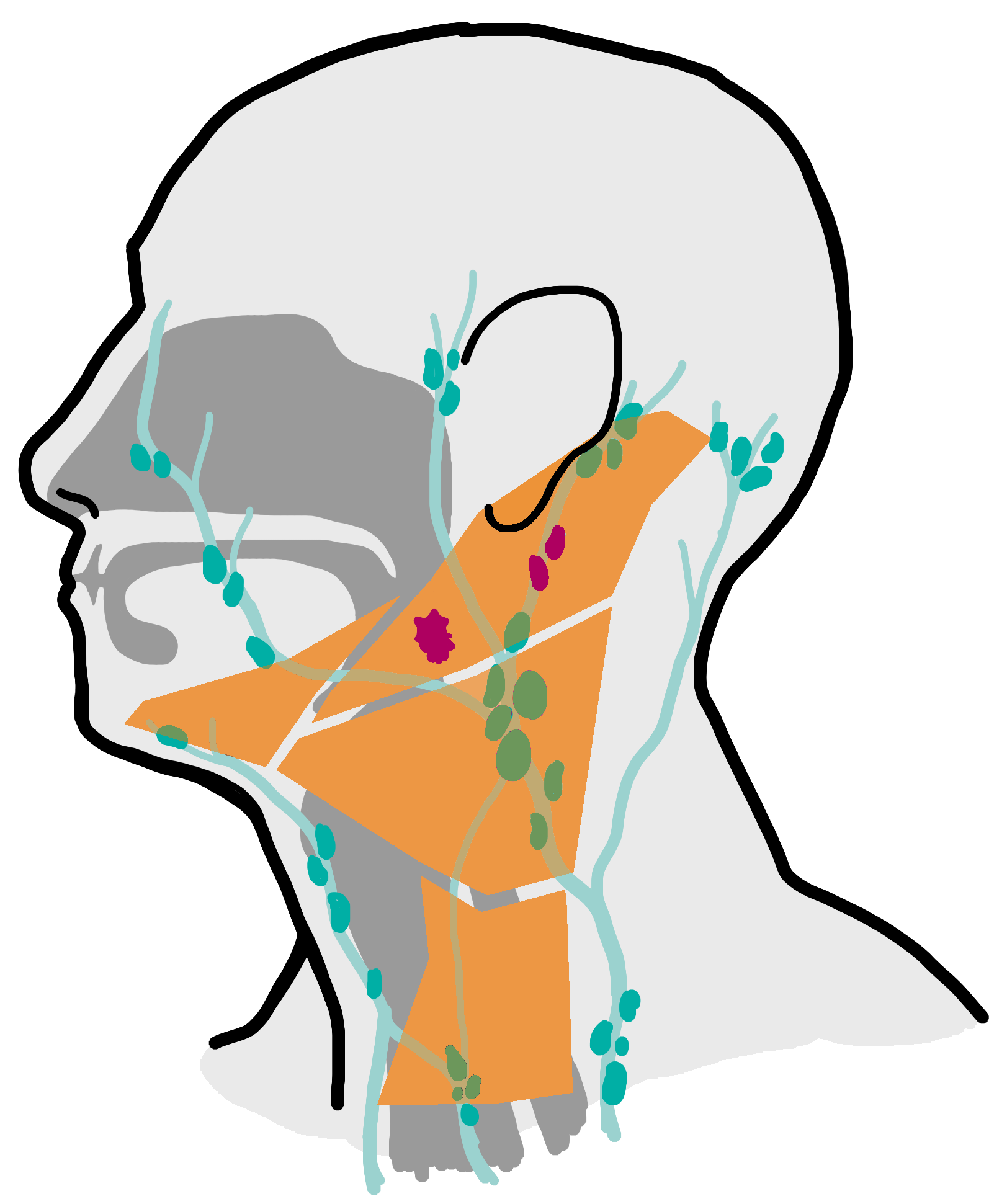}
    \end{minipage}%
    \begin{minipage}{0.48\linewidth}
        \includegraphics[width=\linewidth]{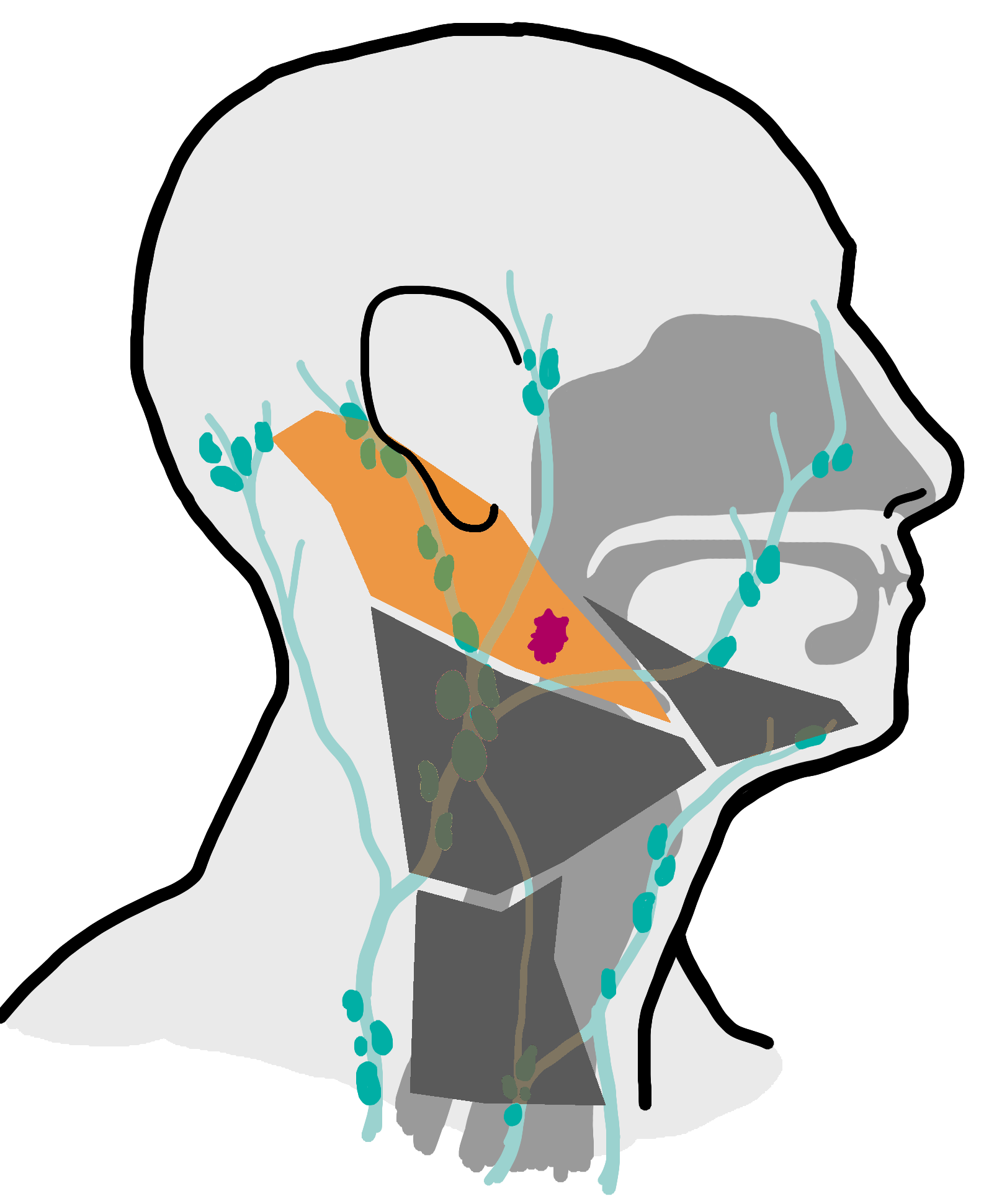}
    \end{minipage}
    
\end{subfigure}

\vspace{1mm}

\begin{subfigure}{0.45\linewidth}
    \centering
    \caption{Model-based for 5\% Risk Threshold}
    \label{fig:Treatment_Protocols_N+_II_5}
    \begin{minipage}{0.48\linewidth}
        \includegraphics[width=\linewidth]{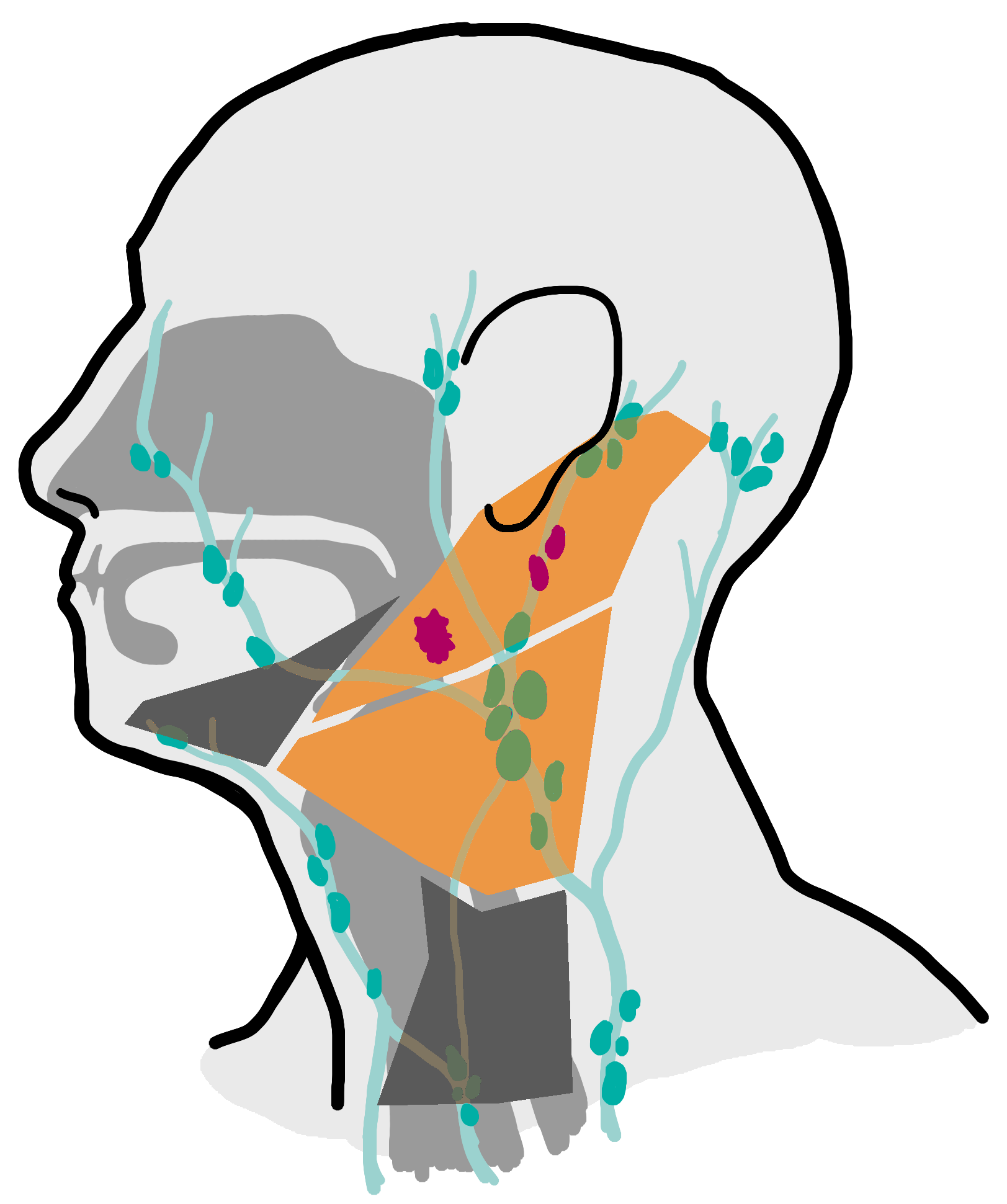}
    \end{minipage}%
    \begin{minipage}{0.48\linewidth}
        \includegraphics[width=\linewidth]{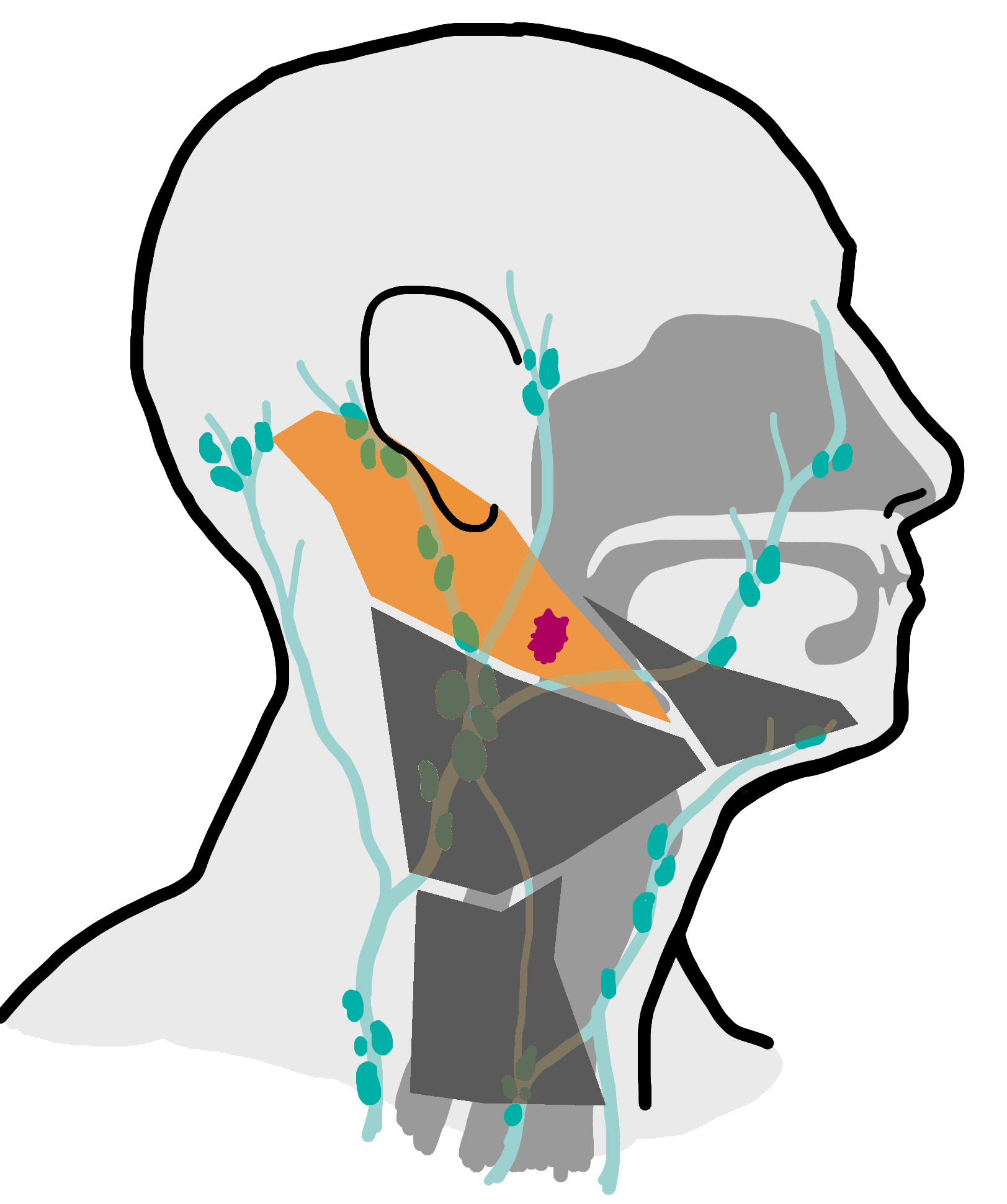}
    \end{minipage}
    
\end{subfigure}
\hfill
\begin{subfigure}{0.45\linewidth}
    \centering
    \caption{Model-based for 8\% Risk Threshold}
    \label{fig:Treatment_Protocols_N+_II_8}
    \begin{minipage}{0.48\linewidth}
        \includegraphics[width=\linewidth]{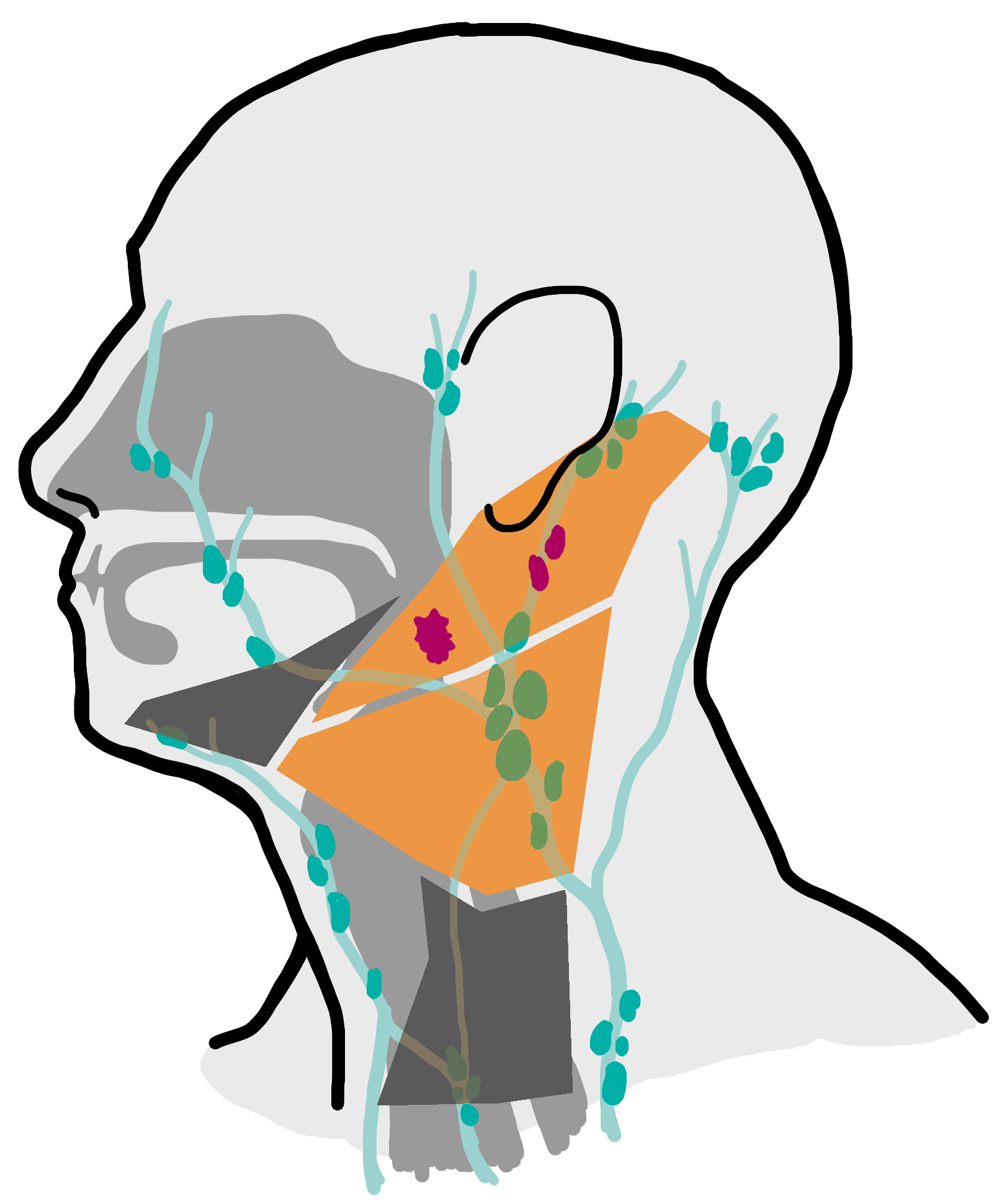}
    \end{minipage}%
    \begin{minipage}{0.48\linewidth}
        \includegraphics[width=\linewidth]{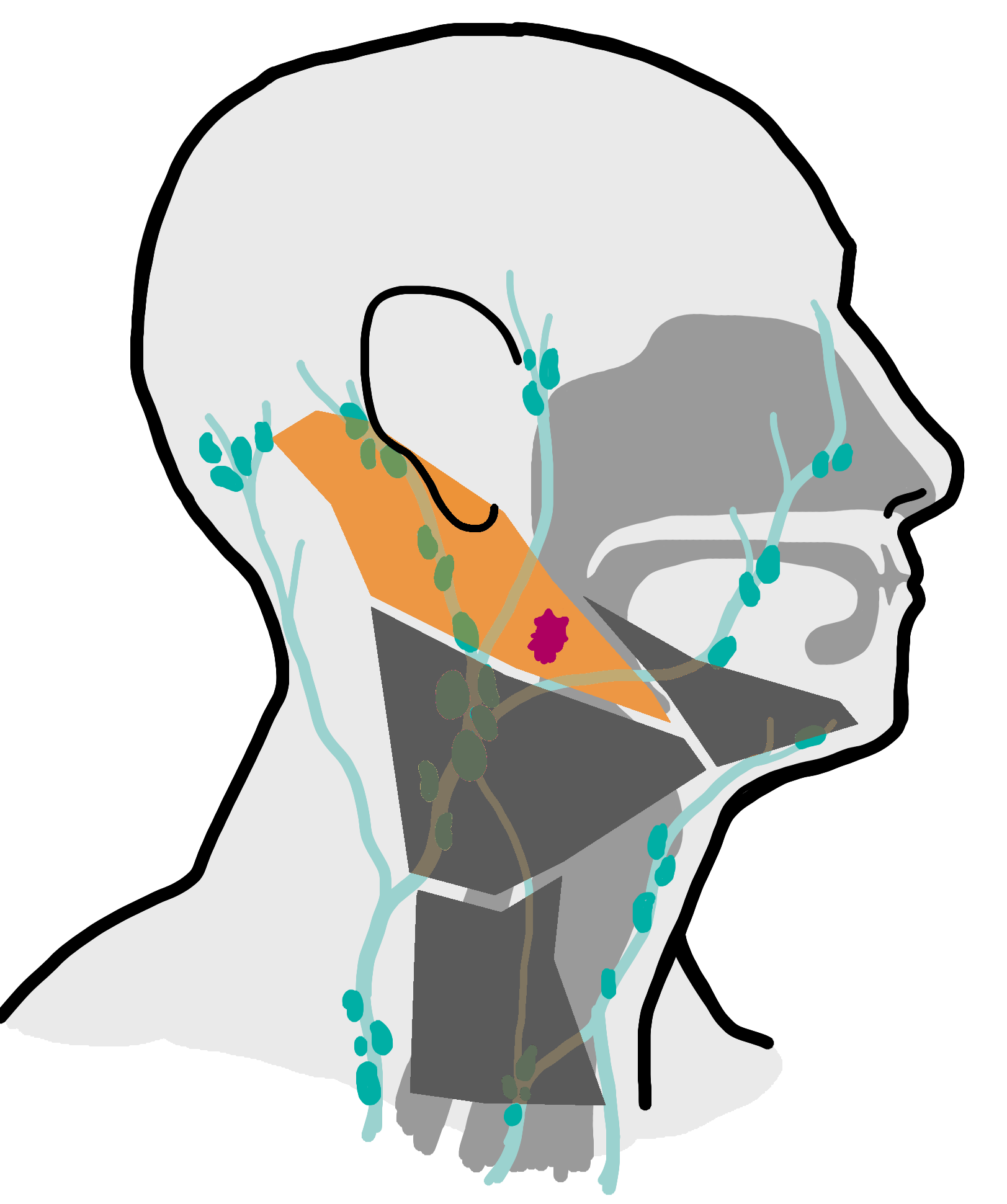}
    \end{minipage}
    
\end{subfigure}

\vspace{1mm}

\begin{subfigure}{0.45\linewidth}
    \centering
    \caption{Model-based for 10\% Risk Threshold}
    \label{fig:Treatment_Protocols_N+_II_10}
    \begin{minipage}{0.48\linewidth}
        \includegraphics[width=\linewidth]{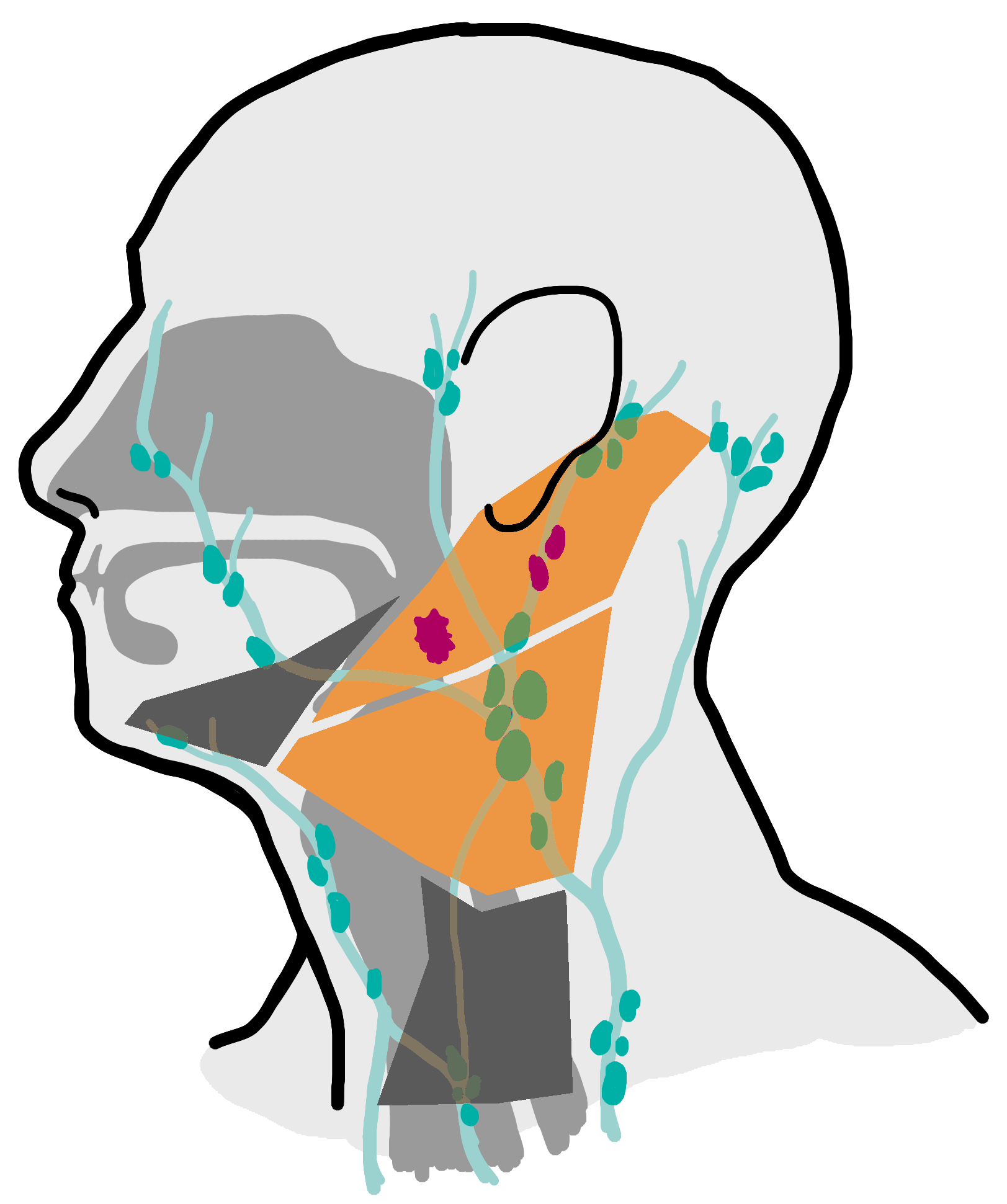}
    \end{minipage}%
    \begin{minipage}{0.48\linewidth}
        \includegraphics[width=\linewidth]{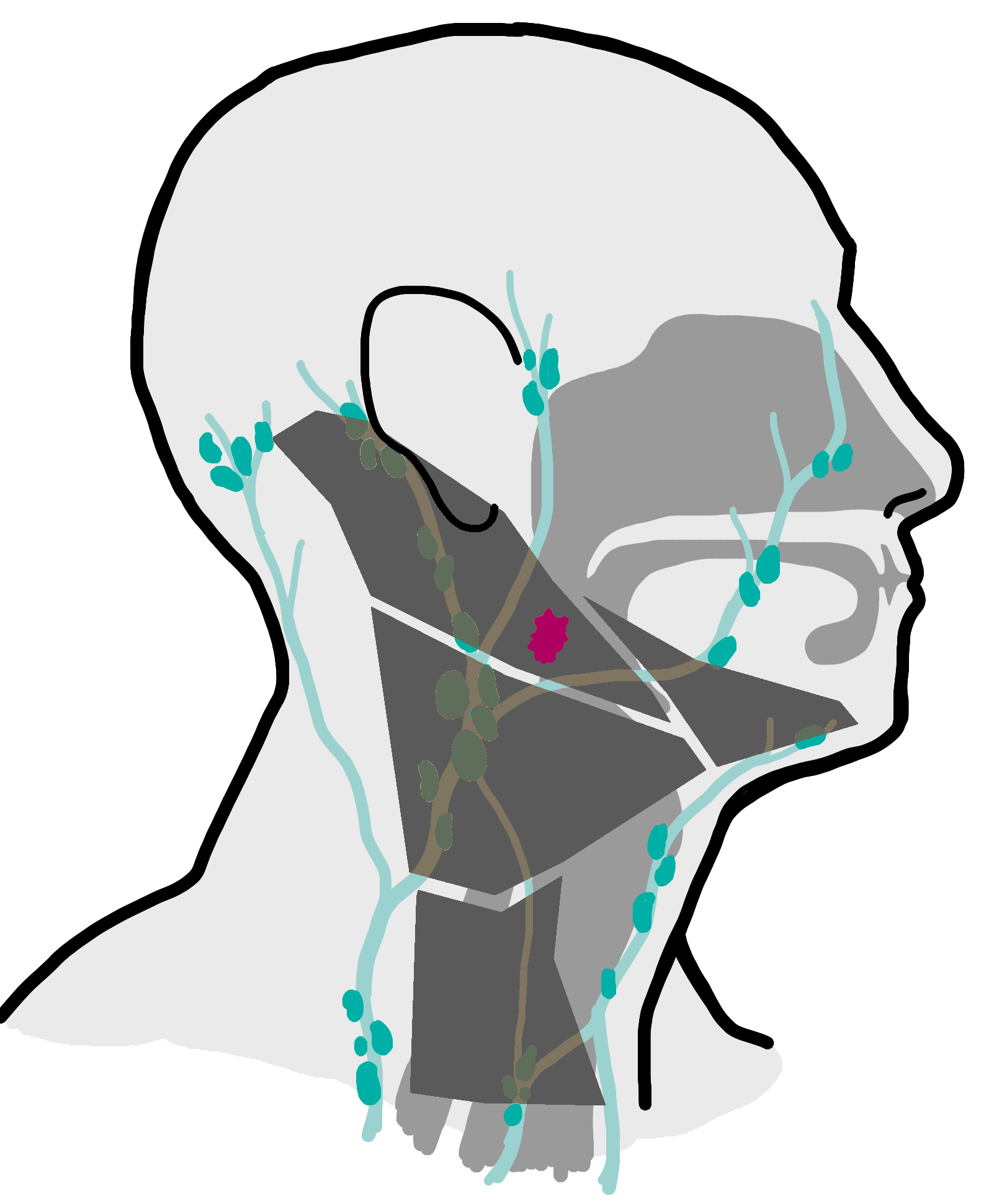}
    \end{minipage}
    
\end{subfigure}
\hfill
\begin{subfigure}{0.45\linewidth}
    \centering
    \caption{Model-based for 12\% Risk Threshold}
    \label{fig:Treatment_Protocols_N+_II_12}
    \begin{minipage}{0.48\linewidth}
        \includegraphics[width=\linewidth]{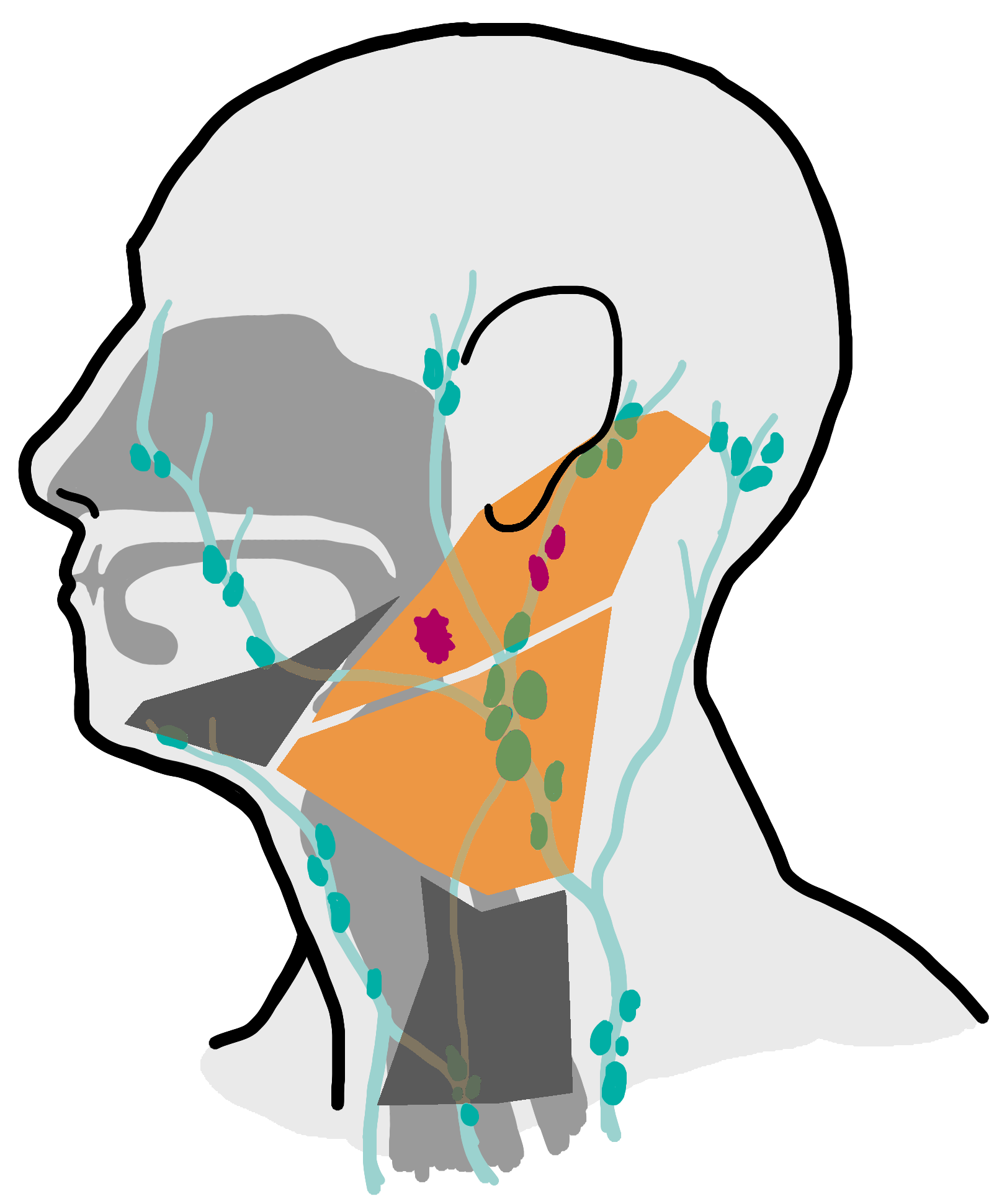}
    \end{minipage}%
    \begin{minipage}{0.48\linewidth}
        \includegraphics[width=\linewidth]{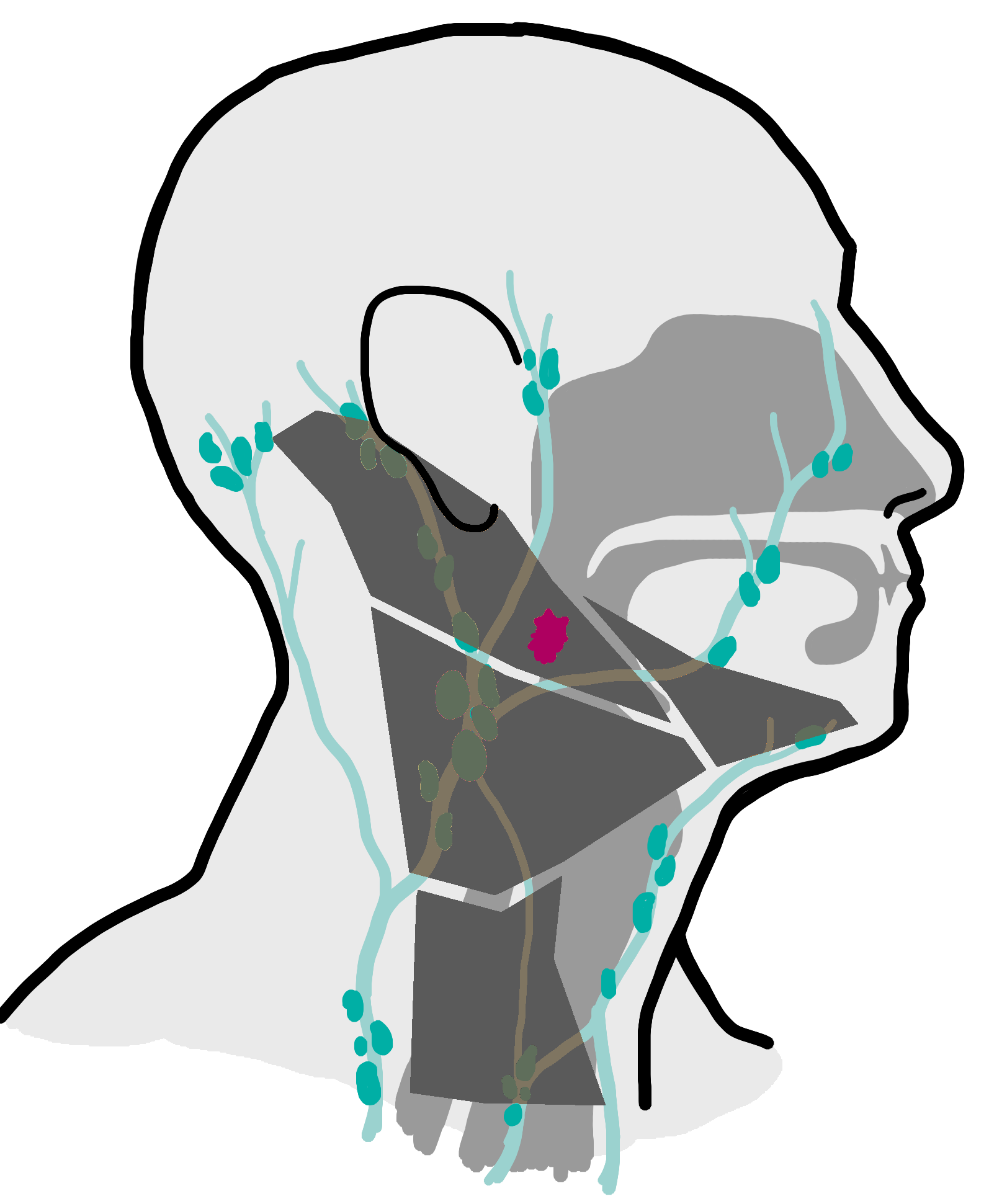}
    \end{minipage}
    
\end{subfigure}

\vspace{1mm}

\begin{subfigure}{0.45\linewidth}
    \centering
    \caption{Model-based for 15\% Risk Threshold}
    \label{fig:Treatment_Protocols_N+_II_15}
    \begin{minipage}{0.48\linewidth}
        \includegraphics[width=\linewidth]{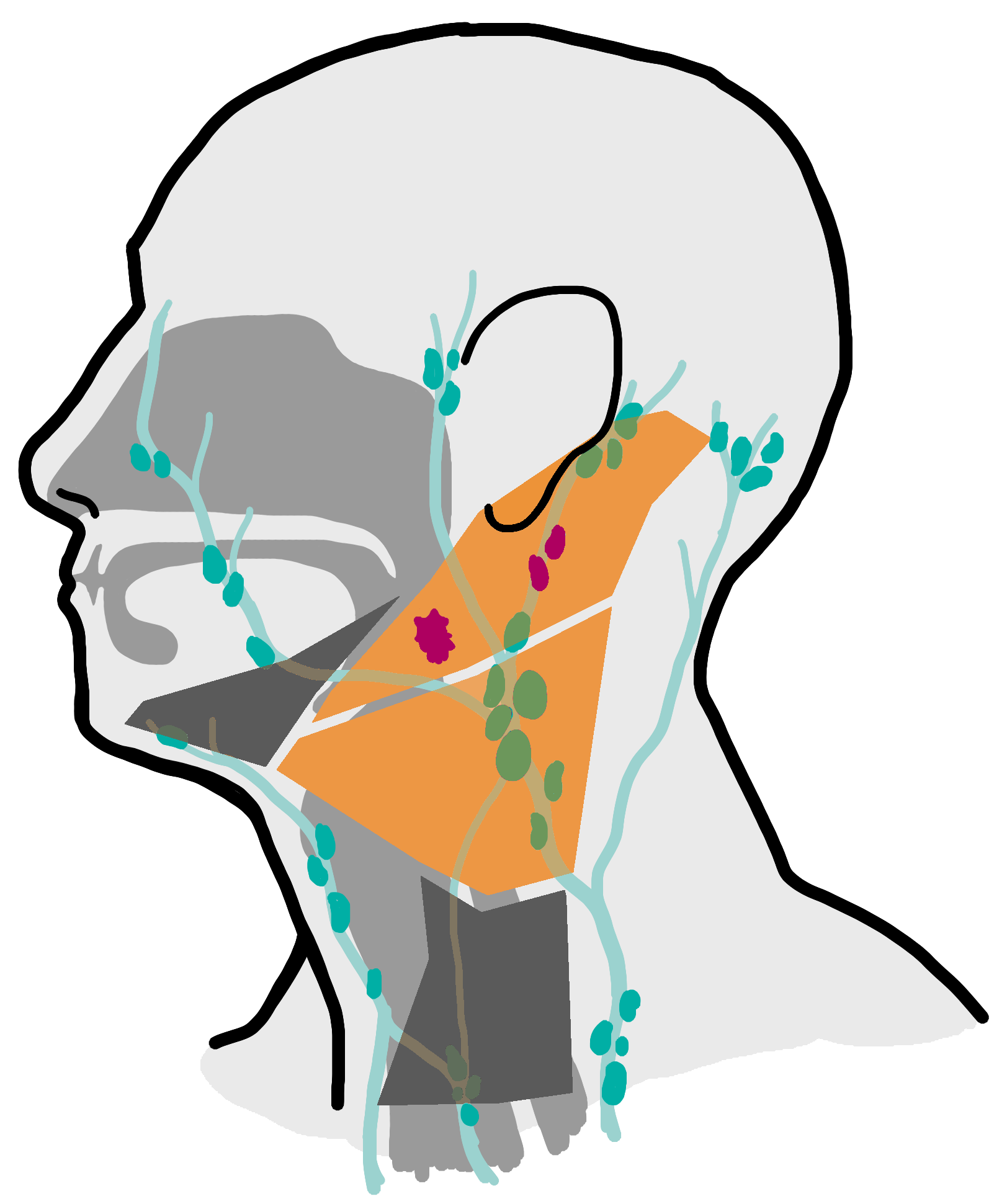}
    \end{minipage}%
    \begin{minipage}{0.48\linewidth}
        \includegraphics[width=\linewidth]{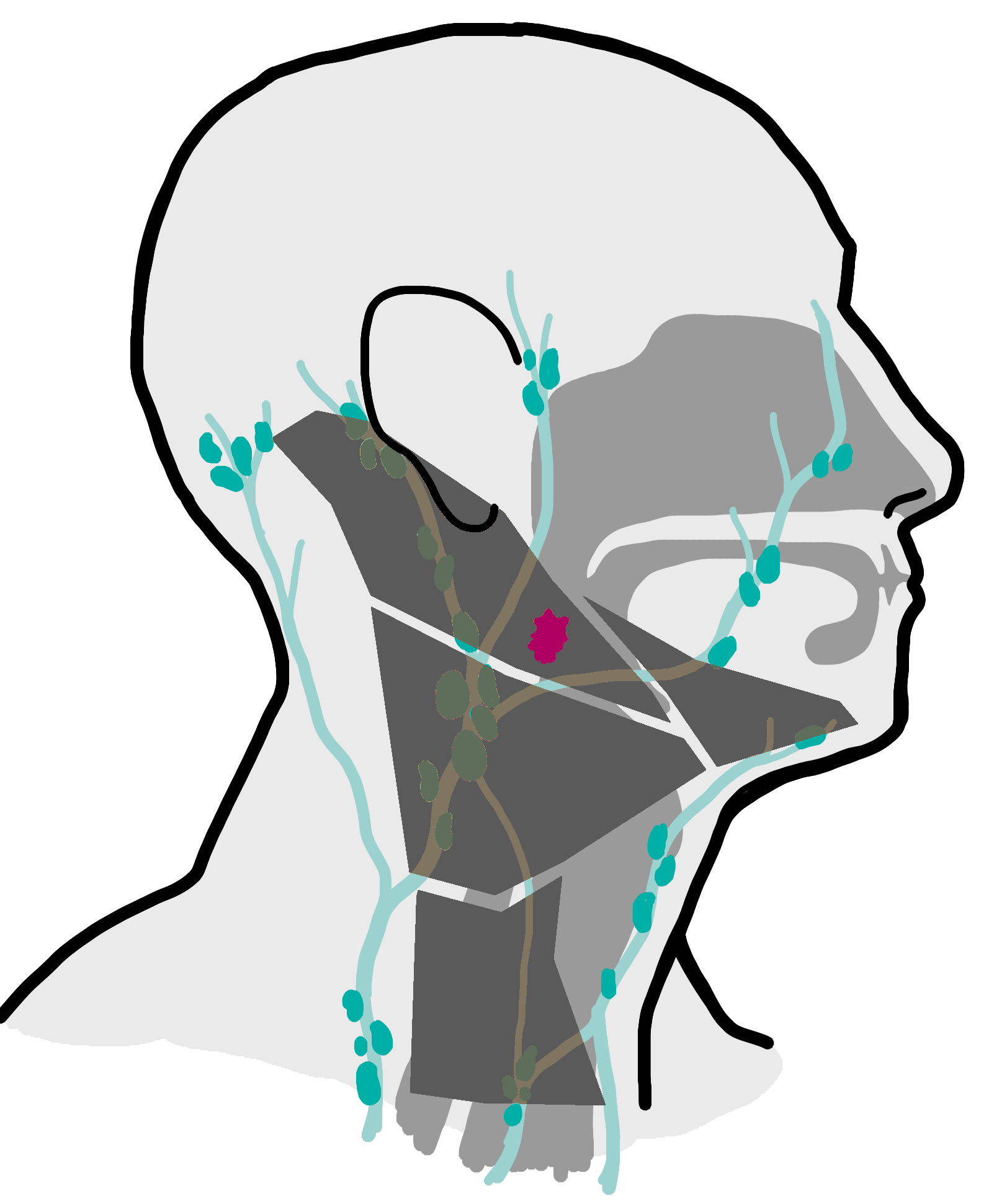}
    \end{minipage}
    
\end{subfigure}
\hfill
\begin{subfigure}{0.45\linewidth}
    \centering
    \caption{Model-based for 20\% Risk Threshold}
    \label{fig:Treatment_Protocols_N+_II_20}
    \begin{minipage}{0.48\linewidth}
        \includegraphics[width=\linewidth]{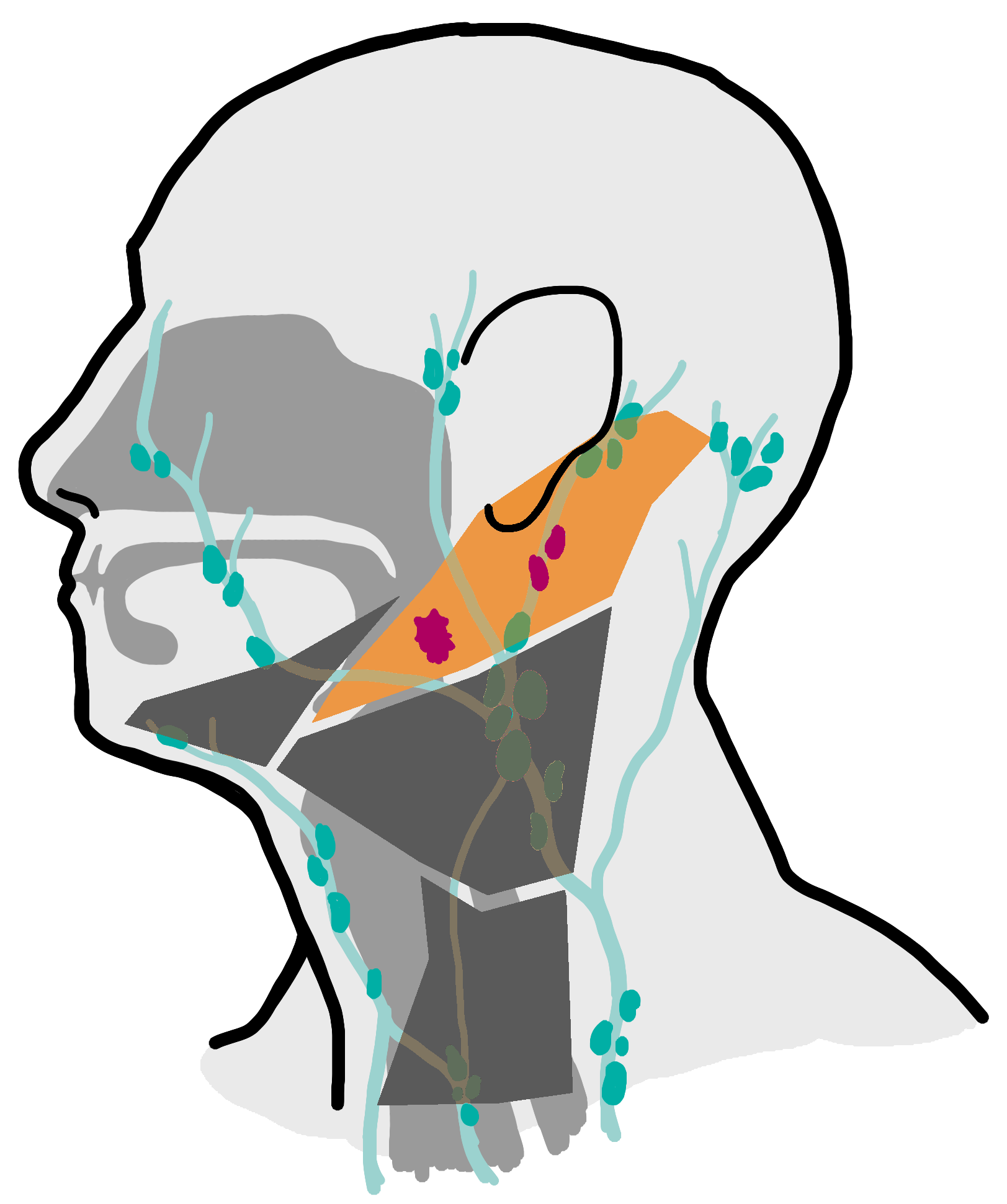}
    \end{minipage}%
    \begin{minipage}{0.48\linewidth}
        \includegraphics[width=\linewidth]{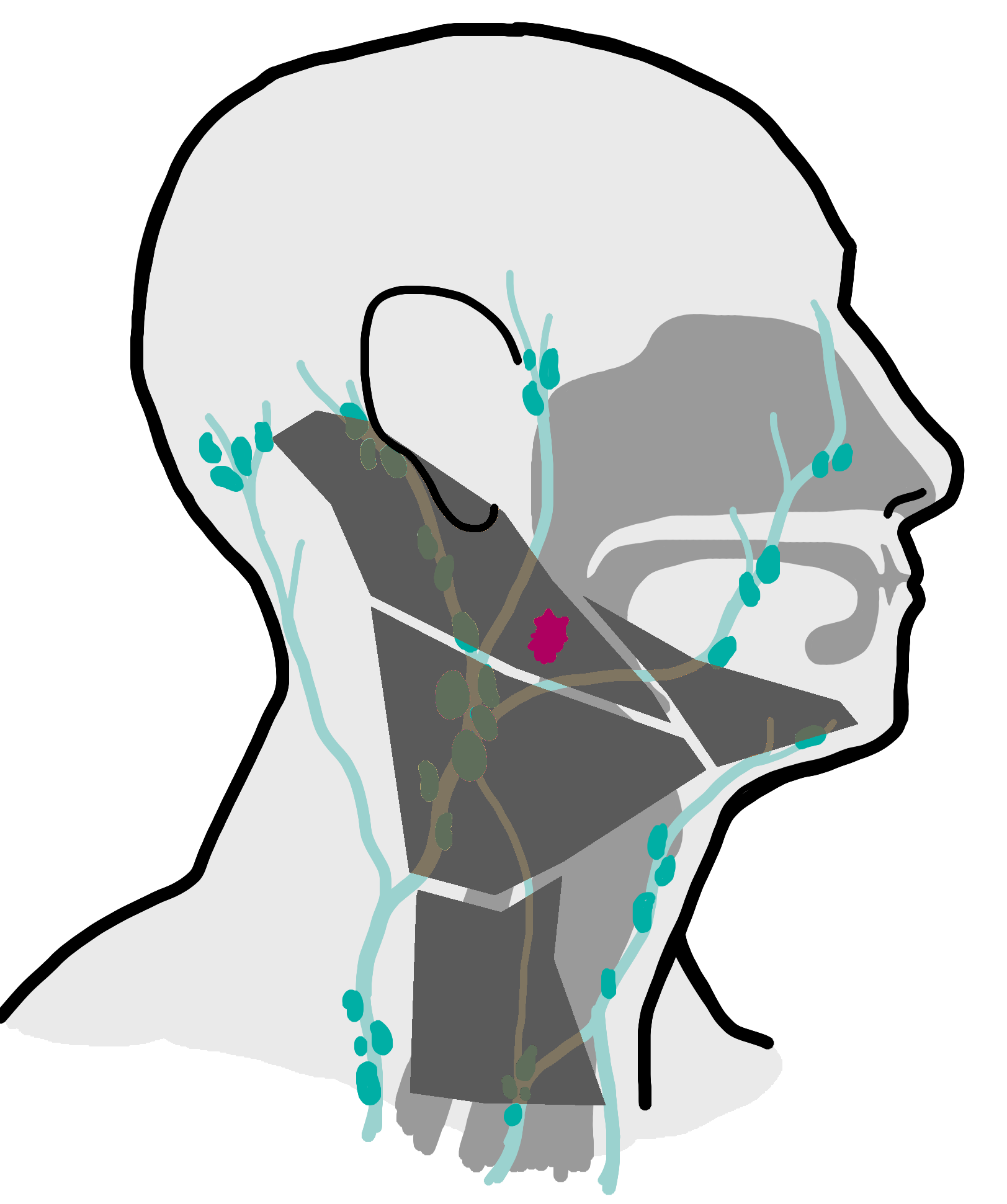}
    \end{minipage}
    
\end{subfigure}

\caption{Estimated elective nodal irradiation strategies for a patient with an early-stage central oropharyngeal tumour and N+ disease in ipsilateral LNL II under varying risk thresholds. LNLs above the inclusion threshold are shown in orange (included into the CTV-E), while LNLs below the threshold are shown in gray (excluded from treatment)}
\label{fig:Treatment_Protocols_N+_II}
\end{figure}

Figure \ref{fig:Treatment_Protocols_N+_II} shows the CTV-E as defined by the model-based, risk-adaptive strategy versus the standard guideline-based method. For this patient with observed nodal involvement limited to ipsilateral LNL II, Danish standard clinical guidelines typically recommend elective irradiation of ipsilateral LNLs II–IV and contralateral II and III (figure \ref{fig:Treatment_Protocols_N+_II_Clinical}).

With a threshold of 2\%, the model suggests expanding the CTV-E to include ipsilateral LNLs I–IV and contralateral LNL II (figure \ref{fig:Treatment_Protocols_N+_II_2}). In this scenario, the treatment field on the ipsilateral neck is broadened beyond current practice, reducing the risk of undetected nodal involvement in any omitted LNLs to just 0.8\%. The cost, of course, is a significant increase in irradiated volume. If the threshold is increased to 5\%, the recommended coverage can be narrowed to ipsilateral LNLs II and III, and contralateral LNL II (figure \ref{fig:Treatment_Protocols_N+_II_5}), with a residual risk of 3.5\%. In the scenario where the risk threshold is increased to 10\%, elective irradiation may be limited to the ipsilateral LNLs II and III alone, omitting all contralateral LNLs from the treatment (figure \ref{fig:Treatment_Protocols_N+_II_10}). Here, all contralateral LNLs can be omitted, if a residual risk of 7.5\% is accepted in the remaining LNLs. Treatment scenarios corresponding to the other risk thresholds are presented in figures \ref{fig:Treatment_Protocols_N+_II_2}-\ref{fig:Treatment_Protocols_N+_II_20}.


\subsection{Patient 3: N+ Ipsilateral LNL II+III, Early-stage (T1-T2)}

\begin{figure}[H]
\centering
\scriptsize

\begin{subfigure}{0.45\linewidth}
    \centering
    \caption{Current Clinical Practice (DAHANCA)}
    \label{fig:Treatment_Protocols_N+_II_III_Clinical}
    \begin{minipage}{0.48\linewidth}
        \includegraphics[width=\linewidth]{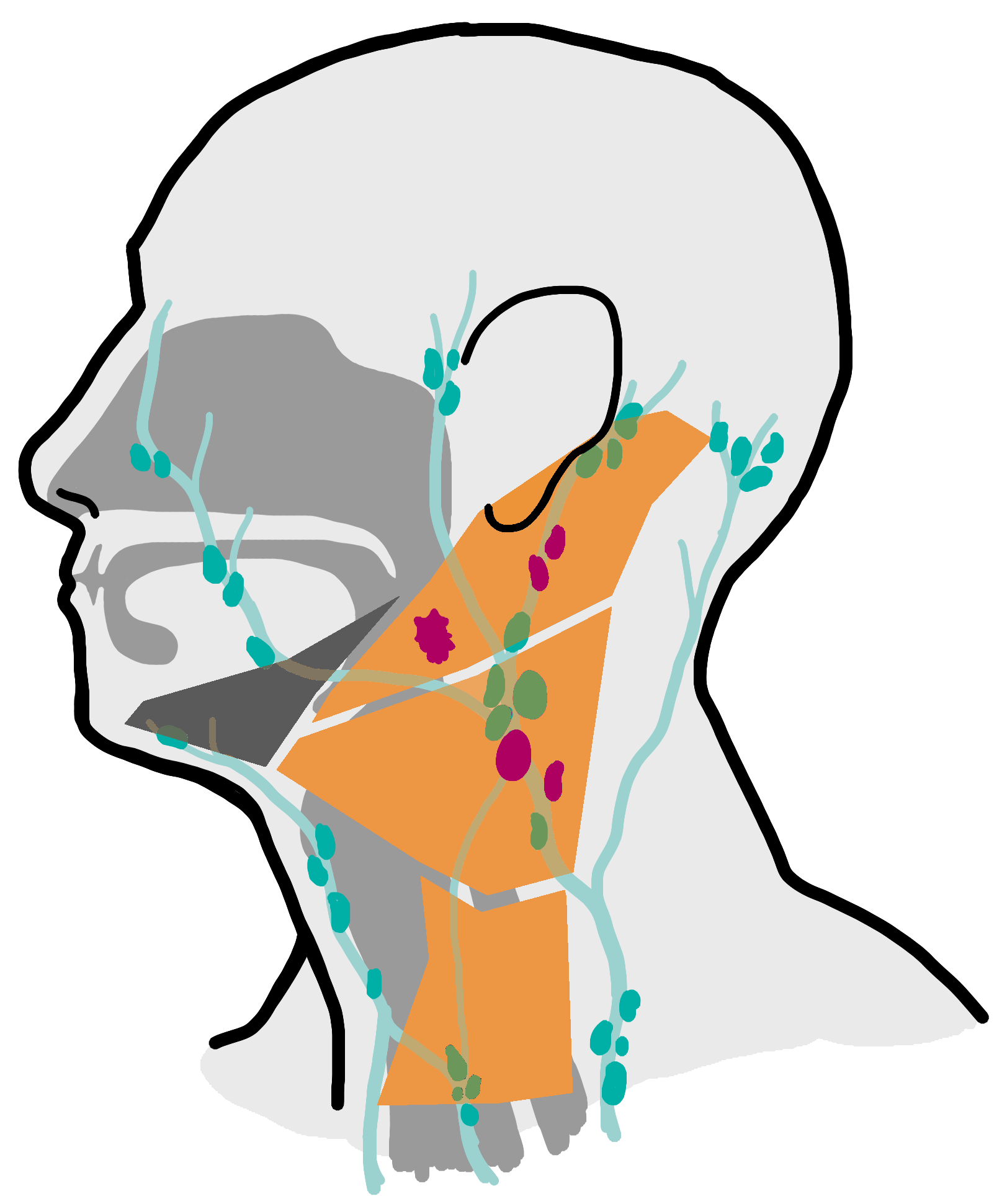}
    \end{minipage}%
    \begin{minipage}{0.48\linewidth}
        \includegraphics[width=\linewidth]{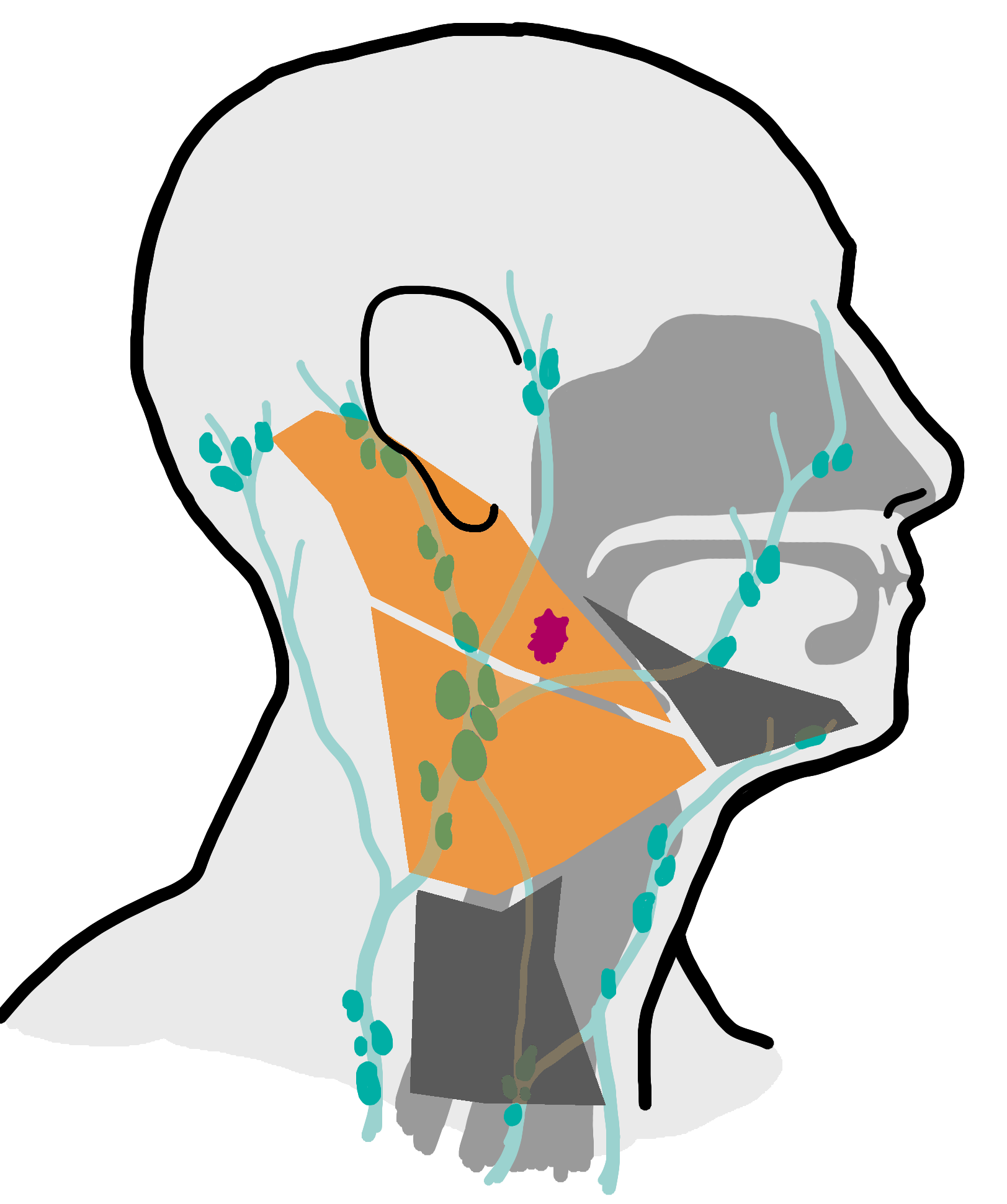}
    \end{minipage}
    
\end{subfigure}
\hfill
\begin{subfigure}{0.45\linewidth}
    \centering
    \caption{Model-based for 2\% Risk Threshold}
    \label{fig:Treatment_Protocols_N+_II_III_2}
    \begin{minipage}{0.48\linewidth}
        \includegraphics[width=\linewidth]{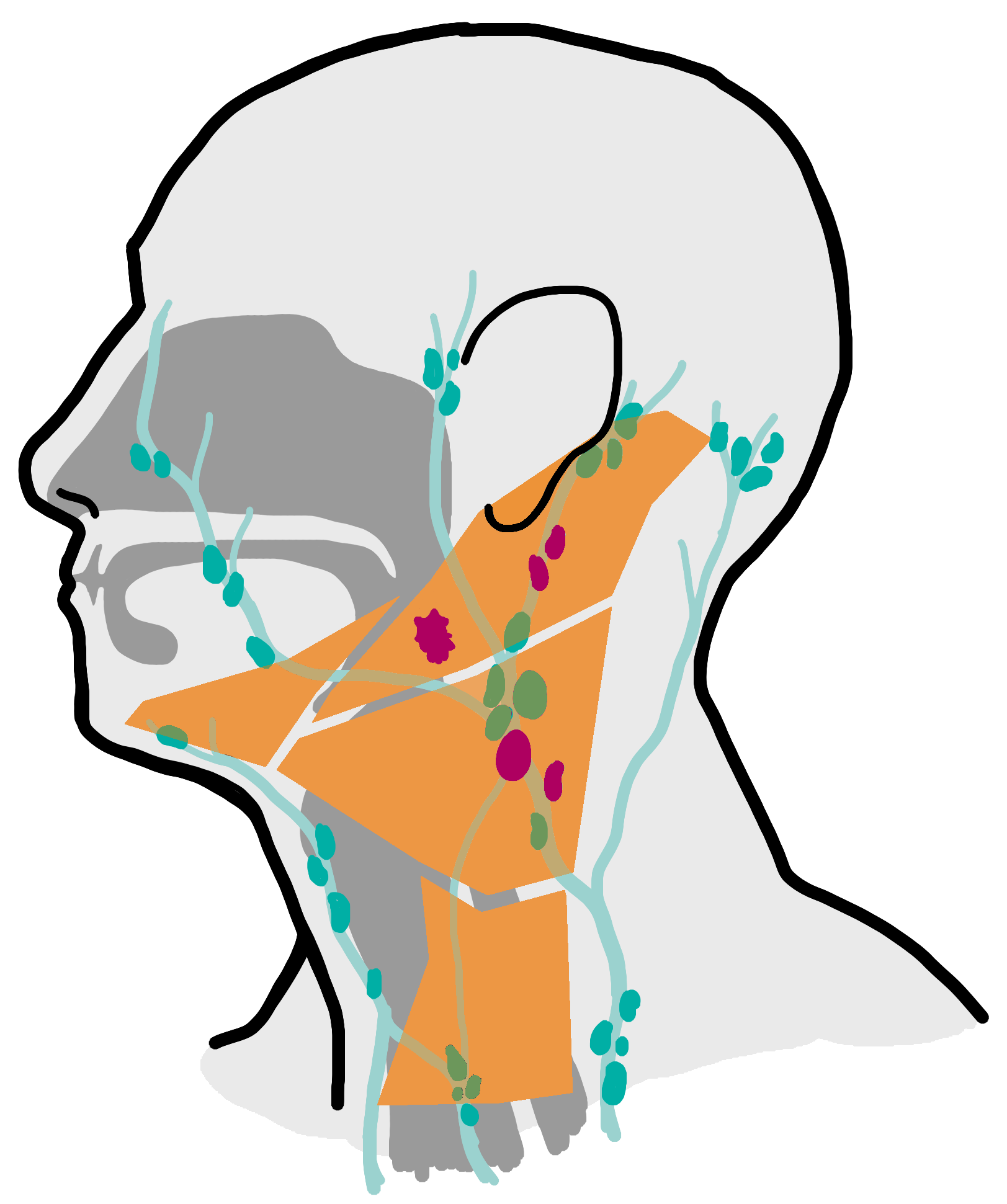}
    \end{minipage}%
    \begin{minipage}{0.48\linewidth}
        \includegraphics[width=\linewidth]{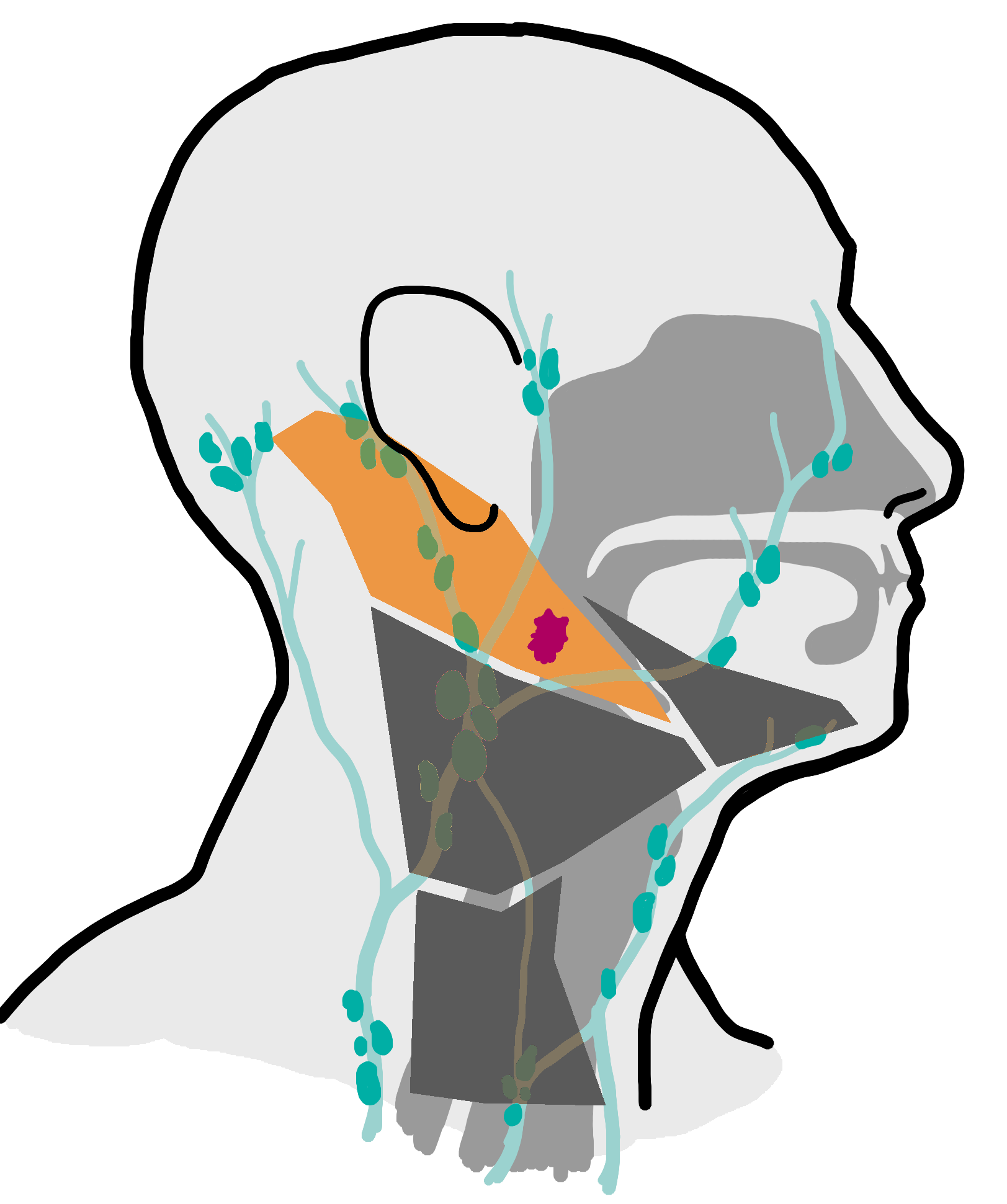}
    \end{minipage}
    
\end{subfigure}

\vspace{1mm}

\begin{subfigure}{0.45\linewidth}
    \centering
    \caption{Model-based for 5\% Risk Threshold}
    \label{fig:Treatment_Protocols_N+_II_III_5}
    \begin{minipage}{0.48\linewidth}
        \includegraphics[width=\linewidth]{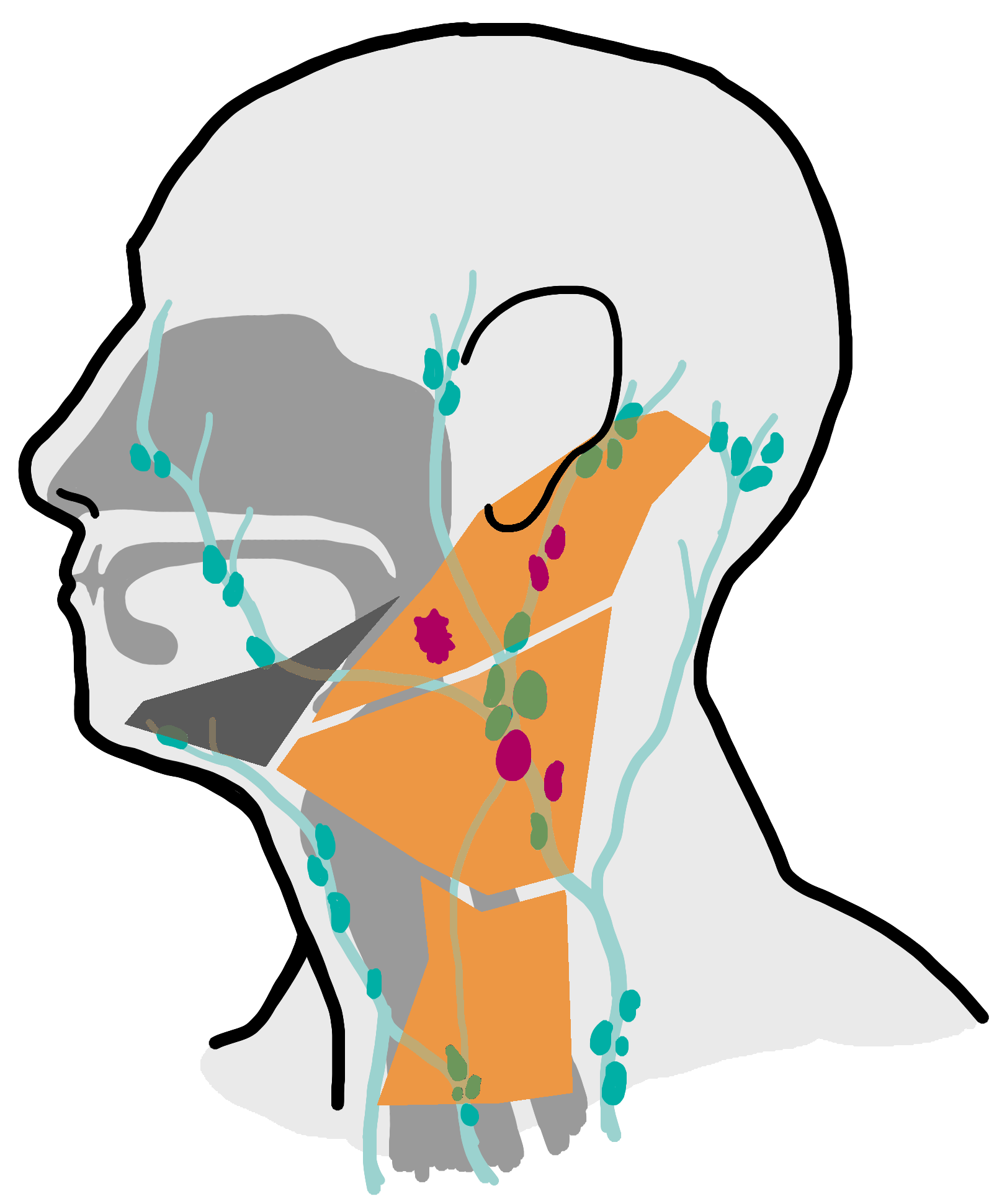}
    \end{minipage}%
    \begin{minipage}{0.48\linewidth}
        \includegraphics[width=\linewidth]{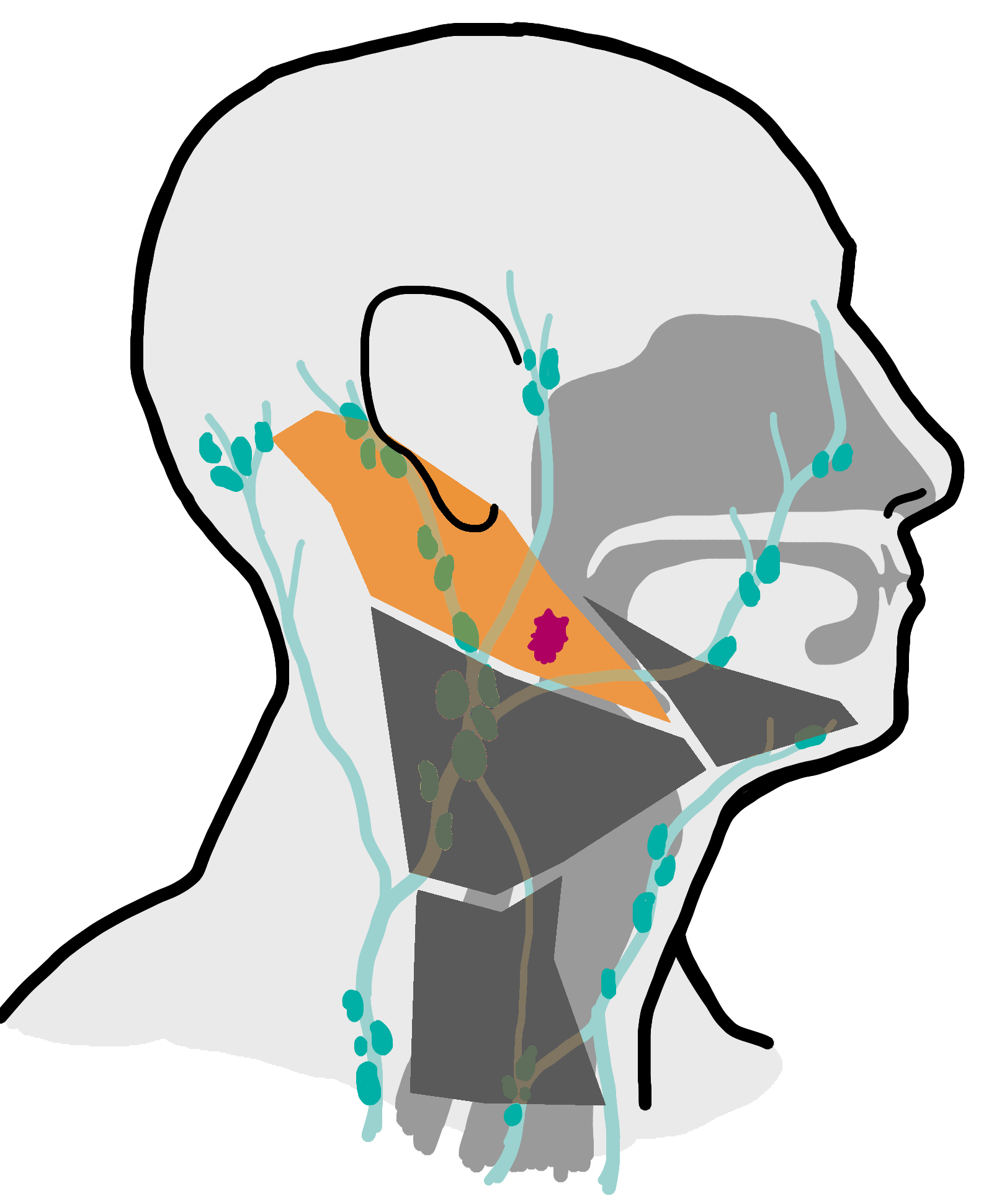}
    \end{minipage}
    
\end{subfigure}
\hfill
\begin{subfigure}{0.45\linewidth}
    \centering
    \caption{Model-based for 8\% Risk Threshold}
    \label{fig:Treatment_Protocols_N+_II_III_8}
    \begin{minipage}{0.48\linewidth}
        \includegraphics[width=\linewidth]{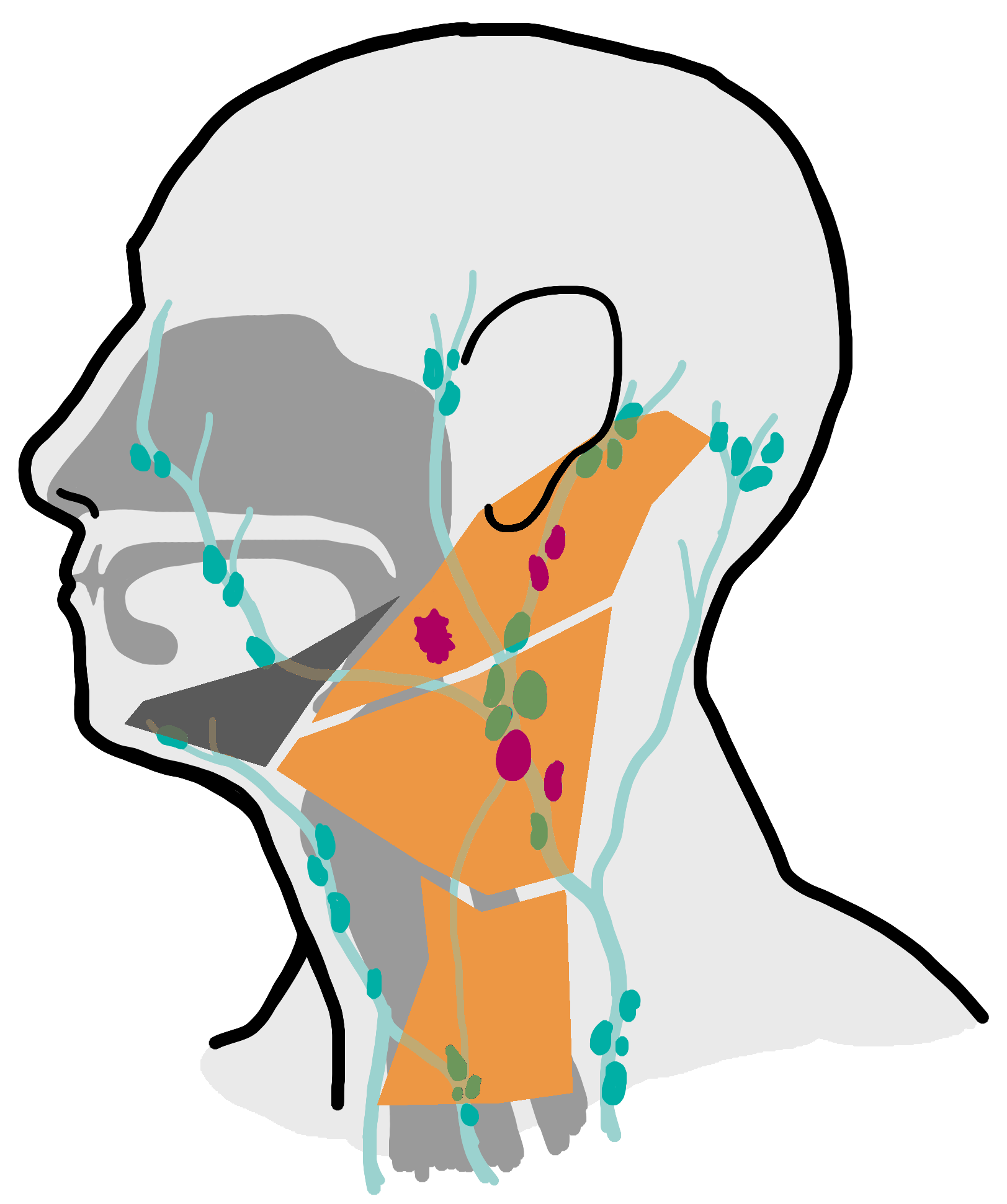}
    \end{minipage}%
    \begin{minipage}{0.48\linewidth}
        \includegraphics[width=\linewidth]{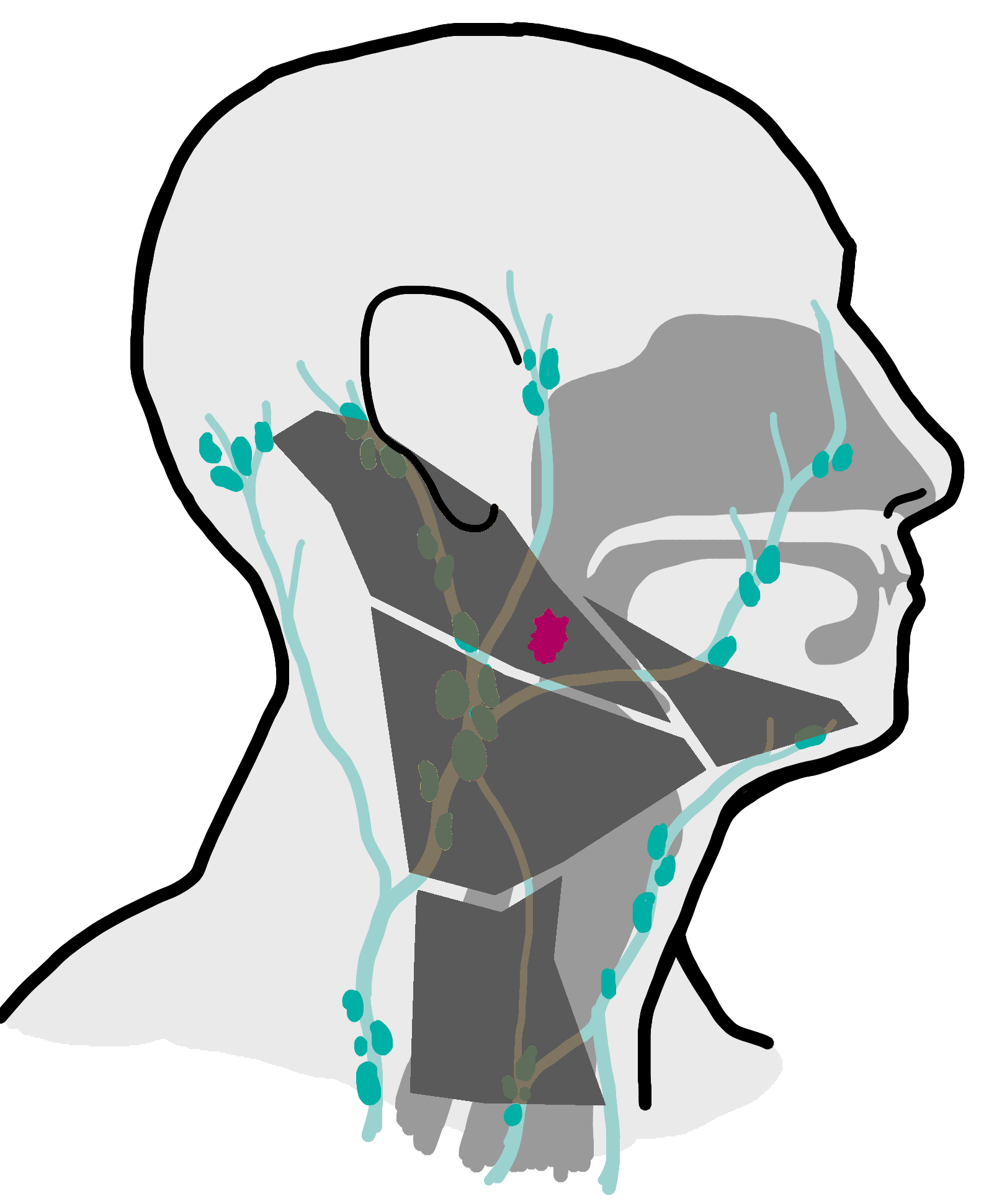}
    \end{minipage}
    
\end{subfigure}

\vspace{1mm}

\begin{subfigure}{0.45\linewidth}
    \centering
    \caption{Model-based for 10\% Risk Threshold}
    \label{fig:Treatment_Protocols_N+_II_III_10}
    \begin{minipage}{0.48\linewidth}
        \includegraphics[width=\linewidth]{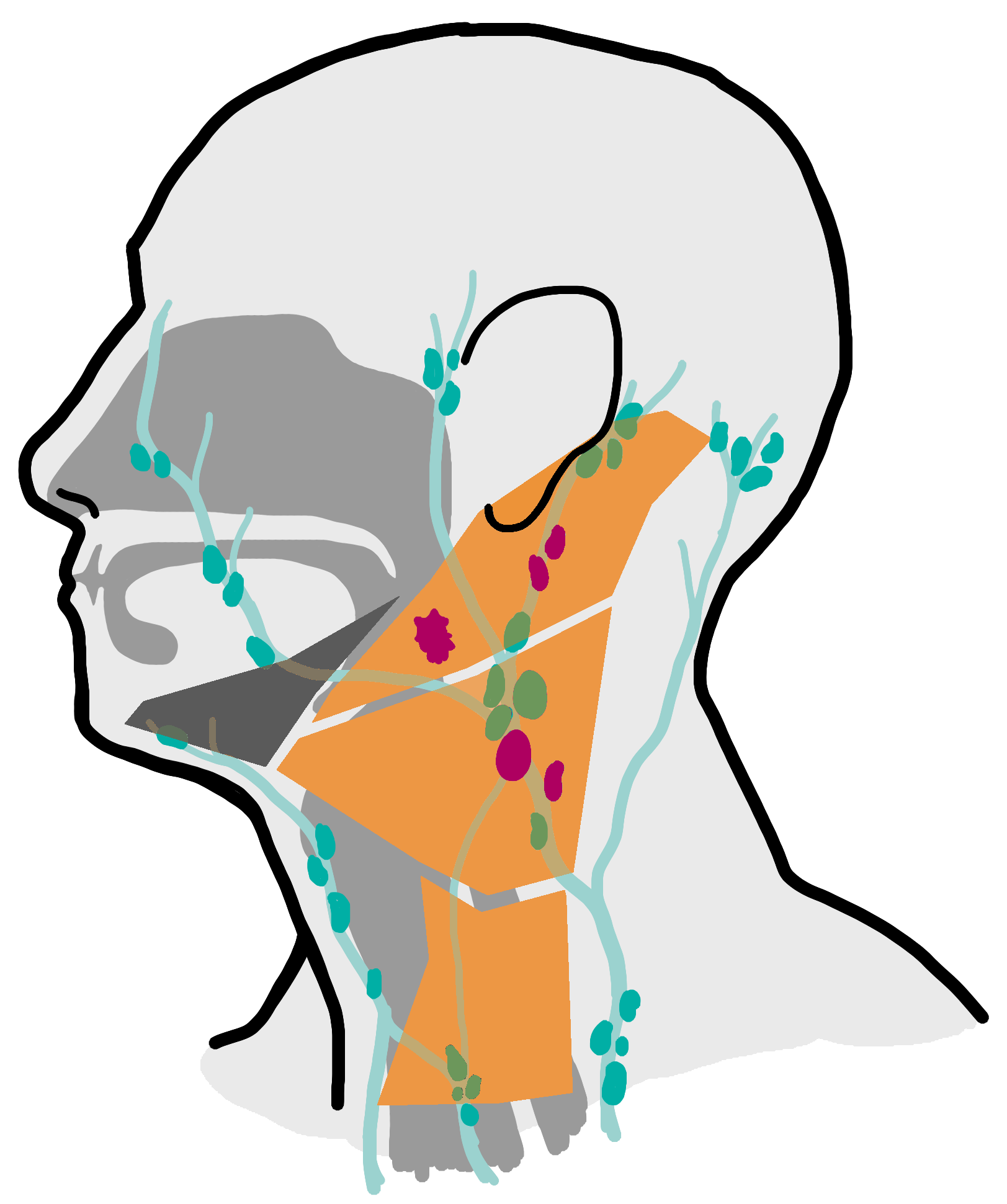}
    \end{minipage}%
    \begin{minipage}{0.48\linewidth}
        \includegraphics[width=\linewidth]{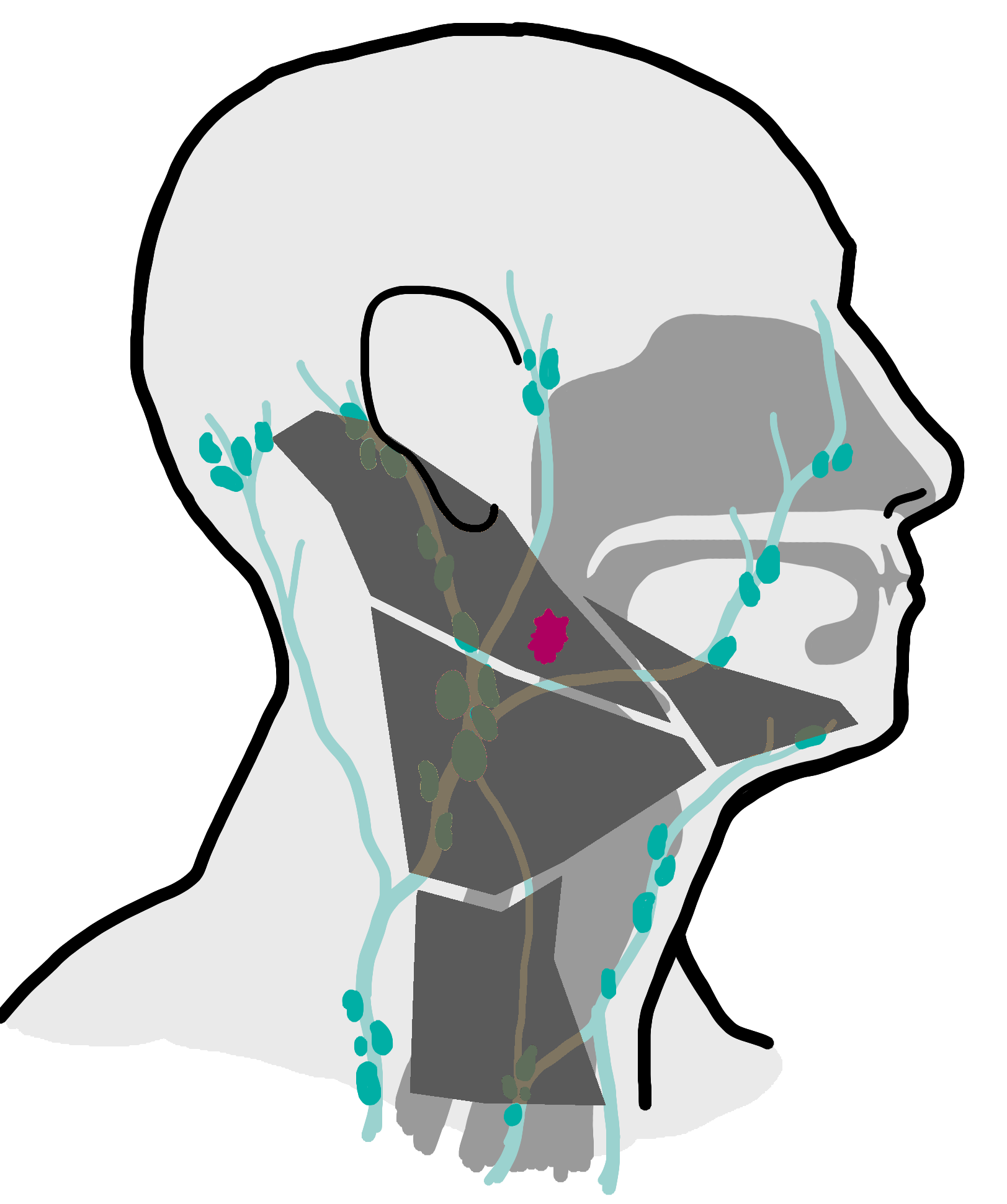}
    \end{minipage}
    
\end{subfigure}
\hfill
\begin{subfigure}{0.45\linewidth}
    \centering
    \caption{Model-based for 12\% Risk Threshold}
    \label{fig:Treatment_Protocols_N+_II_III_12}
    \begin{minipage}{0.48\linewidth}
        \includegraphics[width=\linewidth]{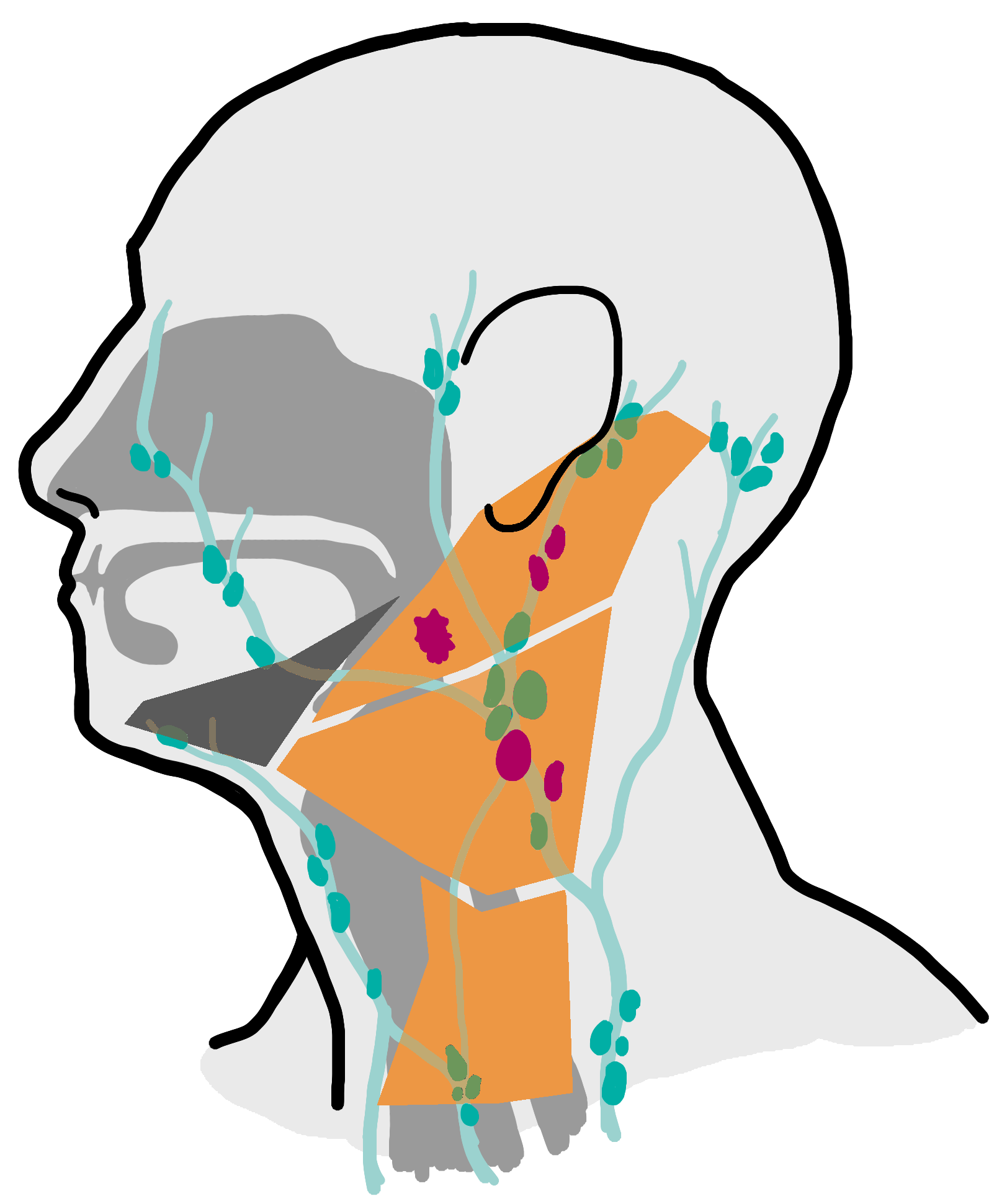}
    \end{minipage}%
    \begin{minipage}{0.48\linewidth}
        \includegraphics[width=\linewidth]{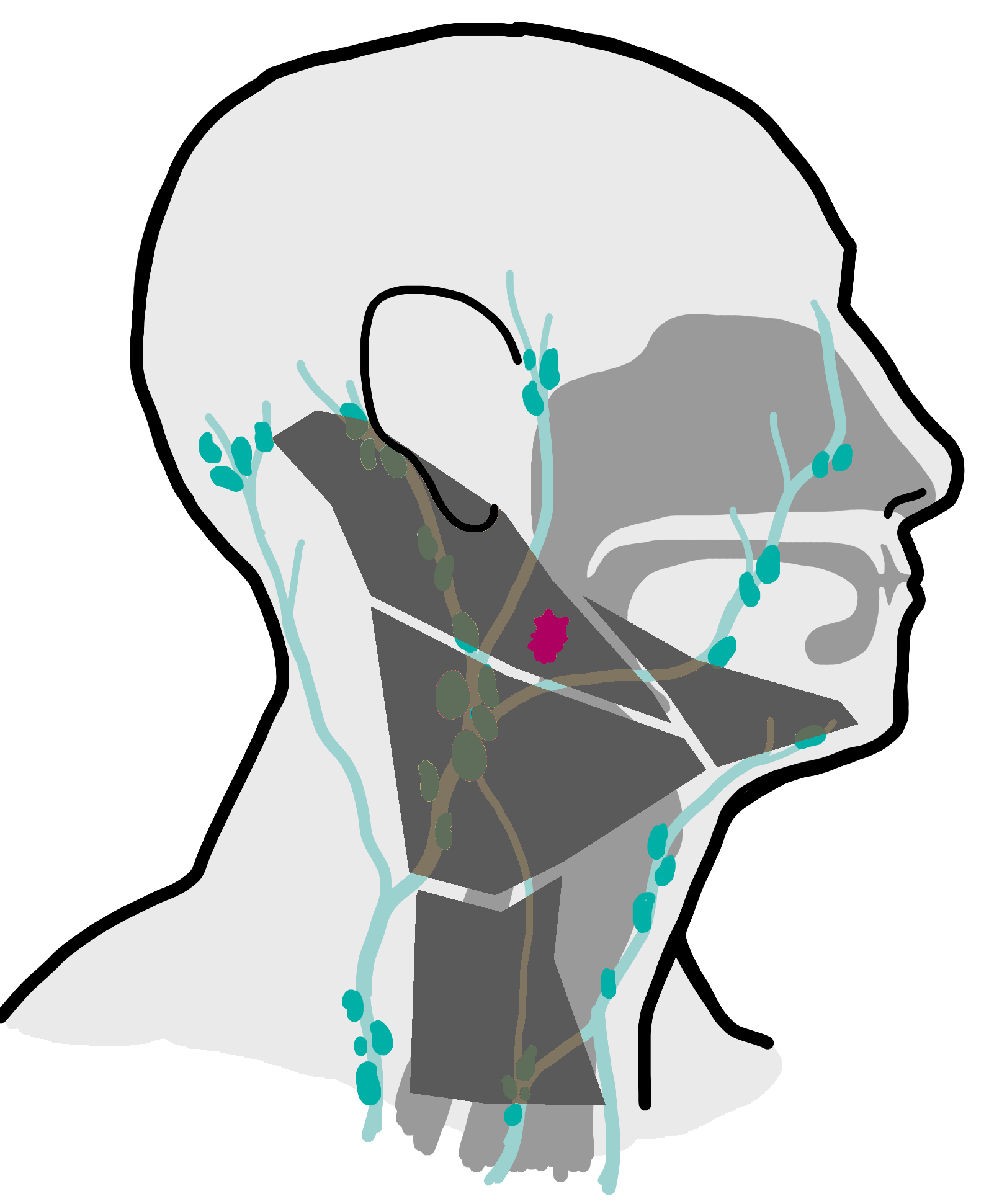}
    \end{minipage}
    
\end{subfigure}

\vspace{1mm}

\begin{subfigure}{0.45\linewidth}
    \centering
    \caption{Model-based for 15\% Risk Threshold}
    \label{fig:Treatment_Protocols_N+_II_III_15}
    \begin{minipage}{0.48\linewidth}
        \includegraphics[width=\linewidth]{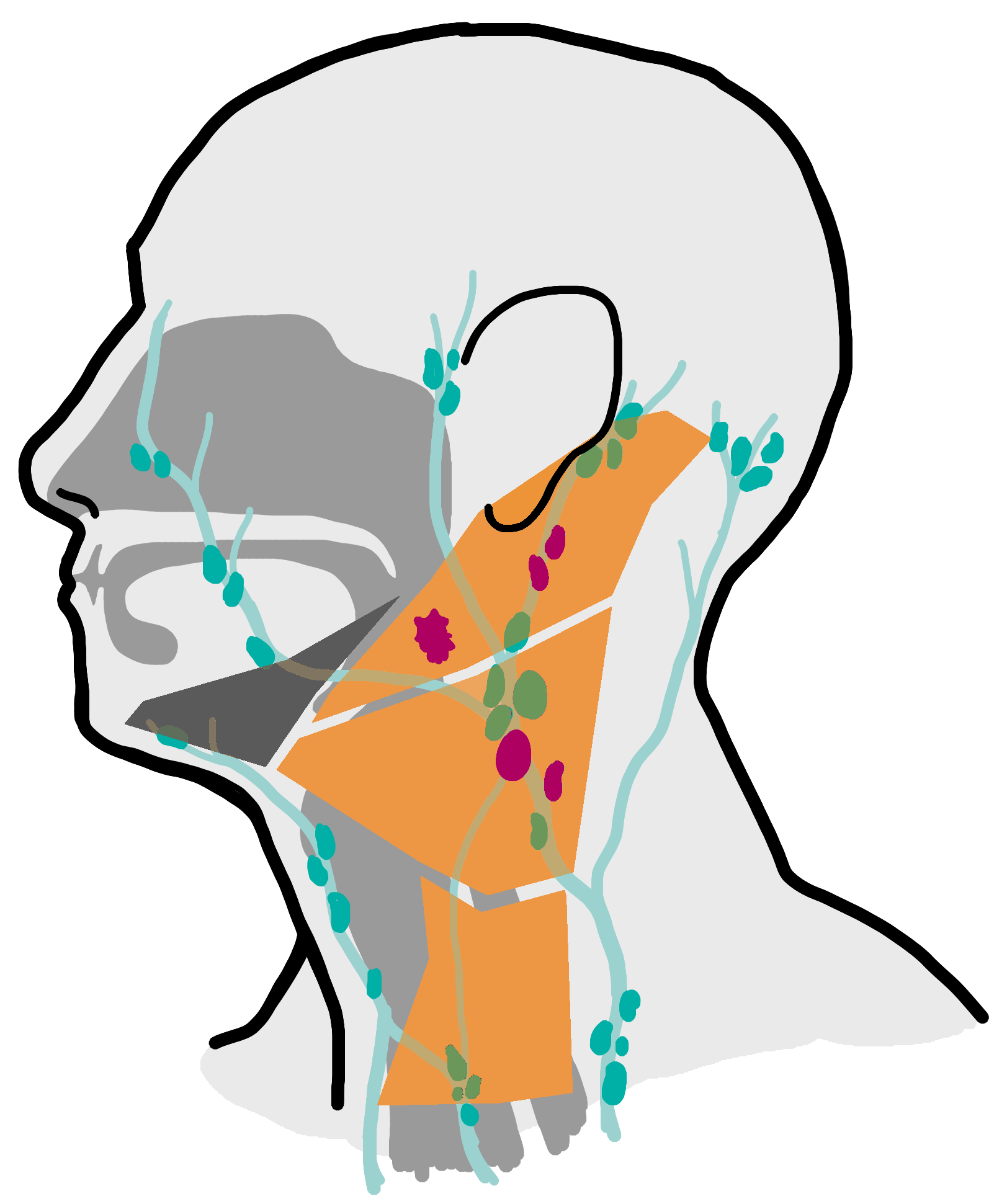}
    \end{minipage}%
    \begin{minipage}{0.48\linewidth}
        \includegraphics[width=\linewidth]{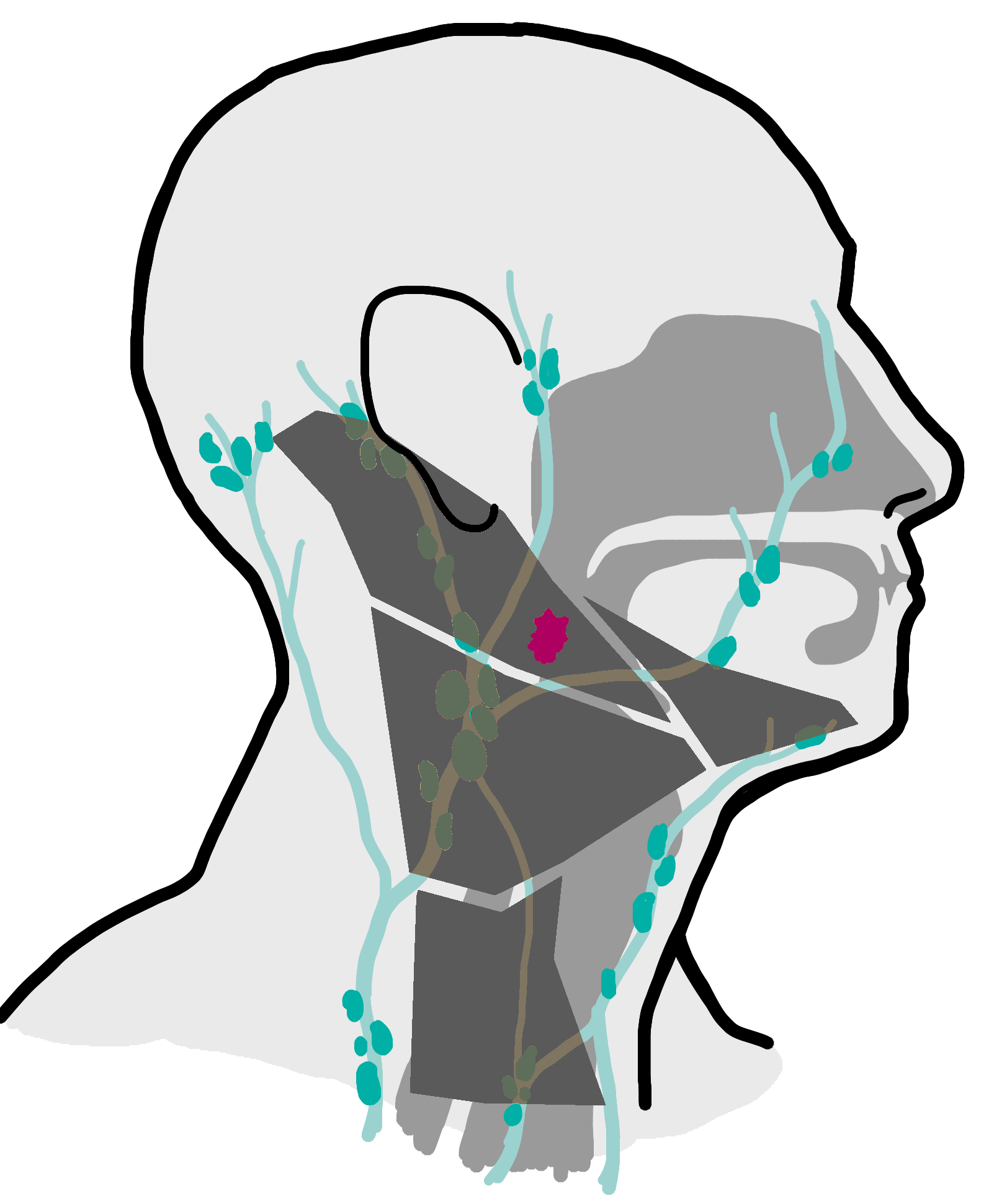}
    \end{minipage}
    
\end{subfigure}
\hfill
\begin{subfigure}{0.45\linewidth}
    \centering
    \caption{Model-based for 20\% Risk Threshold}
    \label{fig:Treatment_Protocols_N+_II_III_20}
    \begin{minipage}{0.48\linewidth}
        \includegraphics[width=\linewidth]{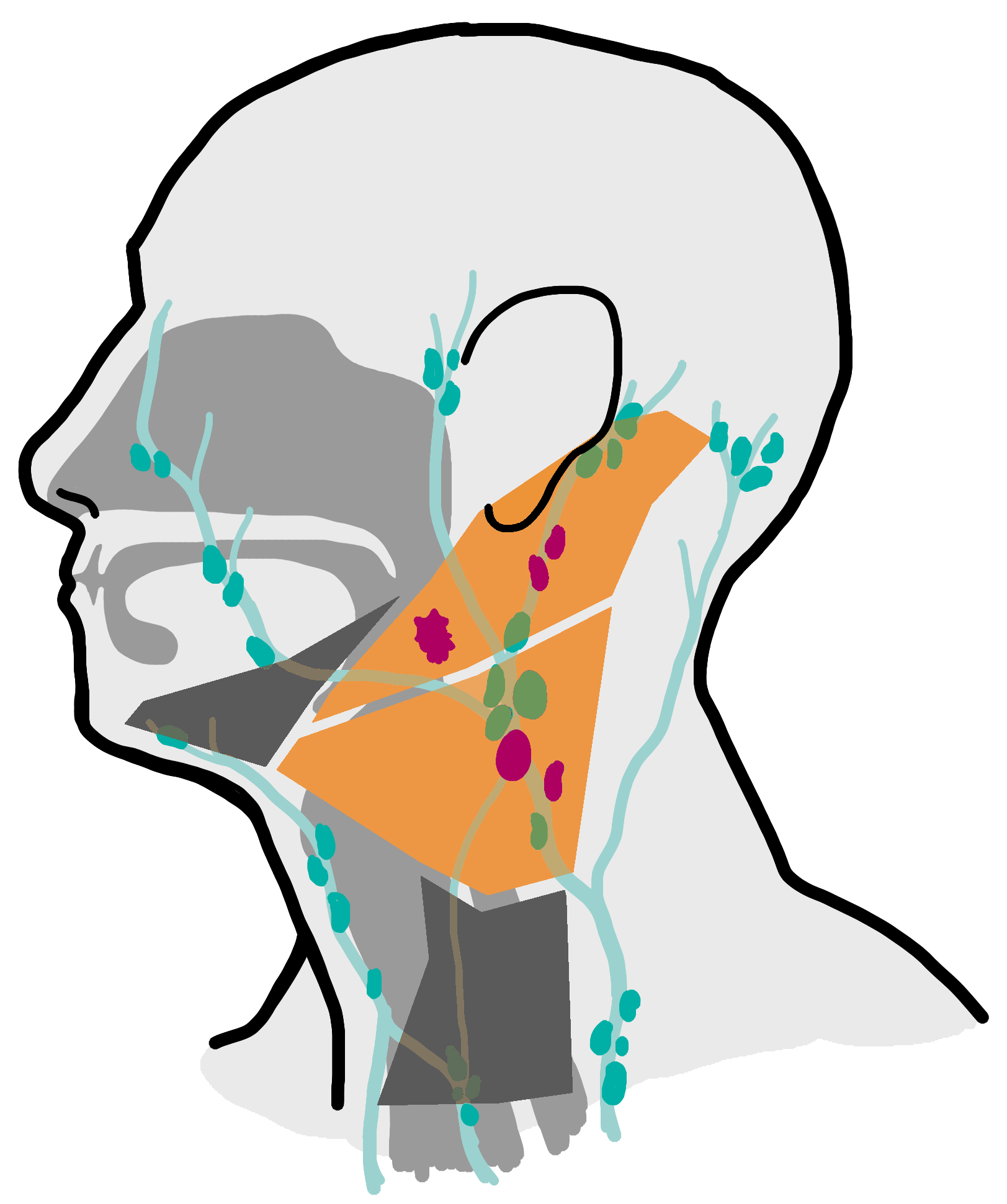}
    \end{minipage}%
    \begin{minipage}{0.48\linewidth}
        \includegraphics[width=\linewidth]{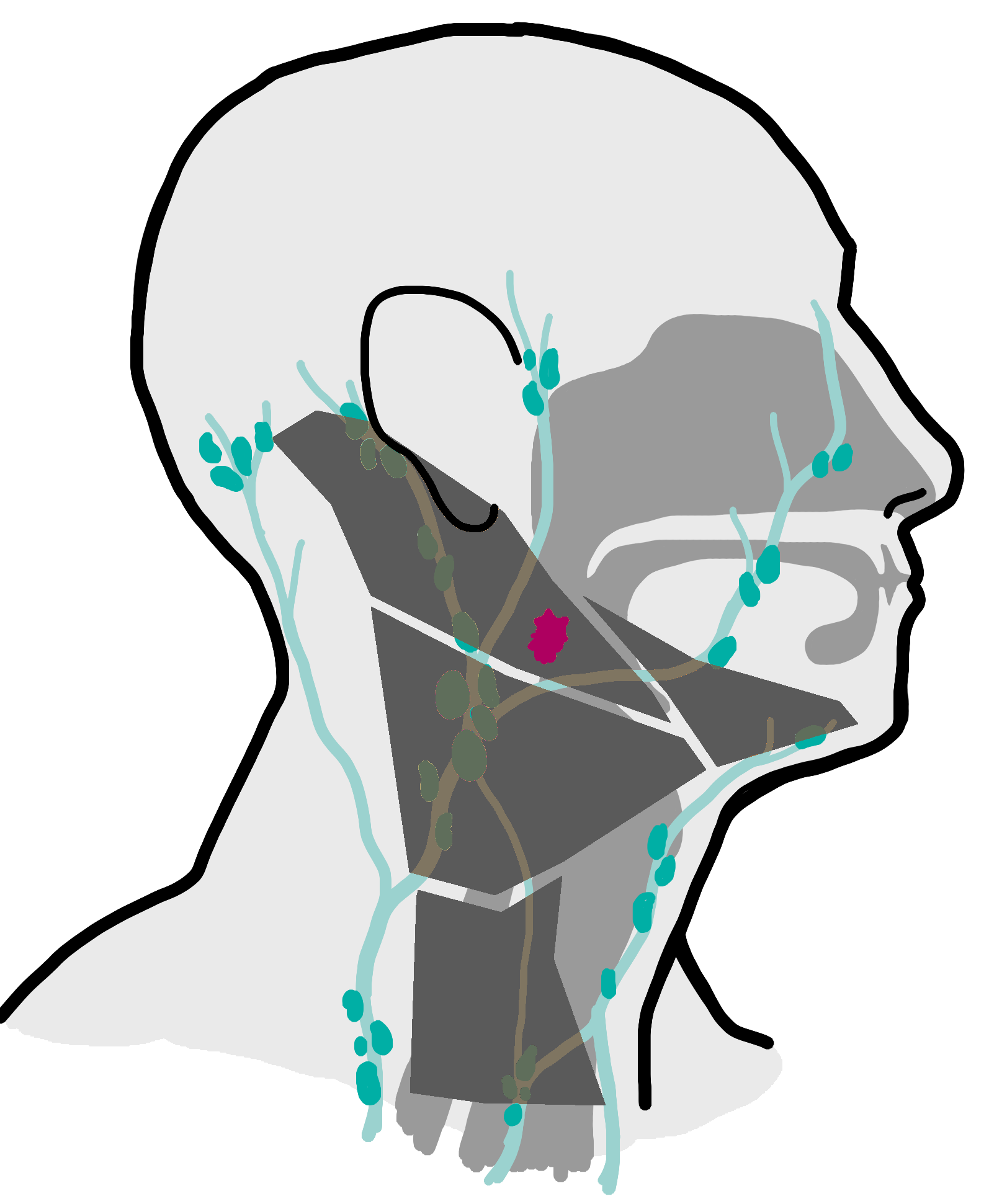}
    \end{minipage}
    
\end{subfigure}

\caption{Estimated elective nodal irradiation strategies for a patient with an early-stage central oropharyngeal tumour and N+ disease in ipsilateral LNL II and III under varying risk thresholds. LNLs above the inclusion threshold are shown in orange (included into the CTV-E), while LNLs below the threshold are shown in gray (excluded from treatment)}
\label{fig:Treatment_Protocols_N+_II_III}
\end{figure}

Figure \ref{fig:Treatment_Protocols_N+_II_III} offers a side-by-side visualization of CTV-E delineation according to standard guidelines versus the model-based, risk-adaptive approach. This case involves a patient with confirmed involvement of both ipsilateral LNL II and III, a scenario where, by default, Danish conventional guidelines include irradiation of ipsilateral LNLs II-IV and contralateral LNLs II and III (figure \ref{fig:Treatment_Protocols_N+_II_III_Clinical}).

In contrast, when the risk threshold is set to 2\%, the resulting CTV-E is expanded beyond standard practice on the ipsilateral side, now covering LNLs I–IV (figure \ref{fig:Treatment_Protocols_N+_II_III_2}). At the same time, this risk threshold enables greater sparing on the contralateral side, as only LNL II is included, with other contralateral LNLs omitted. This threshold limits the probability of undetected nodal involvement in any unirradiated LNLs to 0.8\%, reflecting a near-zero tolerance for missed disease. With a risk threshold of 5\%, elective coverage becomes more focused, omitting ipsilateral LNL I, while the ipsilateral LNLs II–IV and contralateral LNL II remain included (figure \ref{fig:Treatment_Protocols_N+_II_III_5}), resulting in a moderately rise in residual risk for the spared LNLs to 2.1\%. Increasing the risk threshold to 10\%, leads to the CTV-E only covering ipsilateral LNLs II–IV, fully omitting contralateral LNLs (figure \ref{fig:Treatment_Protocols_N+_II_III_10}). In this scenario, the probability of undetected nodal involvement in any omitted LNLs increases to 6.1\%. An overview of the treatment approaches for all additional risk thresholds is presented in figures \ref{fig:Treatment_Protocols_N+_II_III_2}–\ref{fig:Treatment_Protocols_N+_II_III_20}.


\subsection{Patient 4: N+ Ipsilateral LNL II + Contralateral LNL II, Advanced-stage (T3-T4)}

\begin{figure}[H]
\centering
\scriptsize

\begin{subfigure}{0.45\linewidth}
    \centering
    \caption{Current Clinical Practice (DAHANCA)}
    \label{fig:Treatment_Protocols_N+_II_II_Clinical}
    \begin{minipage}{0.48\linewidth}
        \includegraphics[width=\linewidth]{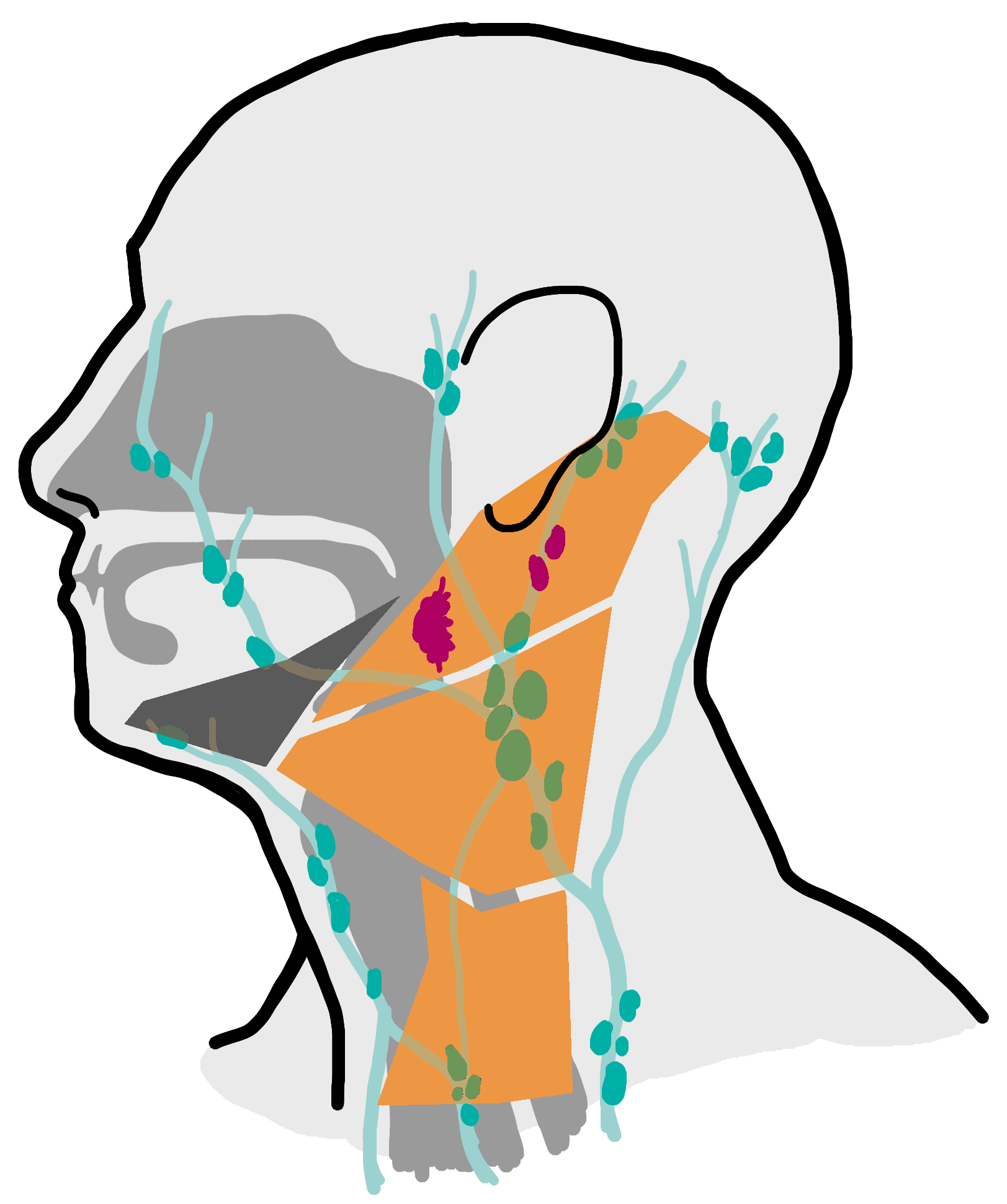}
    \end{minipage}%
    \begin{minipage}{0.48\linewidth}
        \includegraphics[width=\linewidth]{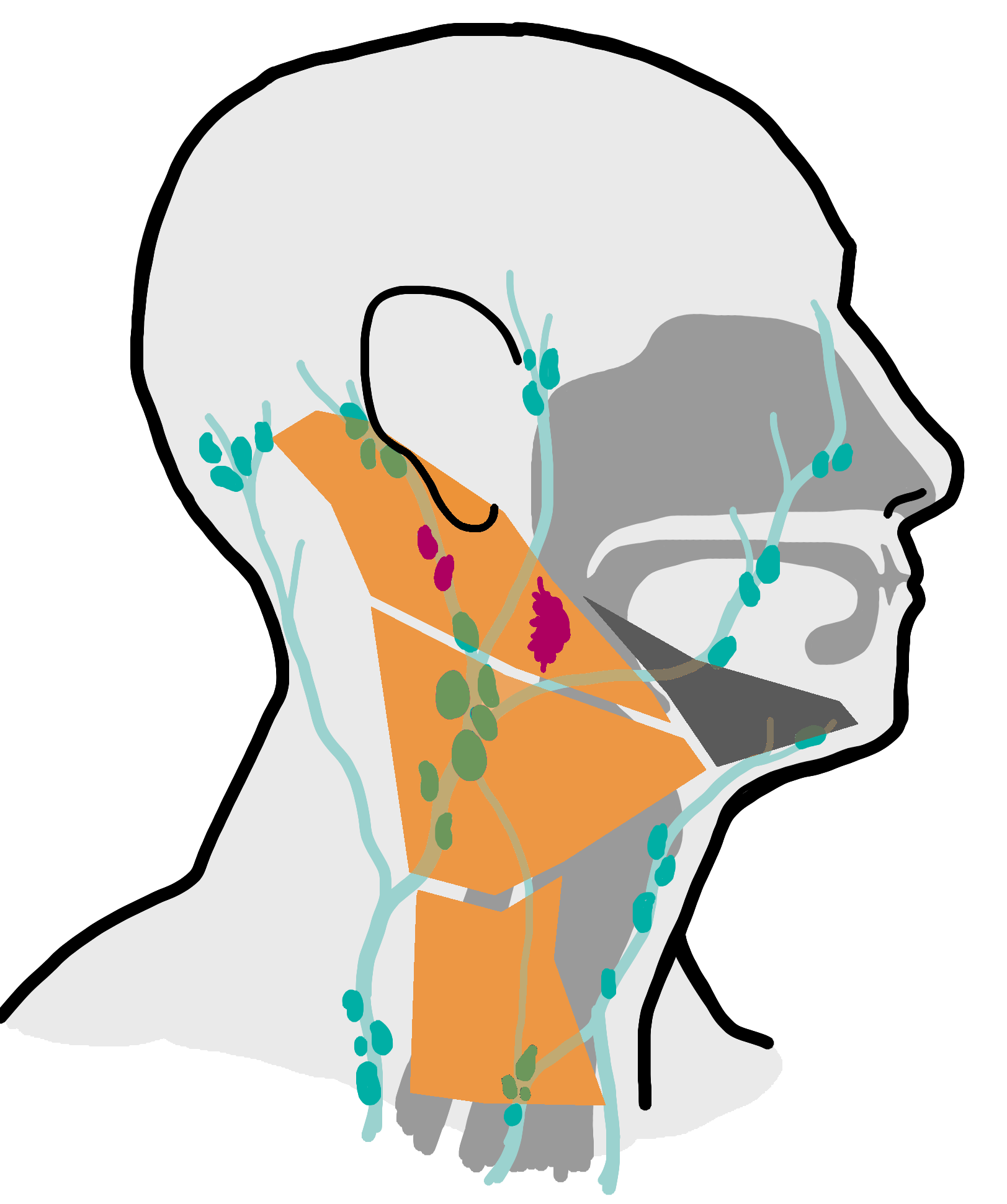}
    \end{minipage}
    
\end{subfigure}
\hfill
\begin{subfigure}{0.45\linewidth}
    \centering
    \caption{Model-based for 2\% Risk Threshold}
    \label{fig:Treatment_Protocols_N+_II_II_2}
    \begin{minipage}{0.48\linewidth}
        \includegraphics[width=\linewidth]{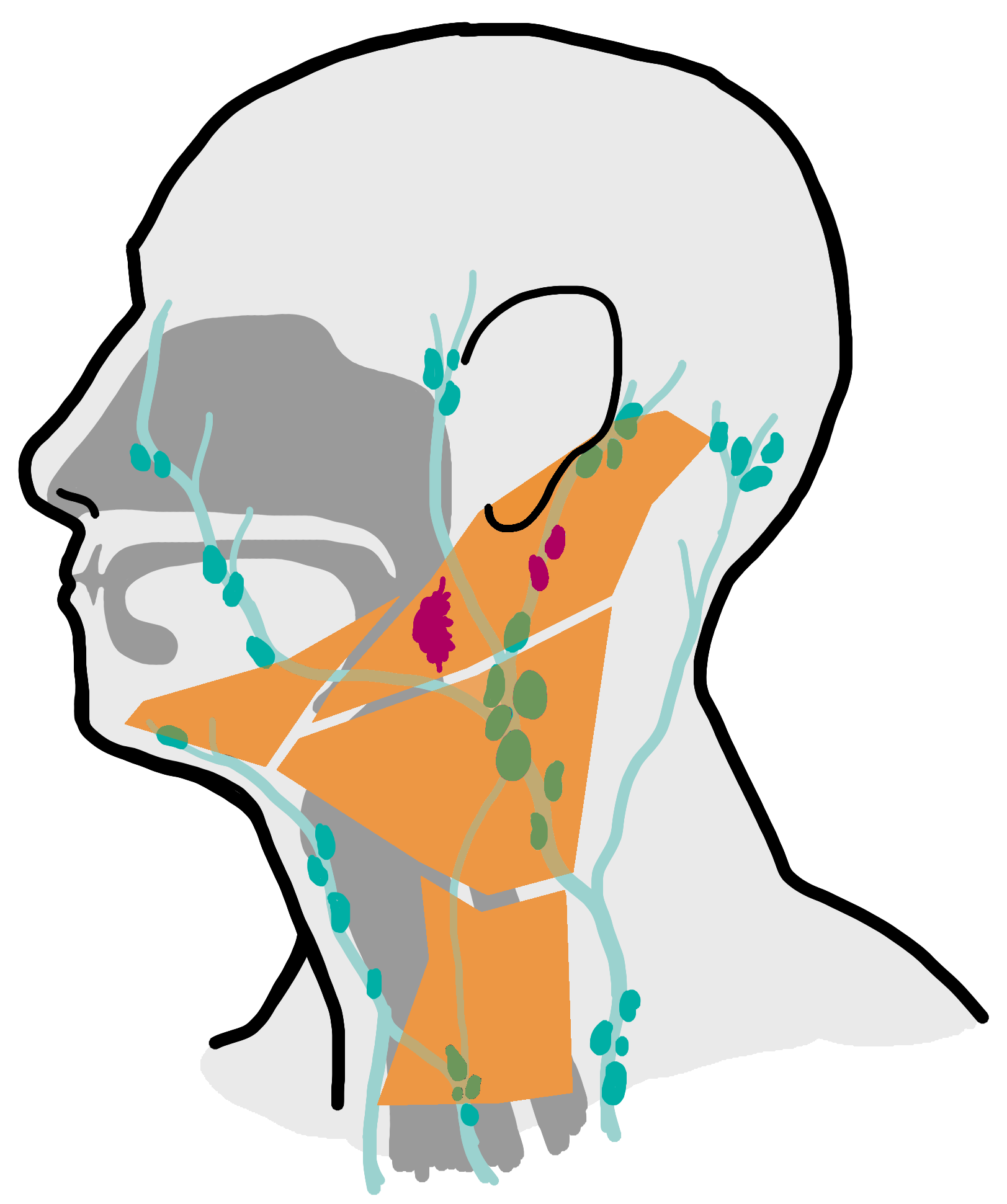}
    \end{minipage}%
    \begin{minipage}{0.48\linewidth}
        \includegraphics[width=\linewidth]{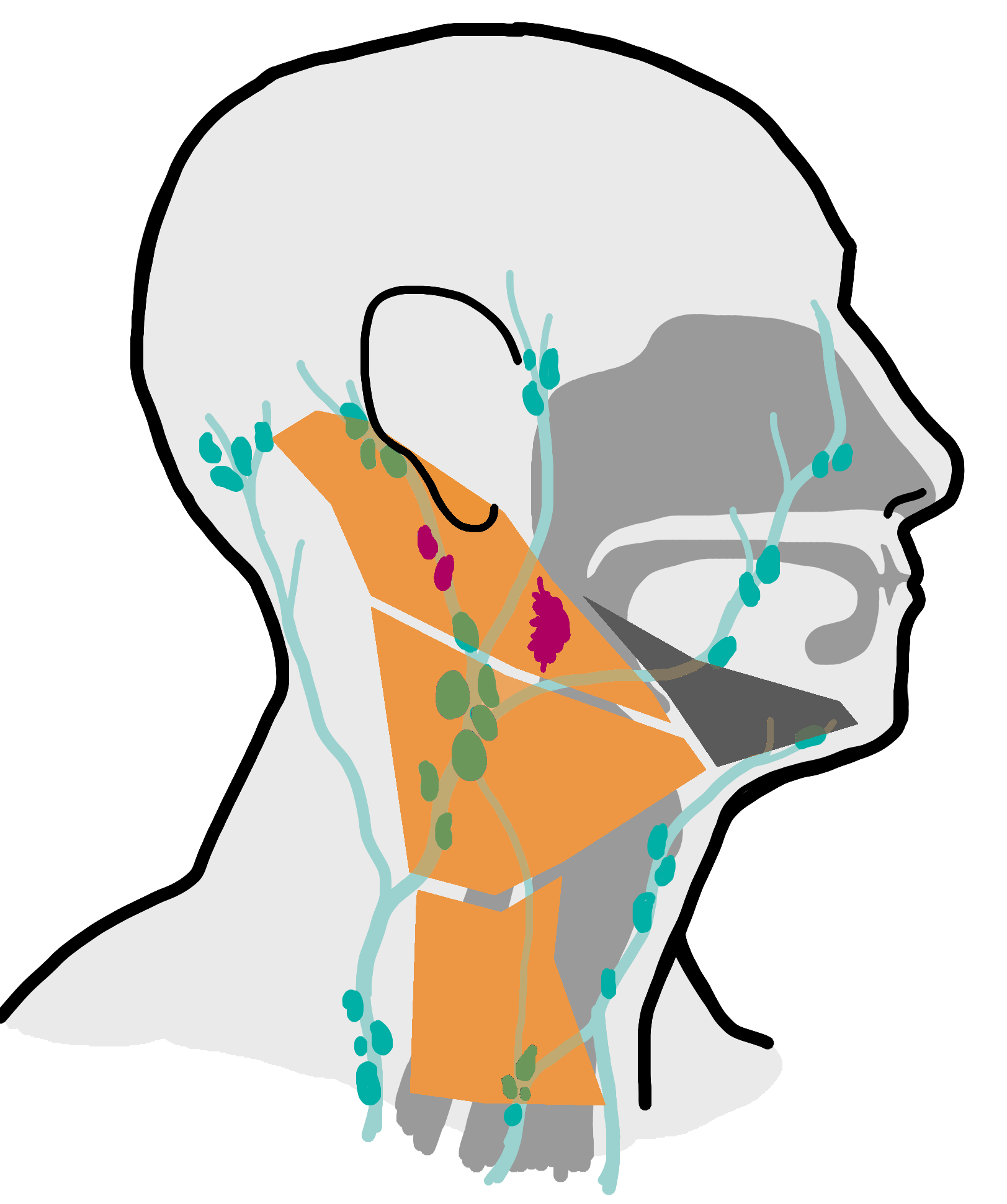}
    \end{minipage}
    
\end{subfigure}

\vspace{1mm}

\begin{subfigure}{0.45\linewidth}
    \centering
    \caption{Model-based for 5\% Risk Threshold}
    \label{fig:Treatment_Protocols_N+_II_II_5}
    \begin{minipage}{0.48\linewidth}
        \includegraphics[width=\linewidth]{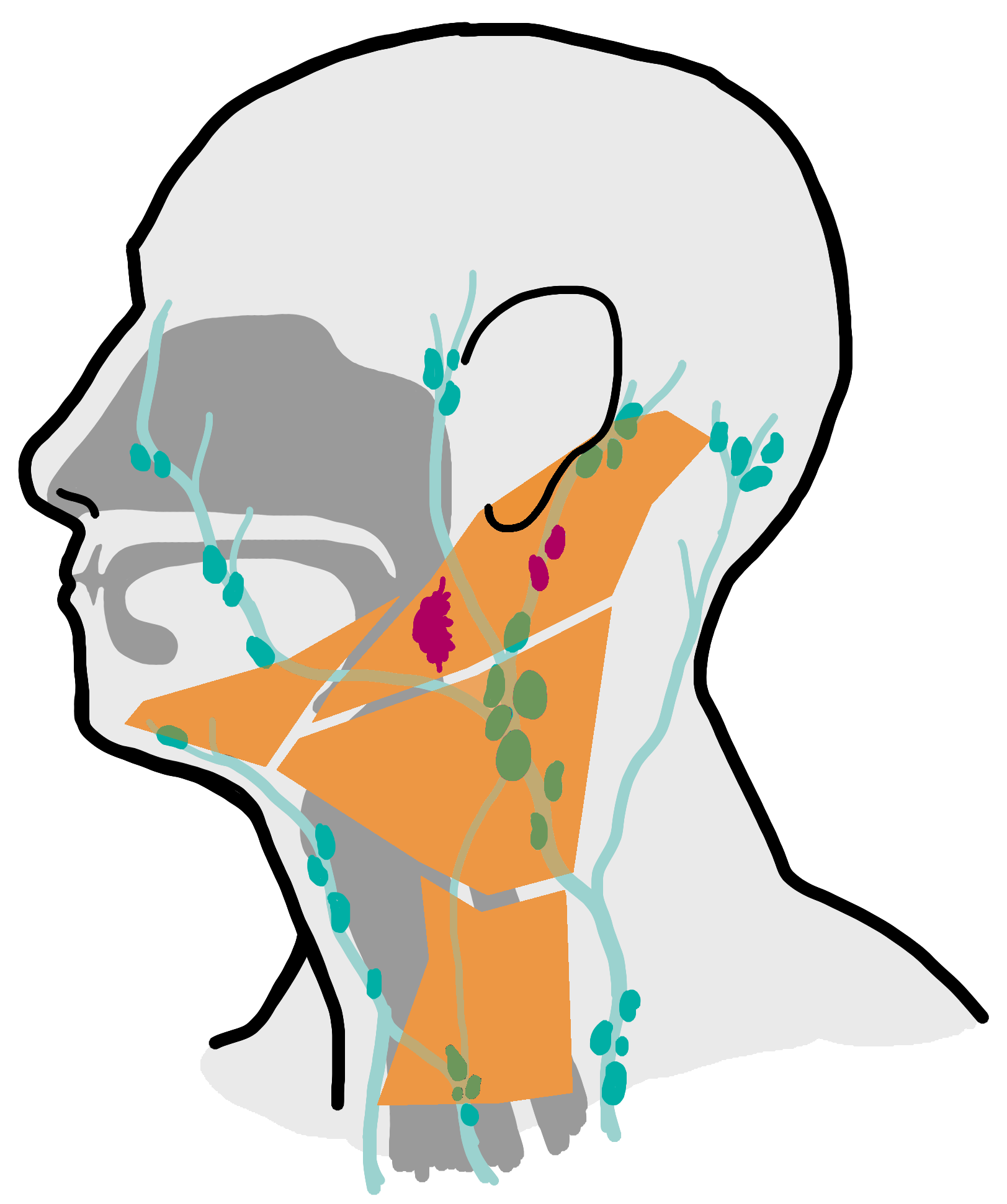}
    \end{minipage}%
    \begin{minipage}{0.48\linewidth}
        \includegraphics[width=\linewidth]{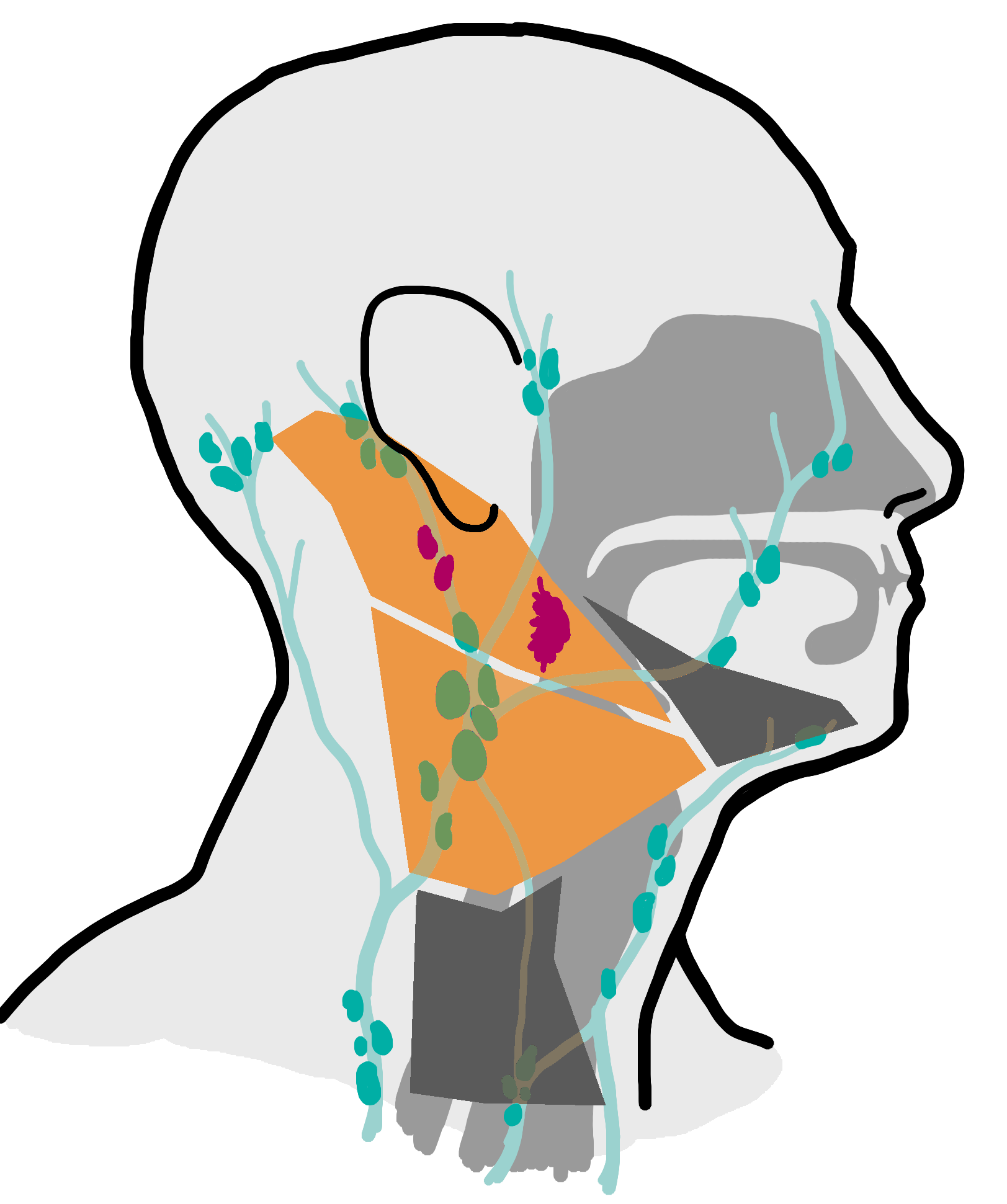}
    \end{minipage}
    
\end{subfigure}
\hfill
\begin{subfigure}{0.45\linewidth}
    \centering
    \caption{Model-based for 8\% Risk Threshold}
    \label{fig:Treatment_Protocols_N+_II_II_8}
    \begin{minipage}{0.48\linewidth}
        \includegraphics[width=\linewidth]{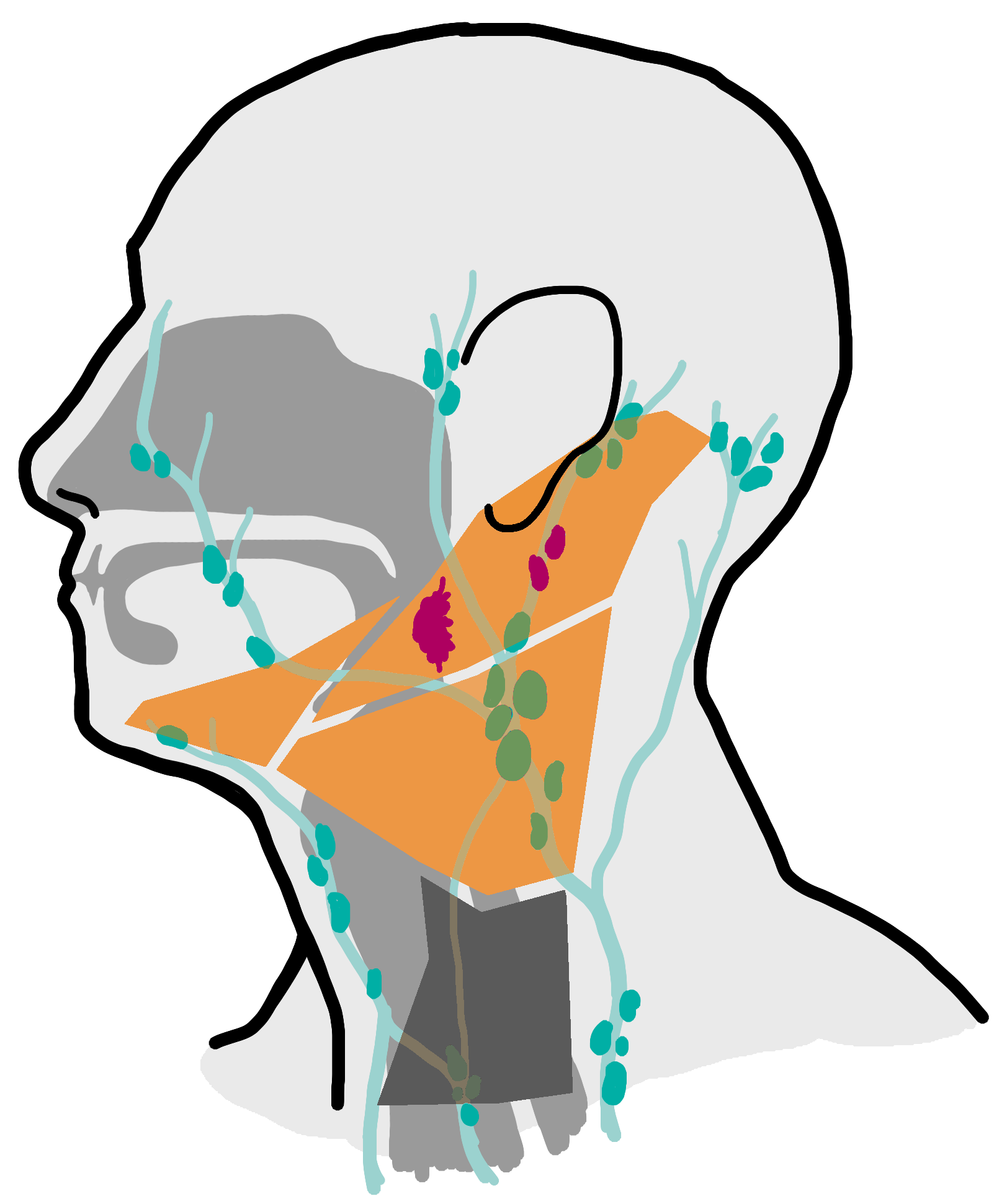}
    \end{minipage}%
    \begin{minipage}{0.48\linewidth}
        \includegraphics[width=\linewidth]{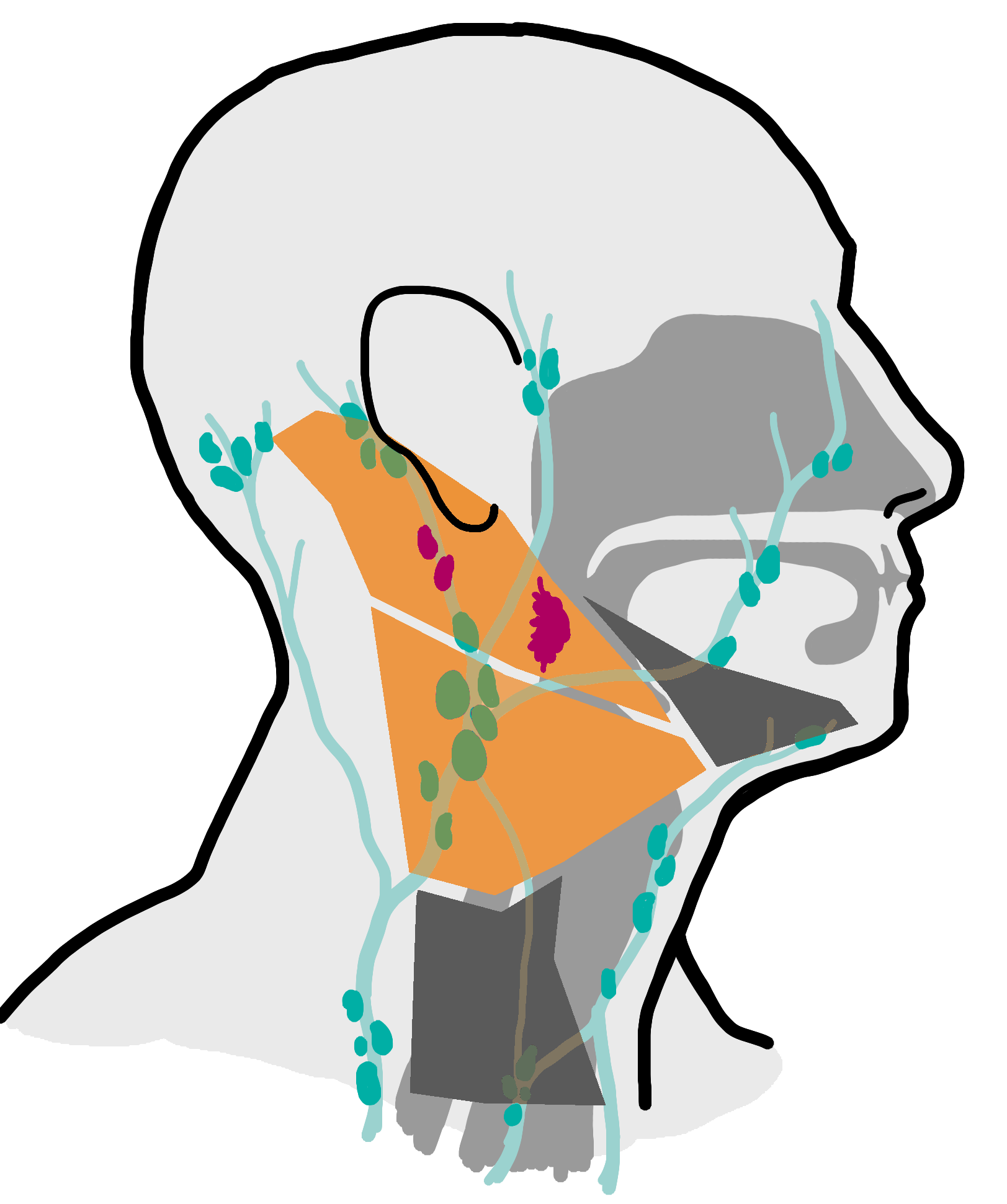}
    \end{minipage}
    
\end{subfigure}

\vspace{1mm}

\begin{subfigure}{0.45\linewidth}
    \centering
    \caption{Model-based for 10\% Risk Threshold}
    \label{fig:Treatment_Protocols_N+_II_II_10}
    \begin{minipage}{0.48\linewidth}
        \includegraphics[width=\linewidth]{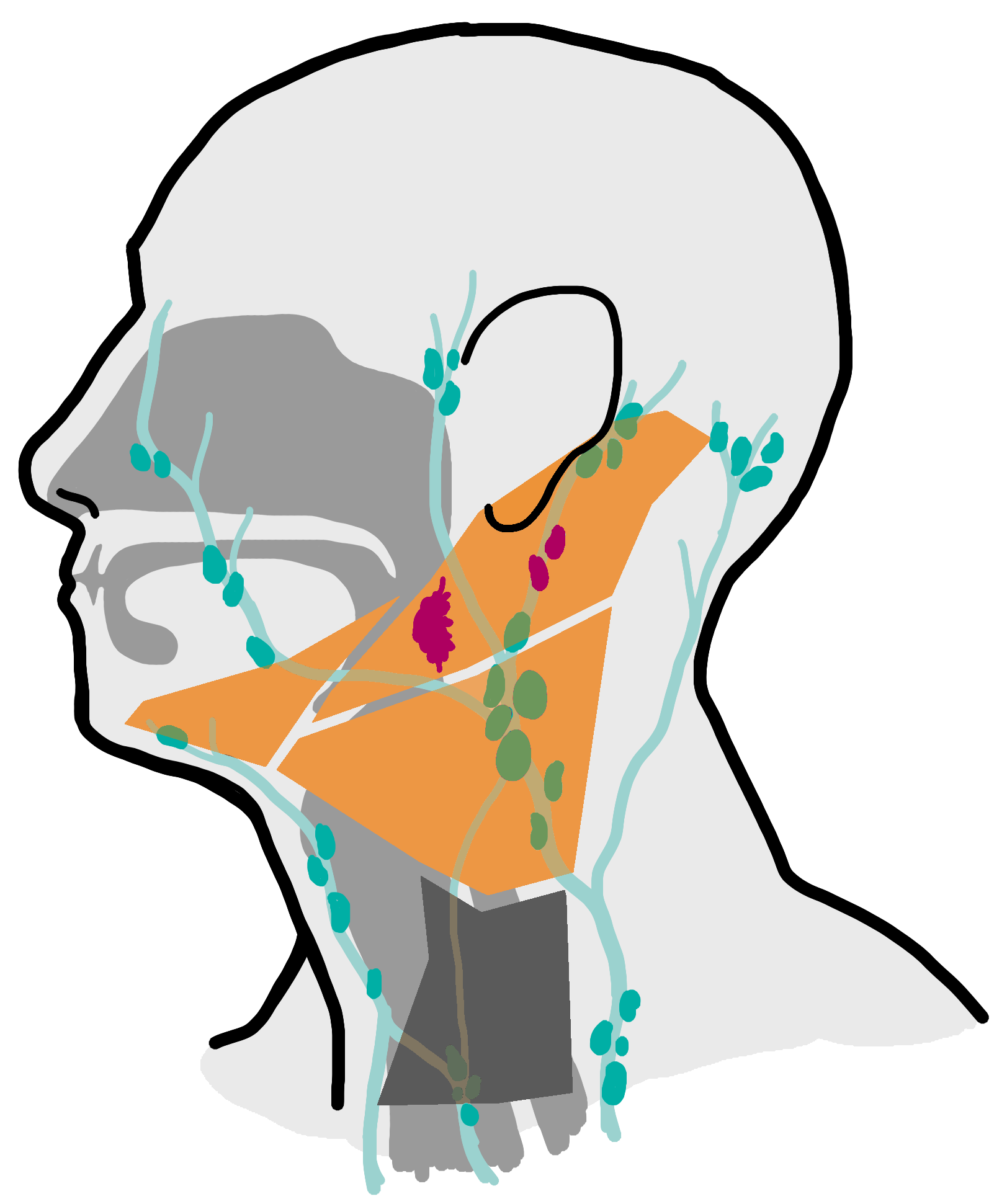}
    \end{minipage}%
    \begin{minipage}{0.48\linewidth}
        \includegraphics[width=\linewidth]{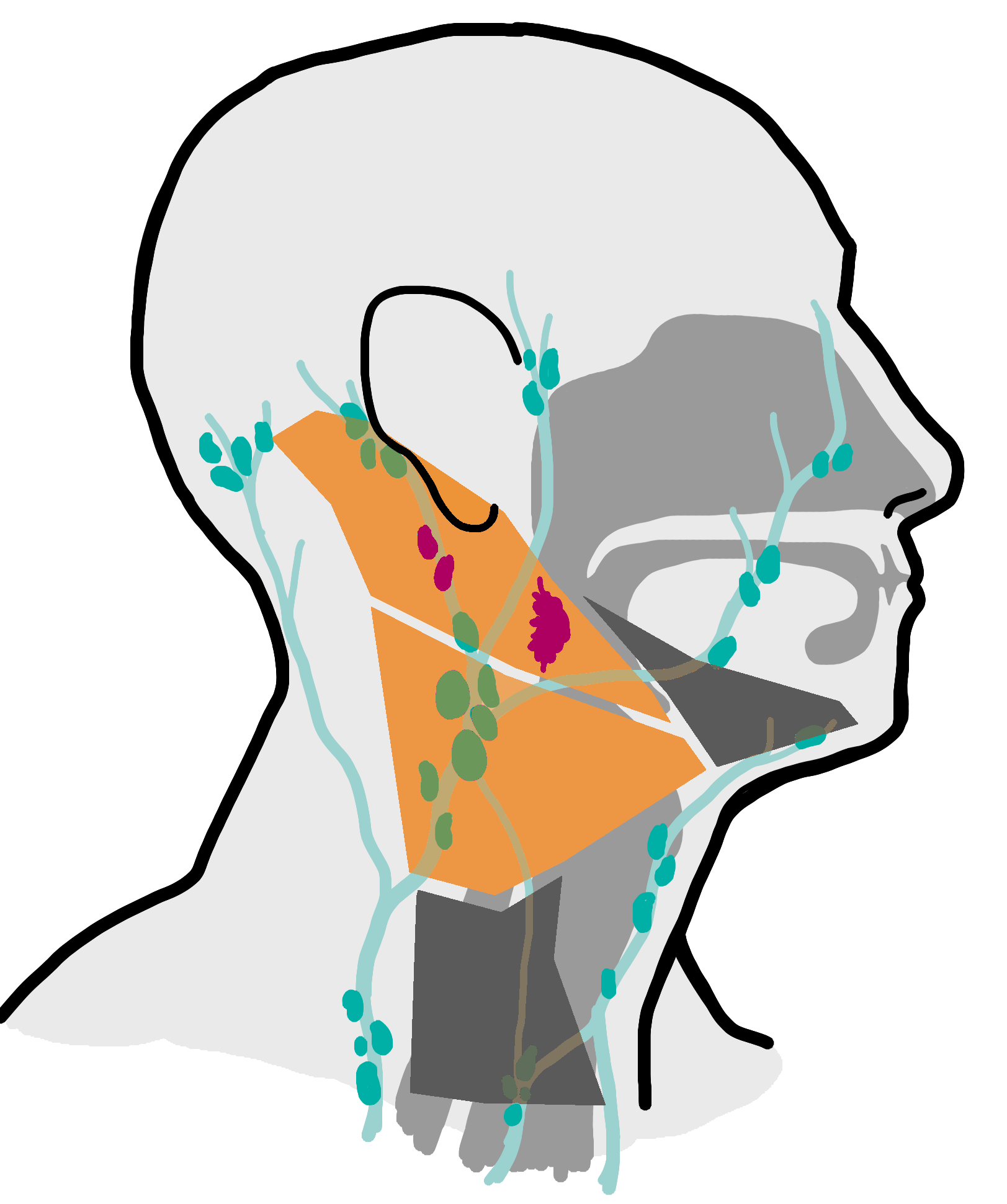}
    \end{minipage}
    
\end{subfigure}
\hfill
\begin{subfigure}{0.45\linewidth}
    \centering
    \caption{Model-based for 12\% Risk Threshold}
    \label{fig:Treatment_Protocols_N+_II_II_12}
    \begin{minipage}{0.48\linewidth}
        \includegraphics[width=\linewidth]{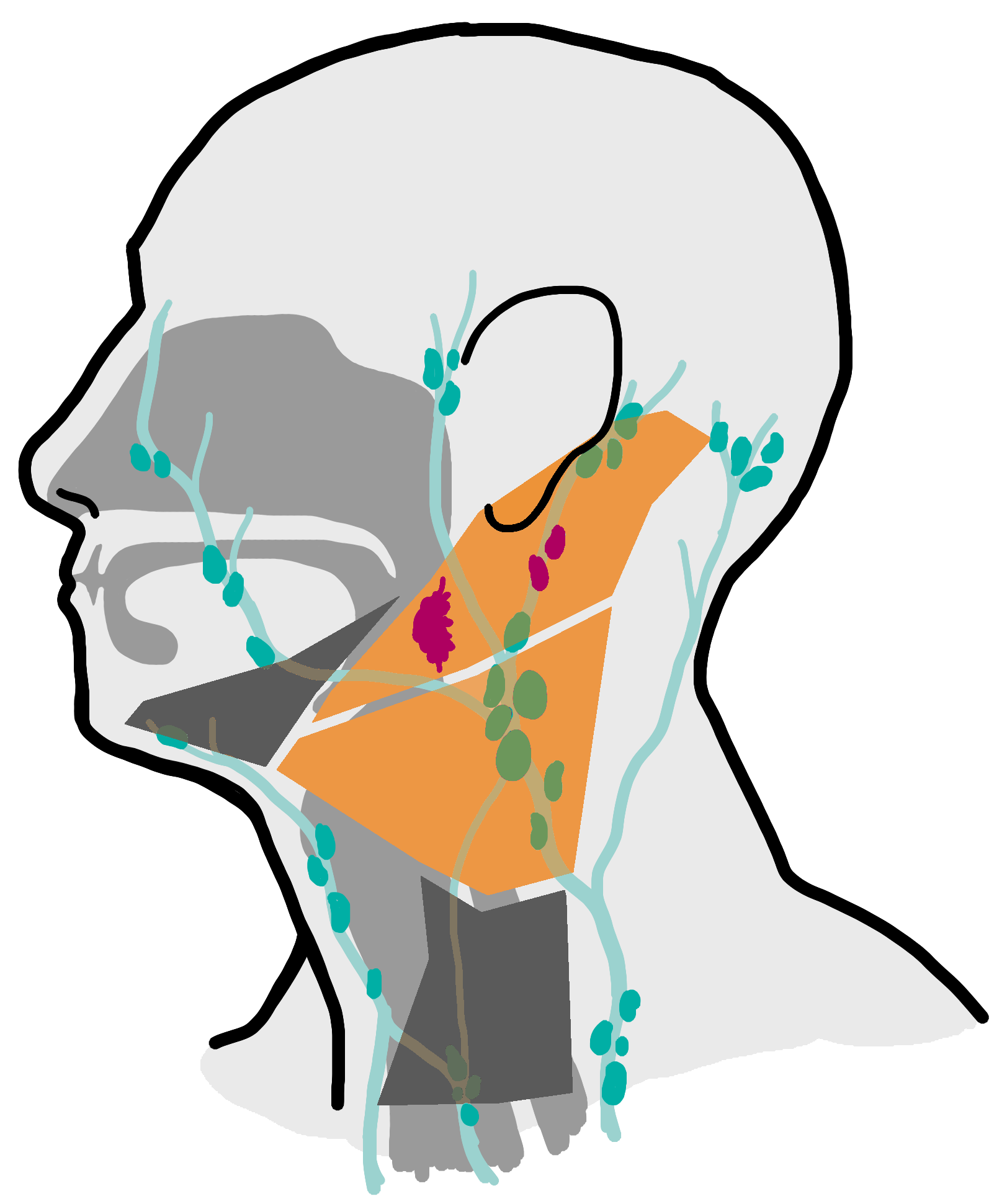}
    \end{minipage}%
    \begin{minipage}{0.48\linewidth}
        \includegraphics[width=\linewidth]{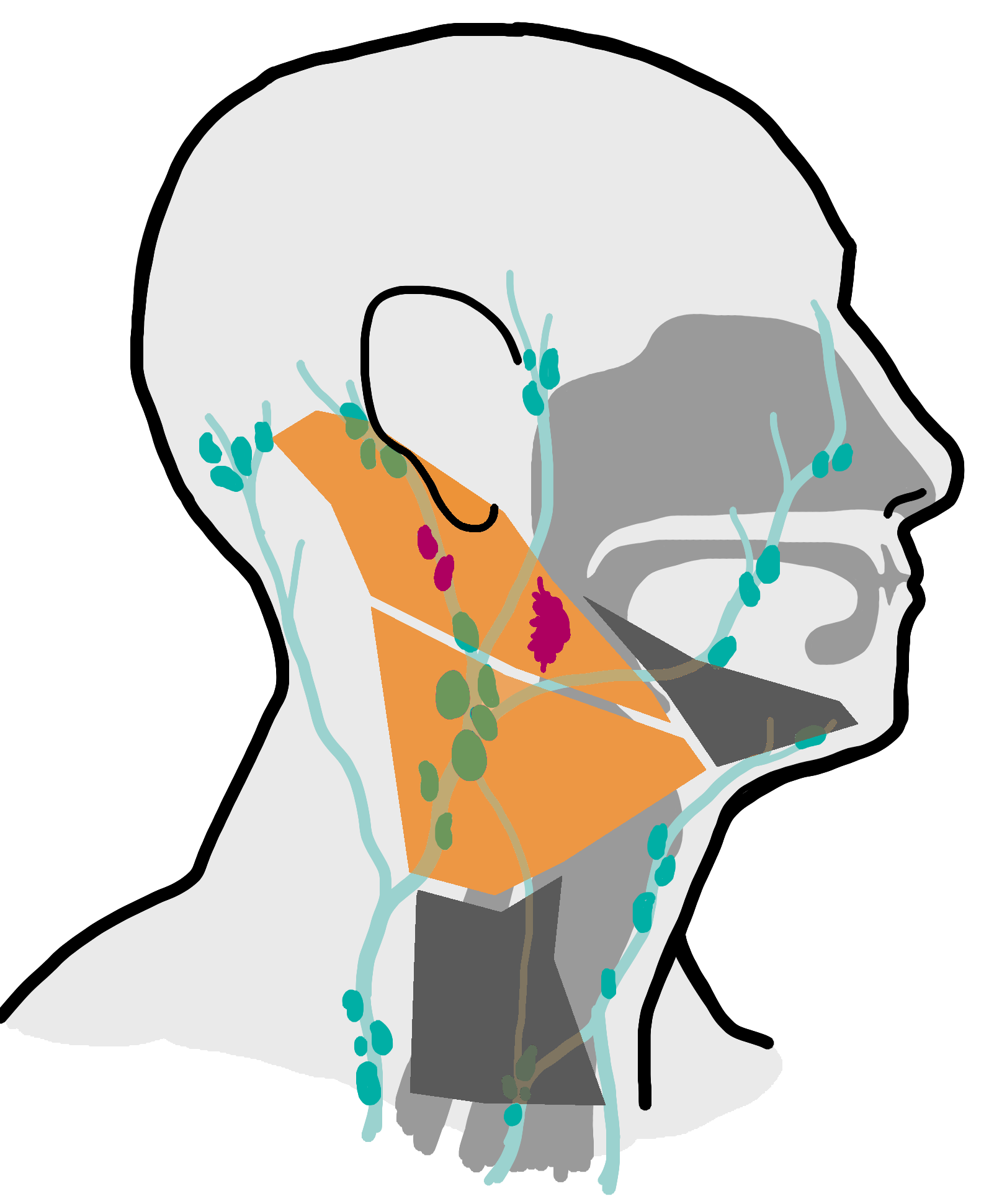}
    \end{minipage}
    
\end{subfigure}

\vspace{1mm}

\begin{subfigure}{0.45\linewidth}
    \centering
    \caption{Model-based for 15\% Risk Threshold}
    \label{fig:Treatment_Protocols_N+_II_II_15}
    \begin{minipage}{0.48\linewidth}
        \includegraphics[width=\linewidth]{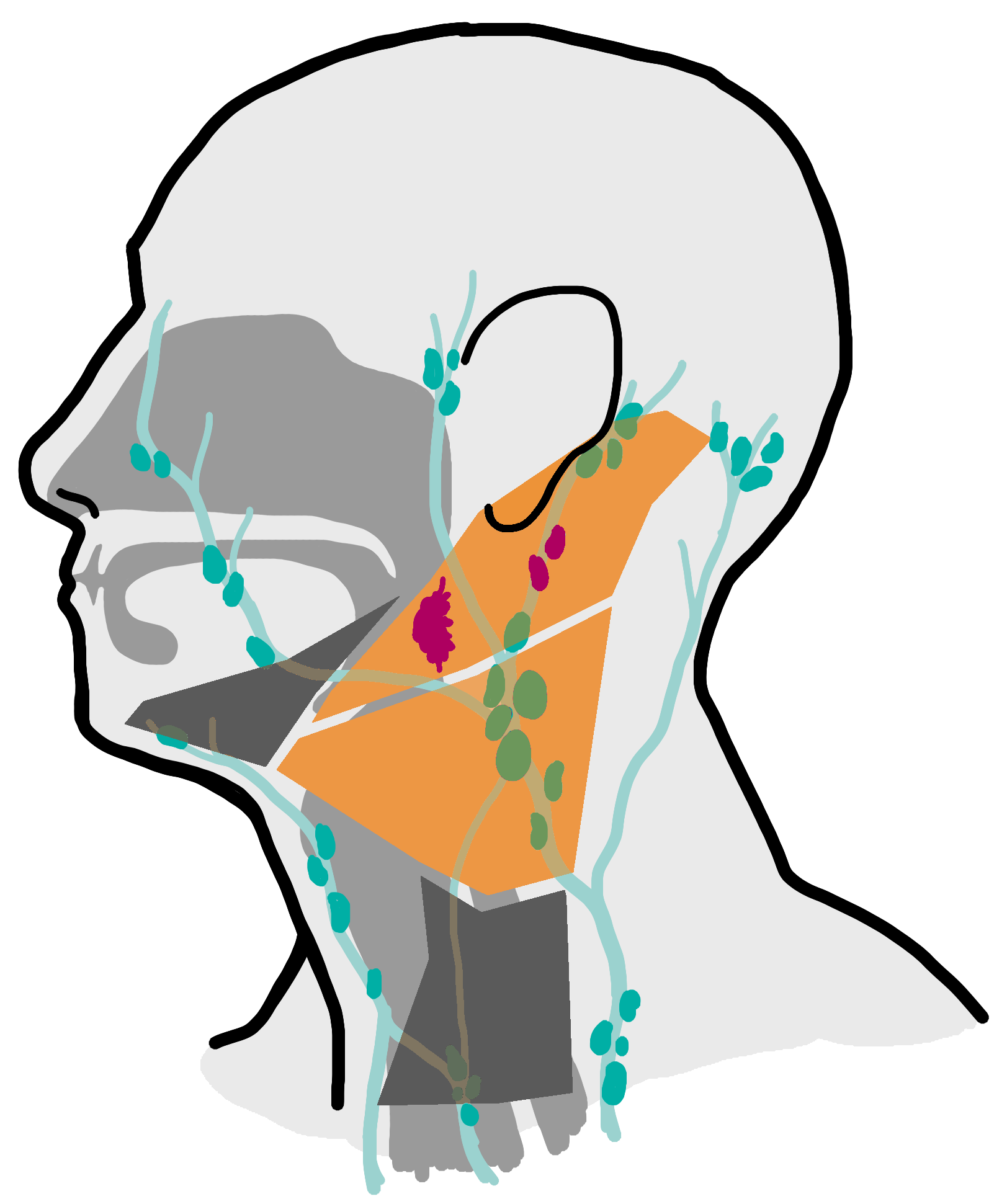}
    \end{minipage}%
    \begin{minipage}{0.48\linewidth}
        \includegraphics[width=\linewidth]{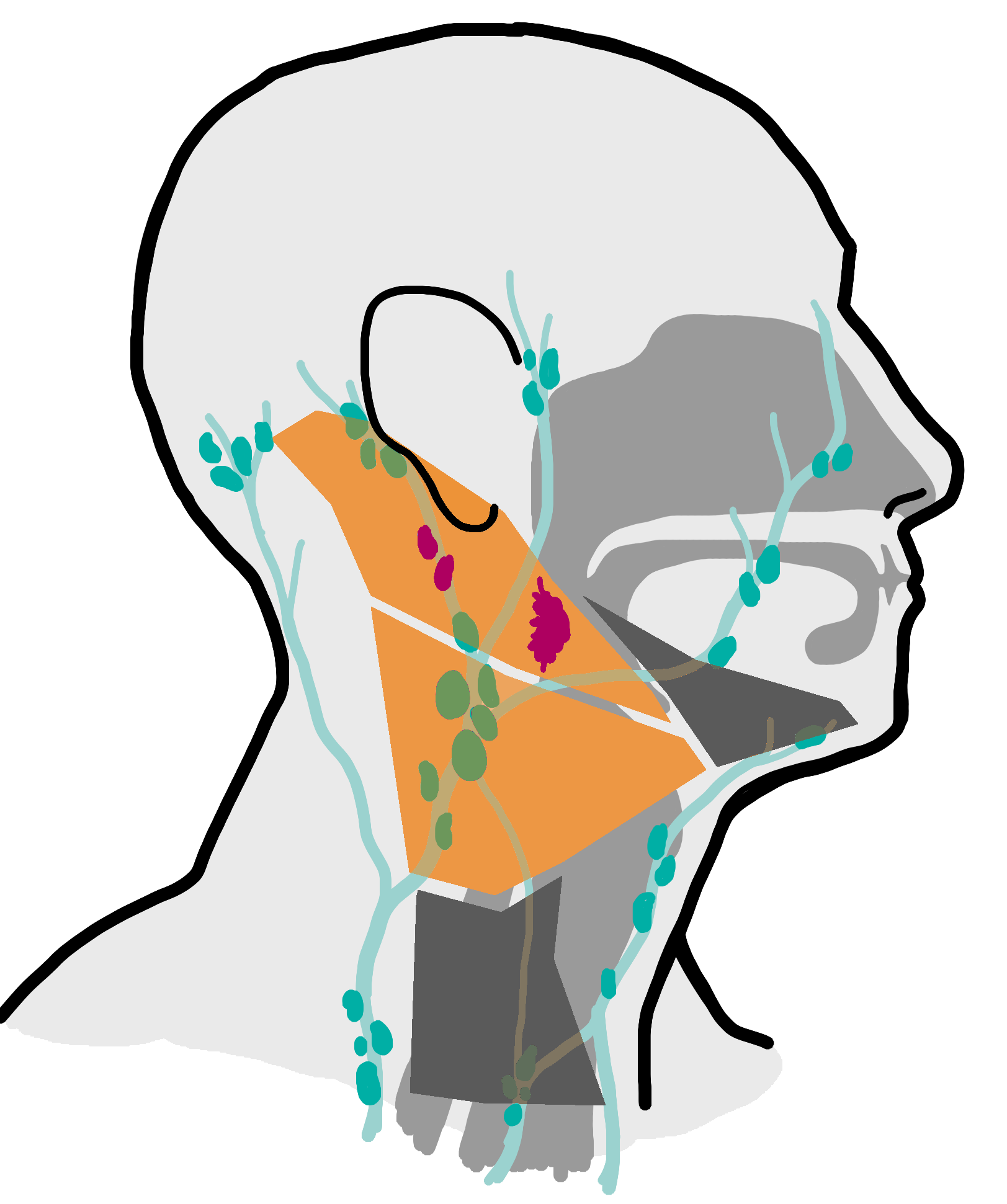}
    \end{minipage}
    
\end{subfigure}
\hfill
\begin{subfigure}{0.45\linewidth}
    \centering
    \caption{Model-based for 20\% Risk Threshold}
    \label{fig:Treatment_Protocols_N+_II_II_20}
    \begin{minipage}{0.48\linewidth}
        \includegraphics[width=\linewidth]{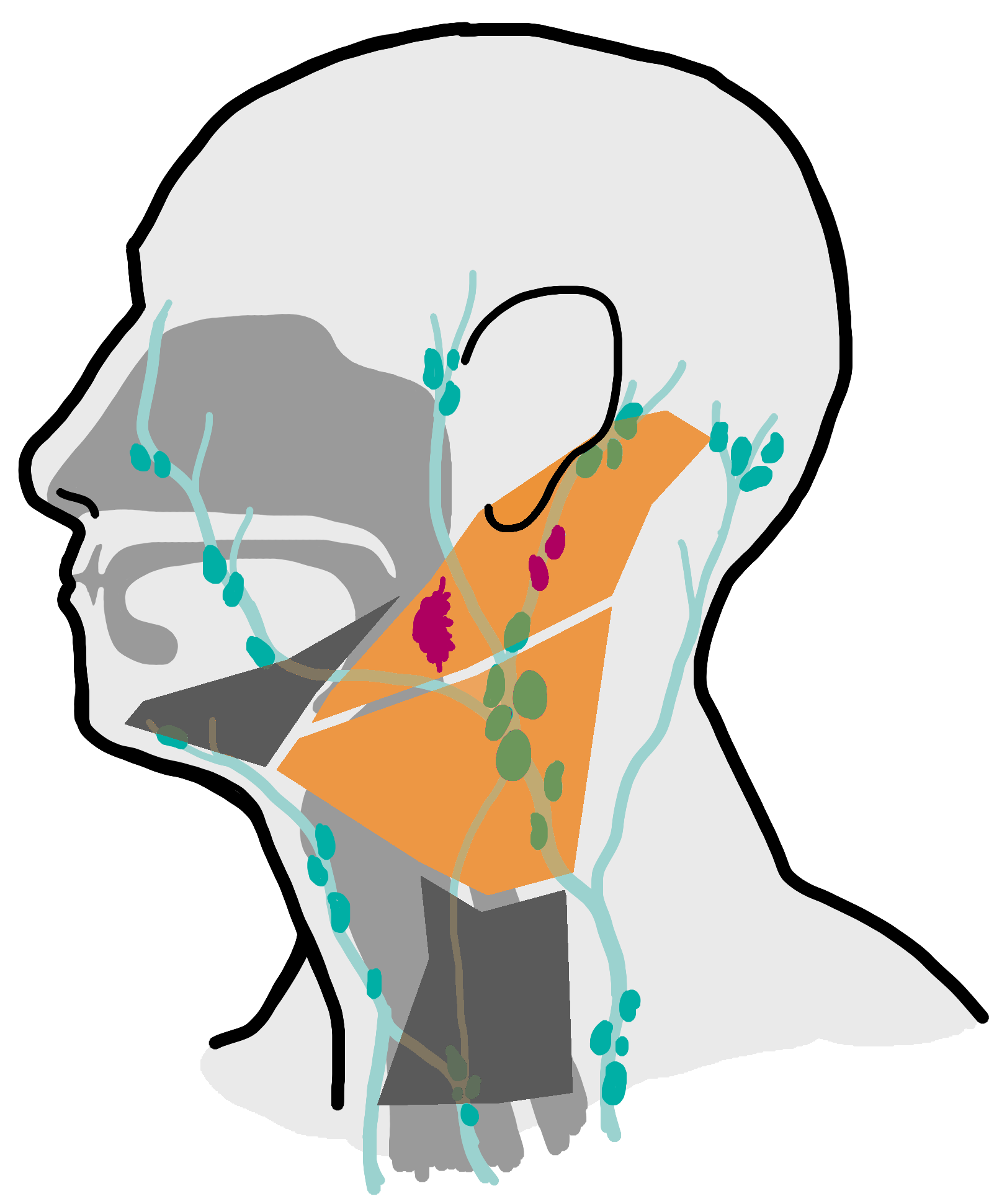}
    \end{minipage}%
    \begin{minipage}{0.48\linewidth}
        \includegraphics[width=\linewidth]{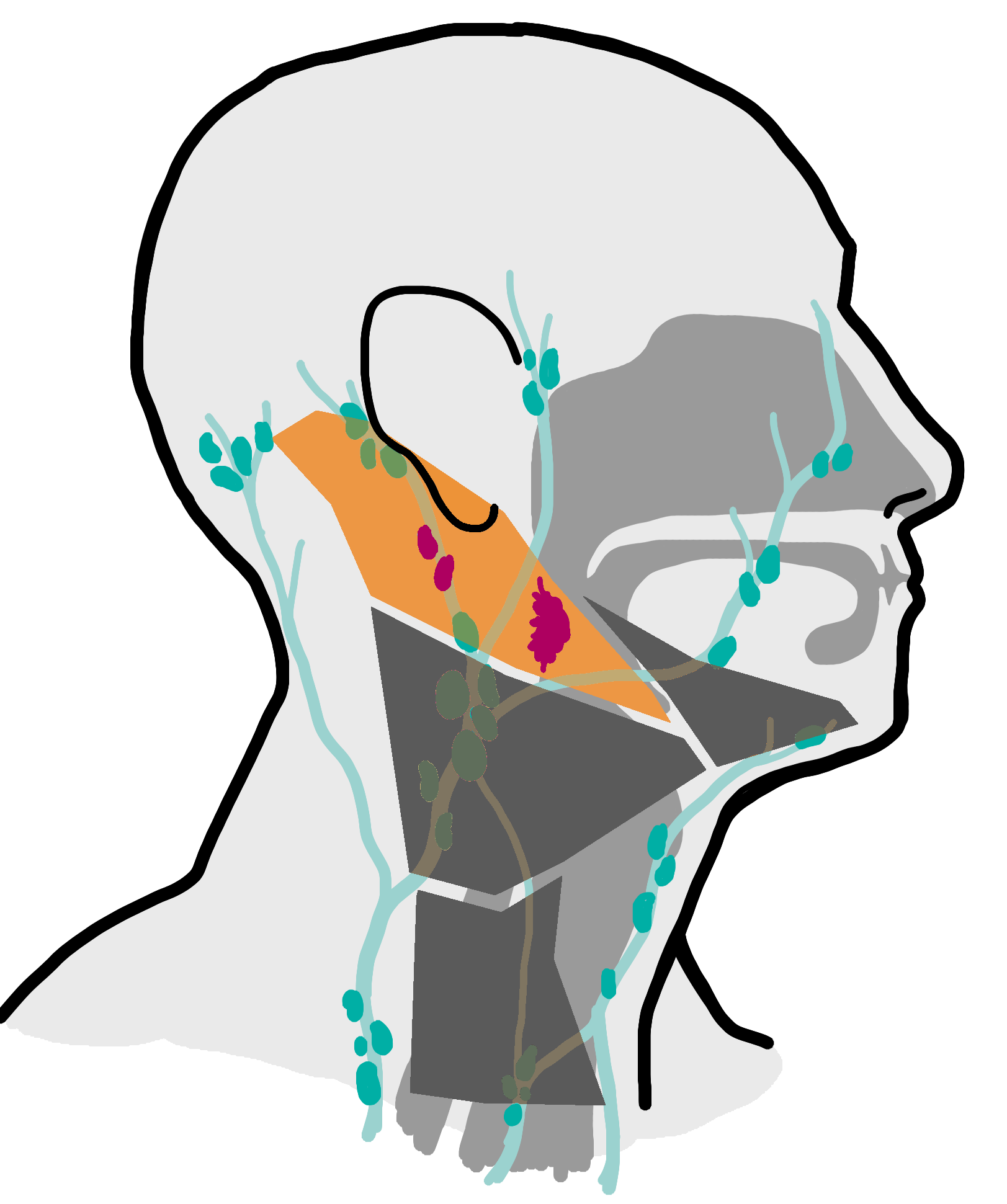}
    \end{minipage}
    
\end{subfigure}

\caption{Estimated elective nodal irradiation strategies for a patient with an advanced-stage central oropharyngeal tumour and N+ disease in ipsilateral and contralateral LNL II under varying risk thresholds. LNLs above the inclusion threshold are shown in orange (included into the CTV-E), while LNLs below the threshold are shown in gray (excluded from treatment)}
\label{fig:Treatment_Protocols_N+_II_II}
\end{figure}

A visual comparison of the conventional clinical guidelines and the risk-adaptive model-based method is provided in figure \ref{fig:Treatment_Protocols_N+_II_II}. In this scenario, where both ipsilateral and contralateral LNL II are involved, DAHANCA guidelines generally recommend treating LNLs II–IV bilaterally (figure \ref{fig:Treatment_Protocols_N+_II_II_Clinical}).

With a 2\% risk threshold, the model recommends a markedly comprehensive elective CTV-E. This encompasses the full spectrum of ipsilateral nodal levels I-IV, while also extending coverage on the contralateral side to include LNLs II-IV (figure \ref{fig:Treatment_Protocols_N+_II_II_2}). The resulting treatment field is notably more extensive than what is typically advised by conventional guidelines. The probability that any omitted nodal region harbors undetected nodal involvement is just 0.4\%. If the risk threshold is increased to 5\%, ipsilateral LNLs I–IV remain included, but contralateral coverage can be limited to LNLs II and III (figure \ref{fig:Treatment_Protocols_N+_II_II_5}), with the residual risk increasing modestly to 1.9\%. When the risk threshold is set at 10\%, further de-escalation is feasible, with elective irradiation restricted to ipsilateral LNLs I–III and contralateral LNLs II and III (figure \ref{fig:Treatment_Protocols_N+_II_II_10}). With this threshold, the estimated residual risk for microscopic involvement in excluded LNLs increases to 4.2\%. All Treatment scenarios, for the other considered risk thresholds, are visualized in figures \ref{fig:Treatment_Protocols_N+_II_II_2}–\ref{fig:Treatment_Protocols_N+_II_II_20}.


\section{Discussion}

\subsection{Main Contributions}

In this study, we propose a strategy for patient-tailored definition of the CTV-E in head and neck cancer by estimating the probability of undetected nodal involvement on a patient-specific basis. By capturing the complex, patient-specific risk of undetected nodal involvement on both sides of the neck, our approach represents an opportunity for personalization beyond conventional clinical protocols, which typically define the CTV-E using coarse, category-based classifications according to observed nodal involvement, rather than individualized risk estimates that capture the nuanced patterns of lymphatic spread.

From a clinical application standpoint, our results demonstrate that individualized risk estimates for undetected nodal involvement can be used to redefine the CTV-E in a patient-specific manner. Standard guideline-based protocols typically recommend broad coverage through large elective volumes, which is recognized as a major contributor to treatment-related toxicity, particularly xerostomia and dysphagia. By selectively including only those LNLs with a clinically meaningful probability of harboring undetected nodal involvement, our approach enables a more refined, data-driven definition of the CTV-E. Depending on the chosen risk threshold, this can allow for substantial volume reduction at a relatively low cost in terms of increased risk of missing LNLs with undetected nodal involvement. In particular, thresholds of 5\% and above were associated with consistent reductions in irradiated volumes compared to conventional practice, highlighting the potential of risk-adaptive CTV-E definitions to reduce unnecessary irradiation of healthy tissues and thereby lower the likelihood of severe side effects.

\subsection{Related Work}

To date, only a handful of studies have attempted to capture lymphatic tumor spread in formal probabilistic models, one addresses bilateral spread and two focusing on unilateral (ipsilateral) involvement \cite{ludwig_hidden_2025, pouymayou_bayesian_2019}. Taken together with the present work, these efforts demonstrate that the probability of occult metastasis can be estimated from detailed clinical information, offering what might be considered a theoretical framework. Yet, these models have yet not been tested in clinical settings, leaving their relevance for real-world decision-making still uncertain. To place our work in context, it is therefore interesting to consider related approaches where volume de-escalation has already been translated into clinical practice. 

A recent study by \citet{weissmann_reduction_2021} proposed a revised elective nodal contouring strategy for definitive radiotherapy in locally advanced head and neck squamous cell carcinoma, omitting LNLs with historically low involvement rates, particularly levels I and IV, based on the primary tumor site and disease extent. The approach was informed by a large institutional dataset on patterns of spread and applied uniformly within each clinical subgroup. In a dosimetric comparison of 30 patients, the volume de-escalated treatment plans maintained target coverage while decreasing the irradiated volume of surrounding normal tissues, leading to lower calculated mean doses to several organs-at-risk. Although this approach can simplify and standardize volume de-escalation, it is inherently limited by its reliance on historically low involvement rates of certain LNLs, and may not sufficiently account for underlying patient-to-patient variability within each subgroup, where the probability of nodal involvement remains clinically relevant.

Another study by \citet{de_veij_mestdagh_spectct_guided_2020} investigated the use of SPECT/CT-guided sentinel lymph node mapping to individualize elective nodal irradiation in head and neck cancer. Their rationale was that standard consensus-based volumes may unnecessarily cover nodal regions that are not at risk for a given patient, while occasionally missing sentinel nodes lying outside these regions. Using SPECT/CT, they observed substantial patient-to-patient differences in lymphatic drainage, and in several cases, sentinel nodes were found outside the regions normally included in guideline-defined target volumes. Incorporating these findings into treatment planning enabled selective omission of uninvolved regions while ensuring coverage of the identified sentinel nodes, thereby reducing irradiated volumes without compromising elective coverage. This demonstrates the clinical feasibility of functionally guided volume de-escalation and highlights the potential of imaging-based approaches to personalize CTV-E definitions. However, this strategy is constrained by the requirement for additional imaging procedures and the fact that drainage patterns may not fully capture the conditional dependencies between LNLs.

Finally, \citet{biau_shifting_2025} provided a comprehensive review of the current evidence on elective nodal irradiation in head and neck squamous cell carcinoma, with a particular focus on strategies to safely reduce or omit elective volumes. Synthesizing data from retrospective series, prospective trials, and emerging model-based approaches, they highlighted that contralateral omission in carefully selected cases has repeatedly demonstrated low rates of regional failure, while substantially decreasing treatment-related morbidity. At the same time, the review underscored the limits of applying population-based rules, pointing out that the safety of de-escalation depends strongly on accurate patient selection and the availability of robust predictive tools. In this context, our probabilistic framework provides a transparent, data-driven basis for elective coverage, data-driven process, where CTV-E definitions are no longer fixed by consensus but tailored to individual risk profiles.

\subsection{Limitations}

A number of limitations should be considered in interpreting our findings. Firstly, the Bayesian Network employed in this study, while interpretable and data-driven, has less modeling flexibility than a Hidden Markov Model in important ways. Specifically, the BN framework imposes conditional independence assumptions that may not fully capture the complex, sequential, or spatial dependencies often present in lymphatic spread. For instance, our BN does not directly encode dependencies between ipsilateral and contralateral nodal involvement, so that macroscopic findings on one side do not affect the probability estimates for the other side. By contrast, HMMs can correlate the ipsi- and contralateral involvement via the progression time. Nevertheless, it is noteworthy that, despite these limitations, the risk estimates and learned parameters produced by our BN are broadly consistent with those obtained from more flexible models, suggesting a degree of robustness in our approach. While HMMs have previously been extended to account for bilateral lymphatic spread, BNs have so far only been applied to model ipsilateral progression. Extending the BN framework to the bilateral setting is important because it provides an alternative probabilistic formulation for joint risk estimation across both sides of the neck. Although both BNs and HMMs can incorporate prior clinical knowledge, represent dependencies, and produce interpretable outputs, their underlying structures differ in ways that may influence model development, validation, and clinical uptake. In particular, BNs are computationally less demanding, as inference requires only a single evaluation of the network, whereas HMMs rely on repeated matrix multiplications for each time step. Introducing a bilateral BN therefore broadens the methodological toolkit available for individualized CTV-E definition, enabling comparison of approaches and offering clinicians and researchers a choice of frameworks suited to their data, preferences, and integration needs

Secondly, our approach relies on sensitivity and specificity estimates to assess observed nodal involvement. These diagnostic metrics are often variably reported across studies, due to heterogeneity in imaging modalities, patient demographics, and diagnostic thresholds, which introduce additional uncertainty into the model’s risk predictions and may limit generalizability. Recent analyses of oropharyngeal cancer datasets, where both pathological and clinical data were available, have demonstrated that sensitivity varies substantially across lymph node levels. While sensitivity for detecting metastases in LNL II aligns with literature values (around 80\%), markedly lower sensitivities are observed in LNL III and LNL IV. These findings suggest that a larger fraction of metastases remains clinically undetected in less frequently involved LNLs, and that statistical models assuming a uniform sensitivity value across all LNLs may systematically underestimate the risk of occult metastases, particularly in these distal LNLs \cite{perez_haas_modelling_2025}. To address this, future work should prioritize standardized and comprehensive reporting of diagnostic accuracy in lymph‐node assessment. Alternatively, probabilistic frameworks that explicitly model diagnostic uncertainty may offer a more robust solution. For instance, \citet{perez_haas_modelling_2025} introduced a trinary‑state HMM that simultaneously accounts for healthy, microscopic, and macroscopic nodal states, explicitly modeling the misclassifications process and capturing transitions between these states. By doing so, the model not only reflects varying confidence in imaging findings but also dynamically integrates this uncertainty into transition probabilities. 

\subsection{Perspectives}

While this work has focused on oropharyngeal cancer as a representative example, the underlying modeling framework is by no means limited to a single tumour site. In principle, the approach can be readily extended to other sub-sites within the head and neck region, or even adapted to entirely different anatomical locations, such as breast cancer, where patterns of lymphatic spread similarly inform target volume definition. The key requirement is access to relevant clinical and imaging data for estimating the models parameters. 

Another important perspective relates to the clinical validation of the proposed models. Although the risk estimates and volume recommendations presented here are derived from population-based data, their real-world clinical utility ultimately depends on prospective validation, including observed recurrence patterns. A multi-center phase II trial (NCT06563362, DeEscO) is currently investigating whether individualized, model-guided reduction of CTV-E in oropharyngeal cancer can effectively spare healthy tissue while maintaining therapeutic efficacy \cite{university_of_zurich_personalized_2025}. The outcomes of this and future studies will be essential for determining the practical feasibility of this approach and for refining model parameters based on actual patient outcomes. Collectively, such prospective validation efforts are essential to establish whether risk-adaptive, individualized target definitions can safely replace conventional guidelines without increasing the risk of recurrence.

\section{Conclusion}

In this study, we propose a data-driven alternative to standard clinical guidelines for the definition of the CTV-E in head and neck cancer. By estimating the probability of undetected nodal involvement on a patient-specific basis, our approach first and foremost enables patient-specific target definition, and depending on the chosen risk threshold, this may also lead to volume de-escalation. This is achieved by estimating the probability of nodal involvement in each LNL and incorporating a user-defined risk threshold, resulting in CTV-E definitions that are individualized to each patient’s diagnostic profile.

By capturing the complex patterns of nodal involvement on both sides of the neck, our approach offers a perspective that may help inform refinements of current practice. The model provides patient-specific risk estimates for nodal involvement, offering a quantitative basis on which clinicians can refine target volumes and adapt treatment strategies. In doing so, it offers a pathway to more personalized radiotherapy with less related morbidity, while maintaining therapeutic efficacy. These findings support a move towards refining existing clinical guidelines to incorporate individualized, risk-based criteria in defining the CTV-E, potentially allowing for volume de-escalation, when a clinically acceptable risk threshold can be met.

\bibliographystyle{unsrtnat}  
\bibliography{references}     

\appendix
\newpage
\section{Configuration for Monte Carlo Markov Sampling}
\label{app:Sampling}


\begin{table}[H]
\centering
\begin{threeparttable}
\caption{The Monte Carlo Markov Chain sampling configurations with the retained sample count}
\label{tab:sampling_config}
\renewcommand{\arraystretch}{1.15}
\setlength{\tabcolsep}{6pt}
\sisetup{group-separator = {\,}, group-minimum-digits = 4} 
\begin{tabularx}{\textwidth}{@{}l l X@{}}
\toprule
\textbf{Setting} & \textbf{Value} & \textbf{Notes} \\
\midrule
Number of walkers & $500$ & Parallel, connected ensemble of Markov chains (“walkers”). \\
Steps per walker  & $9000$   & Total steps generated per walker. \\
Initialization    & $\mathcal{U}([0,1]^d)$ & Each walker initialized at a random position drawn uniformly. \\
Proposal mixture (DE/Snooker)  & 80\% \,/\, 20\%  & Differential–evolution moves and snooker updates to efficiently traverse the parameter space while remaining robust to affine transformations. \\
Burn-in (steps per walker)        & 6000  & First \num{6000} steps of each chain discarded. \\
Retained per walker (steps) & 3000  & Used to form the posterior sample set. \\
\addlinespace[2pt]
\textbf{Total retained samples} & \(\num{3000} \times \num{500} = \num{1500000}\) & Posterior sample set size after burn-in removal. \\
\bottomrule
\end{tabularx}
\begin{tablenotes}
\footnotesize
\item $\mathcal{U}([0,1]^d)$: uniform distribution on the $d$-dimensional unit hypercube. DE: differential–evolution move; “snooker”: snooker update (See \cite{nelson_run_2013, ter_braak_differential_2008} for details).
\end{tablenotes}
\end{threeparttable}
\end{table}

\section{Transition and Base Probabilities}
\label{app:params}


\begin{table}[H]
\centering
\begin{threeparttable}
\caption{Posterior medians with asymmetric bounds for transition ($\mathrm{t}$) and base ($\mathrm{b}$) probabilities. 
$\mathrm{X}^{\mathrm{i}/\mathrm{c}}_{\mathrm{I\text{--}IV}}$ denote ipsilateral/contralateral LNLs I–IV, and $\mathrm{T}$ is the primary tumour. Bounds reflect the reported asymmetric credible intervals.}
\label{tab:BN_Params}
\renewcommand{\arraystretch}{1.15}
\setlength{\tabcolsep}{5.5pt}
\begin{tabularx}{\textwidth}{@{}L
S[table-format=1.4] S[table-format=1.4]@{\text{-} \,}S[table-format=1.4]
S[table-format=1.4] S[table-format=1.4]@{\text{-} \,}S[table-format=1.4]@{}}
\toprule
& \multicolumn{3}{c}{Early-stage} & \multicolumn{3}{c}{Advanced-stage} \\
\cmidrule(lr){2-4}\cmidrule(lr){5-7}
Arcs & {Median} & \multicolumn{2}{c}{Bounds} & {Median} & \multicolumn{2}{c}{Bounds} \\
\midrule

\multicolumn{7}{@{}l}{\textbf{Ipsilateral Transition Probabilities} ($\mathrm{t}^{\mathrm{i}}$)}\\
\addlinespace[2pt]
$\mathrm{X}^{\mathrm{i}}_{\mathrm{I}}\!\to\!\mathrm{X}^{\mathrm{i}}_{\mathrm{II}} \; (\mathrm{t}^{\mathrm{i}}_{1,2})$
  & 0.200 & 0.060 & 0.400 & 0.170 & 0.050 & 0.360 \\
$\mathrm{X}^{\mathrm{i}}_{\mathrm{II}}\!\to\!\mathrm{X}^{\mathrm{i}}_{\mathrm{III}} \; (\mathrm{t}^{\mathrm{i}}_{2,3})$
  & 0.350 & 0.310 & 0.390 & 0.430 & 0.370 & 0.480 \\
$\mathrm{X}^{\mathrm{i}}_{\mathrm{III}}\!\to\!\mathrm{X}^{\mathrm{i}}_{\mathrm{IV}} \; (\mathrm{t}^{\mathrm{i}}_{3,4})$
  & 0.200 & 0.160 & 0.240 & 0.180 & 0.130 & 0.230 \\

\addlinespace[2pt]
\multicolumn{7}{@{}l}{\textbf{Ipsilateral Base Probabilities} ($\mathrm{b}^{\mathrm{i}}$)}\\
\addlinespace[2pt]
$\mathrm{T}\!\to\!\mathrm{X}^{\mathrm{i}}_{\mathrm{I}} \; (\mathrm{b}^{\mathrm{i}}_{1})$
  & 0.035 & 0.026 & 0.045 & 0.120 & 0.100 & 0.150 \\
$\mathrm{T}\!\to\!\mathrm{X}^{\mathrm{i}}_{\mathrm{II}} \; (\mathrm{b}^{\mathrm{i}}_{2})$
  & 0.790 & 0.770 & 0.810 & 0.860 & 0.830 & 0.880 \\
$\mathrm{T}\!\to\!\mathrm{X}^{\mathrm{i}}_{\mathrm{III}} \; (\mathrm{b}^{\mathrm{i}}_{3})$
  & 0.063 & 0.043 & 0.093 & 0.075 & 0.035 & 0.135 \\
$\mathrm{T}\!\to\!\mathrm{X}^{\mathrm{i}}_{\mathrm{IV}} \; (\mathrm{b}^{\mathrm{i}}_{4})$
  & 0.009 & 0.004 & 0.015 & 0.029 & 0.019 & 0.049 \\

\addlinespace[2pt]
\multicolumn{7}{@{}l}{\textbf{Contralateral Transition Probabilities} ($\mathrm{t}^{\mathrm{c}}$)}\\
\addlinespace[2pt]
$\mathrm{X}^{\mathrm{c}}_{\mathrm{I}}\!\to\!\mathrm{X}^{\mathrm{c}}_{\mathrm{II}} \; (\mathrm{t}^{\mathrm{c}}_{1,2})$
  & 0.270 & 0.080 & 0.550 & 0.270 & 0.080 & 0.550 \\
$\mathrm{X}^{\mathrm{c}}_{\mathrm{II}}\!\to\!\mathrm{X}^{\mathrm{c}}_{\mathrm{III}} \; (\mathrm{t}^{\mathrm{c}}_{2,3})$
  & 0.200 & 0.150 & 0.260 & 0.230 & 0.180 & 0.280 \\
$\mathrm{X}^{\mathrm{c}}_{\mathrm{III}}\!\to\!\mathrm{X}^{\mathrm{c}}_{\mathrm{IV}} \; (\mathrm{t}^{\mathrm{c}}_{3,4})$
  & 0.230 & 0.130 & 0.350 & 0.330 & 0.240 & 0.420 \\

\addlinespace[2pt]
\multicolumn{7}{@{}l}{\textbf{Contralateral Base Probabilities} ($\mathrm{b}^{\mathrm{c}}$)}\\
\addlinespace[2pt]
$\mathrm{T}\!\to\!\mathrm{X}^{\mathrm{c}}_{\mathrm{I}} \; (\mathrm{b}^{\mathrm{c}}_{1})$
  & 0.006 & 0.001 & 0.009 & 0.010 & 0.004 & 0.018 \\
$\mathrm{T}\!\to\!\mathrm{X}^{\mathrm{c}}_{\mathrm{II}} \; (\mathrm{b}^{\mathrm{c}}_{2})$
  & 0.140 & 0.120 & 0.160 & 0.450 & 0.410 & 0.490 \\
$\mathrm{T}\!\to\!\mathrm{X}^{\mathrm{c}}_{\mathrm{III}} \; (\mathrm{b}^{\mathrm{c}}_{3})$
  & 0.010 & 0.005 & 0.015 & 0.040 & 0.020 & 0.060 \\
$\mathrm{T}\!\to\!\mathrm{X}^{\mathrm{c}}_{\mathrm{IV}} \; (\mathrm{b}^{\mathrm{c}}_{4})$
  & 0.002 & 0.000 & 0.005 & 0.010 & 0.003 & 0.020 \\

\bottomrule
\end{tabularx}
\begin{tablenotes}
\footnotesize
\item \emph{Note:} Values are reported as median and bounds \([\,\text{lower},\,\text{upper}\,]\) derived from the original asymmetric uncertainties \(x^{+u}_{-l}\) (lower \(= x-l\), upper \(= x+u\)).
\end{tablenotes}
\end{threeparttable}
\end{table}

\includepdf[
  pages=1,
  pagecommand={\section{Treatment Protocols for Early-stage Tumours with Risk Threshold of 2\%}\label{app:proto_early_2}}
]{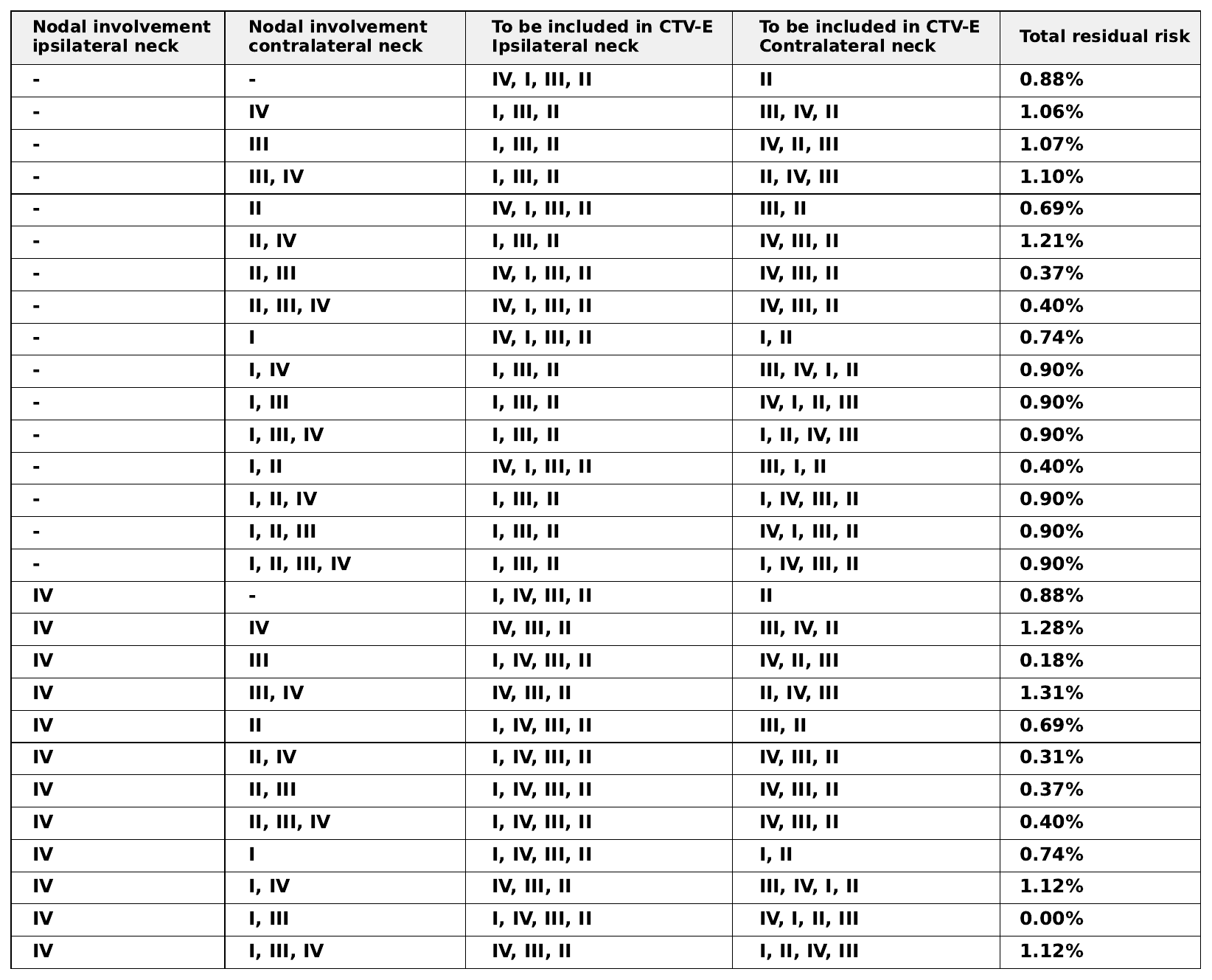}
\includepdf[
  pages=2-,
  pagecommand={}
]{pdf_files/Proposed_Treatment_Protocols_tr_0.02_early}

\includepdf[
  pages=1,
  pagecommand={\section{Treatment Protocols for Advanced-stage Tumours with Risk Threshold of 2\%}}
]{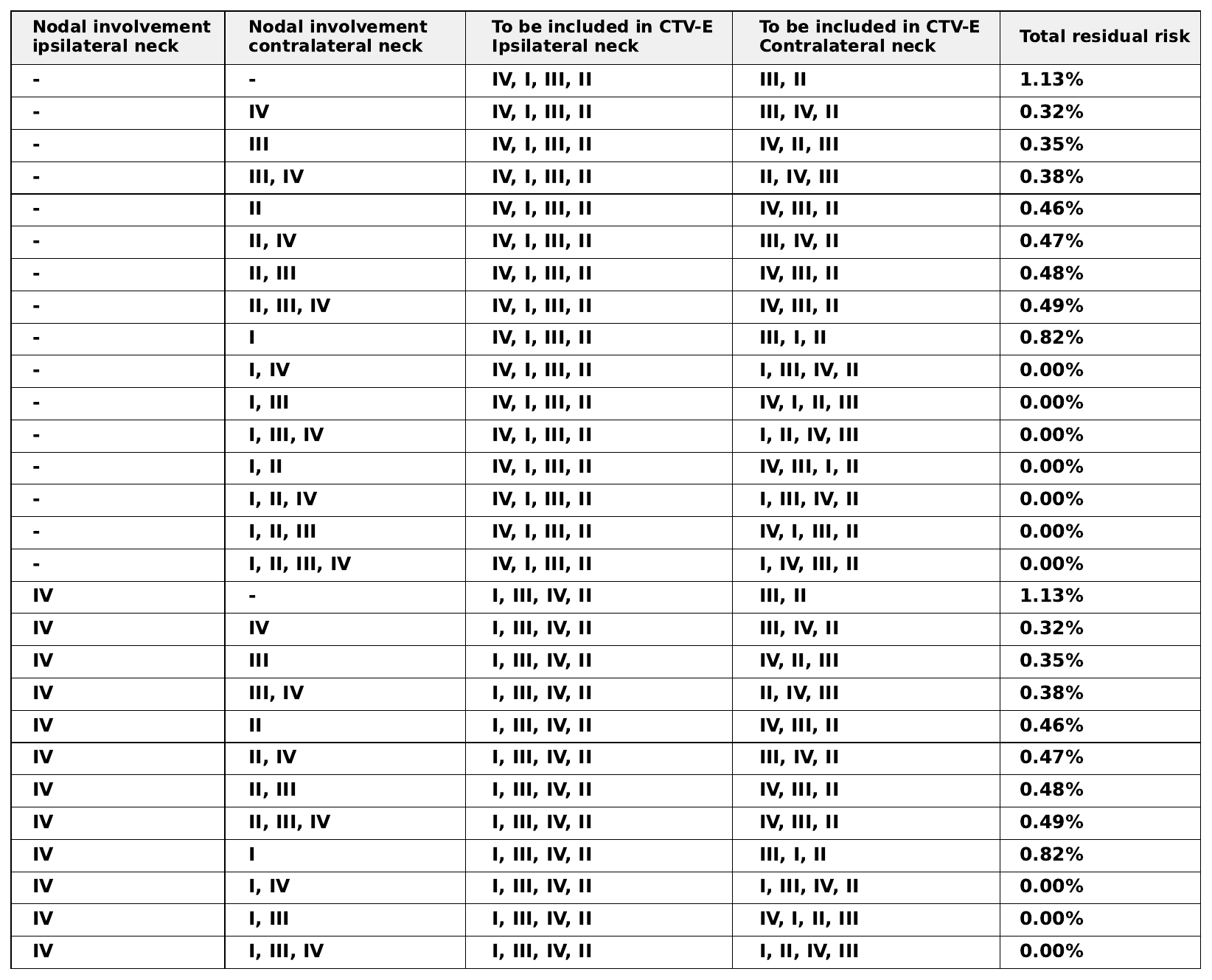}
\includepdf[
  pages=2-,
  pagecommand={}
]{pdf_files/Proposed_Treatment_Protocols_tr_0.02_advanced}

\includepdf[
  pages=1,
  pagecommand={\section{Treatment Protocols for Early-stage Tumours with Risk Threshold of 5\%}}
]{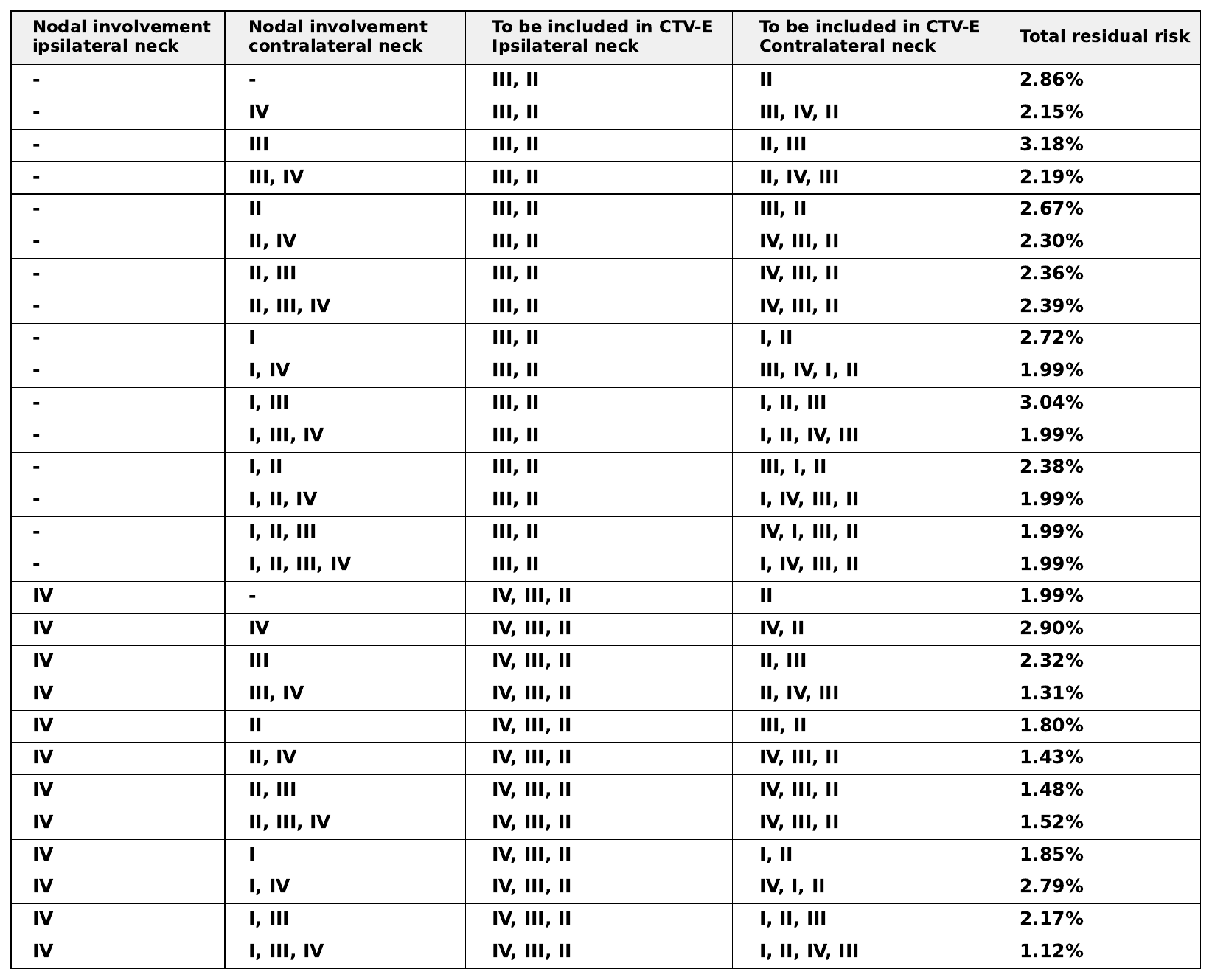}
\includepdf[
  pages=2-,
  pagecommand={}
]{pdf_files/Proposed_Treatment_Protocols_tr_0.05_early}

\includepdf[
  pages=1,
  pagecommand={\section{Treatment Protocols for Advanced-stage Tumours with Risk Threshold of 5\%}}
]{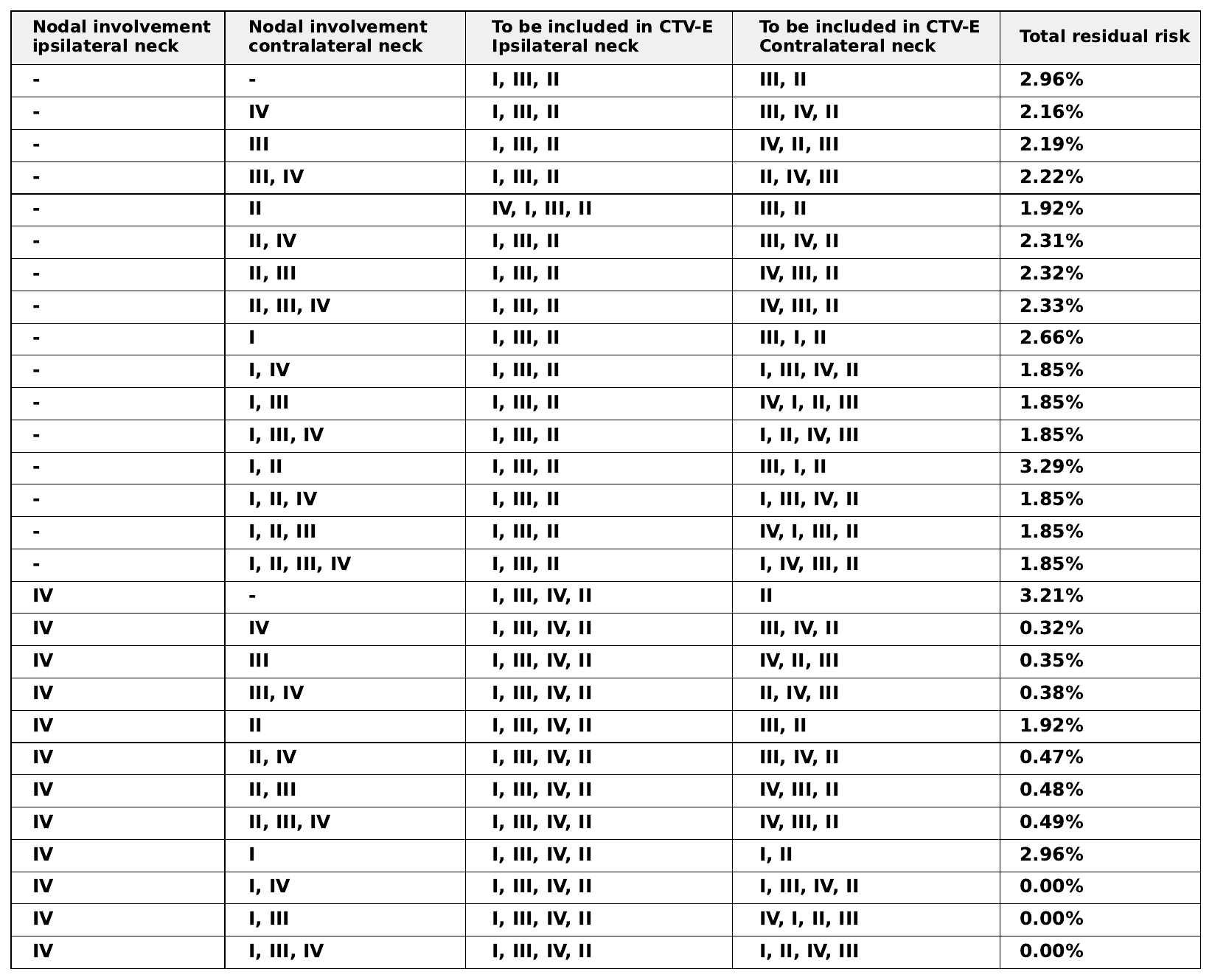}
\includepdf[
  pages=2-,
  pagecommand={}
]{pdf_files/Proposed_Treatment_Protocols_tr_0.05_advanced}

\includepdf[
  pages=1,
  pagecommand={\section{Treatment Protocols for Early-stage Tumours with Risk Threshold of 8\%}}
]{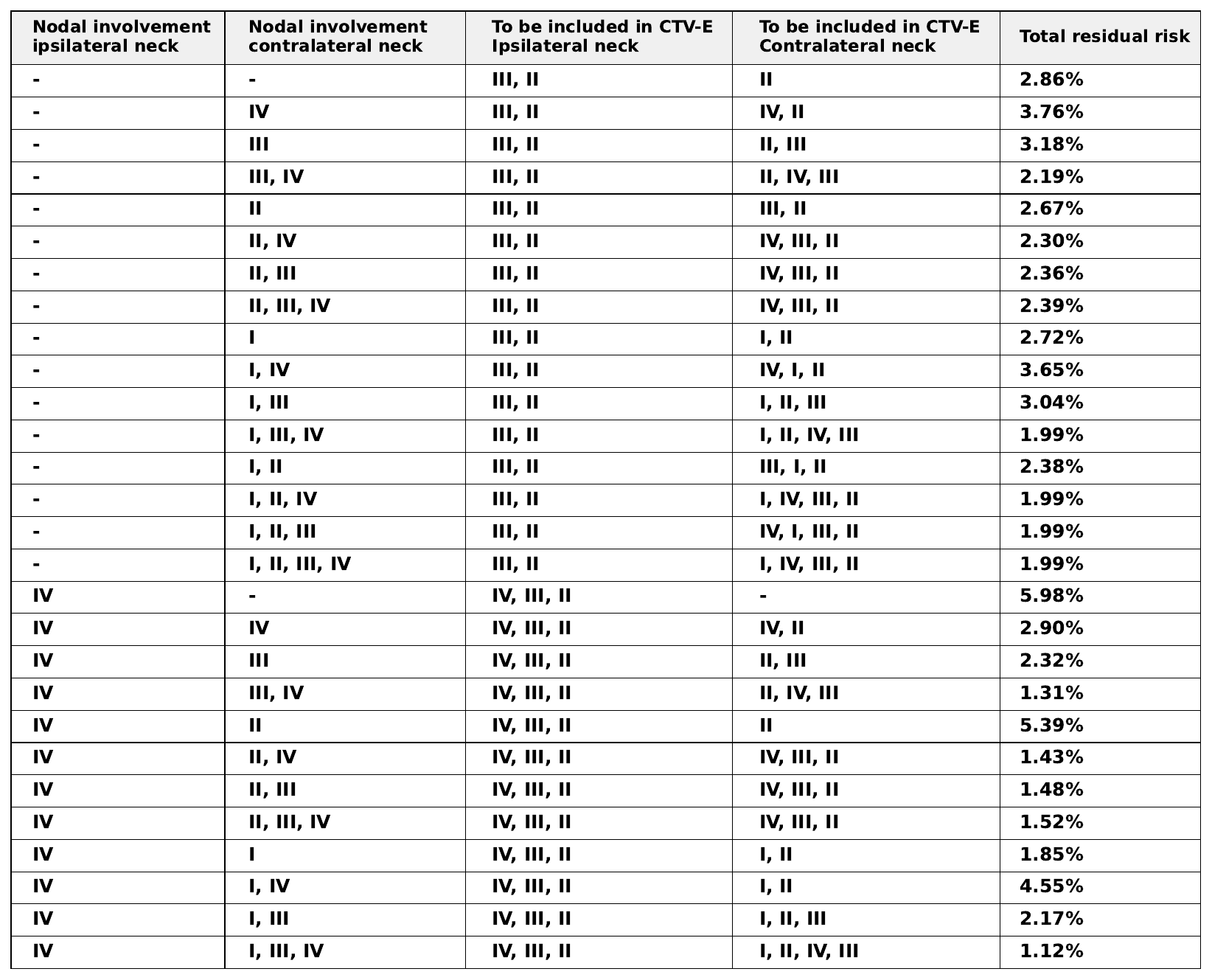}
\includepdf[
  pages=2-,
  pagecommand={}
]{pdf_files/Proposed_Treatment_Protocols_tr_0.08_early}

\includepdf[
  pages=1,
  pagecommand={\section{Treatment Protocols for Advanced-stage Tumours with Risk Threshold of 8\%}}
]{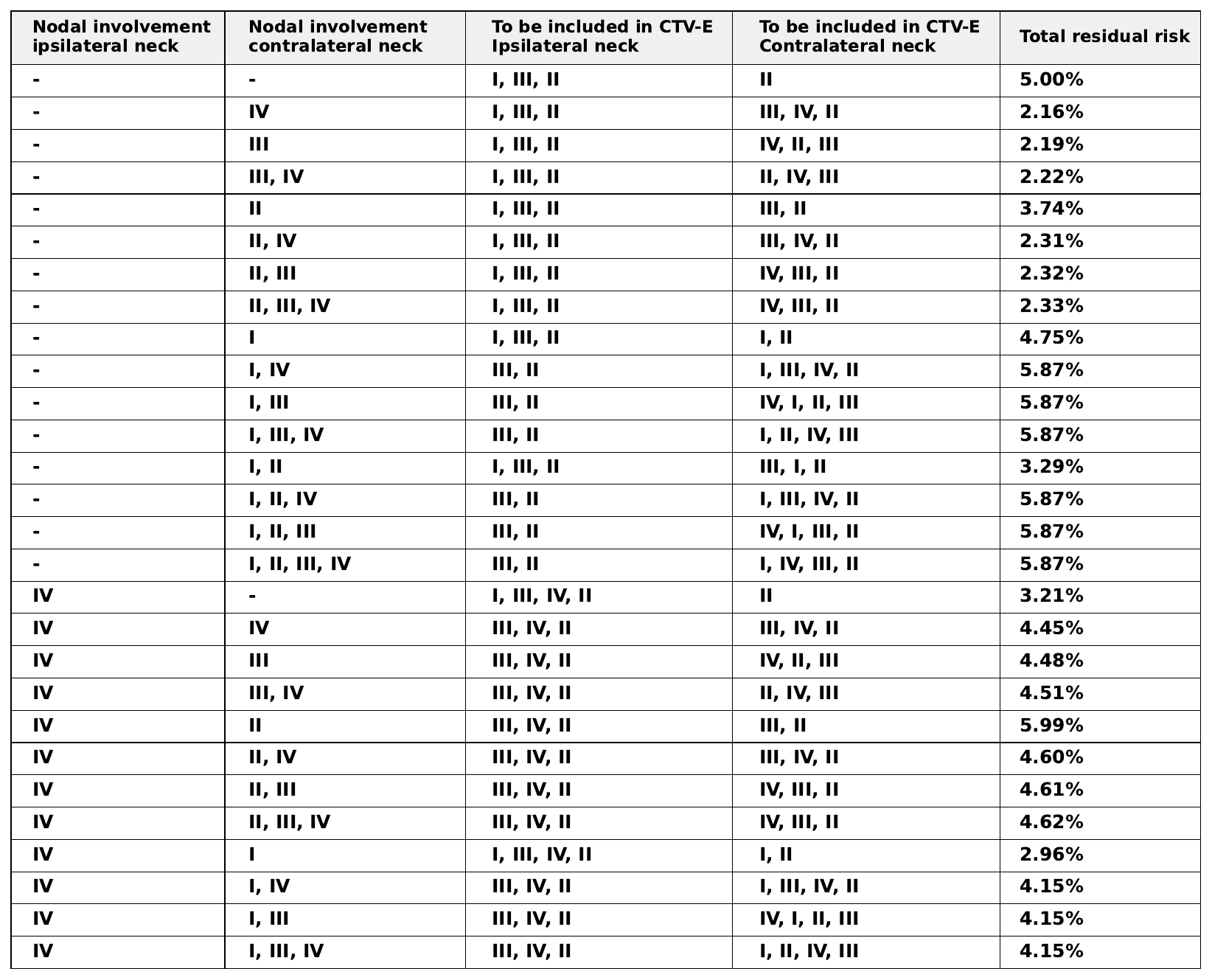}
\includepdf[
  pages=2-,
  pagecommand={}
]{pdf_files/Proposed_Treatment_Protocols_tr_0.08_advanced}

\includepdf[
  pages=1,
  pagecommand={\section{Treatment Protocols for Early-stage Tumours with Risk Threshold of 10\%}}
]{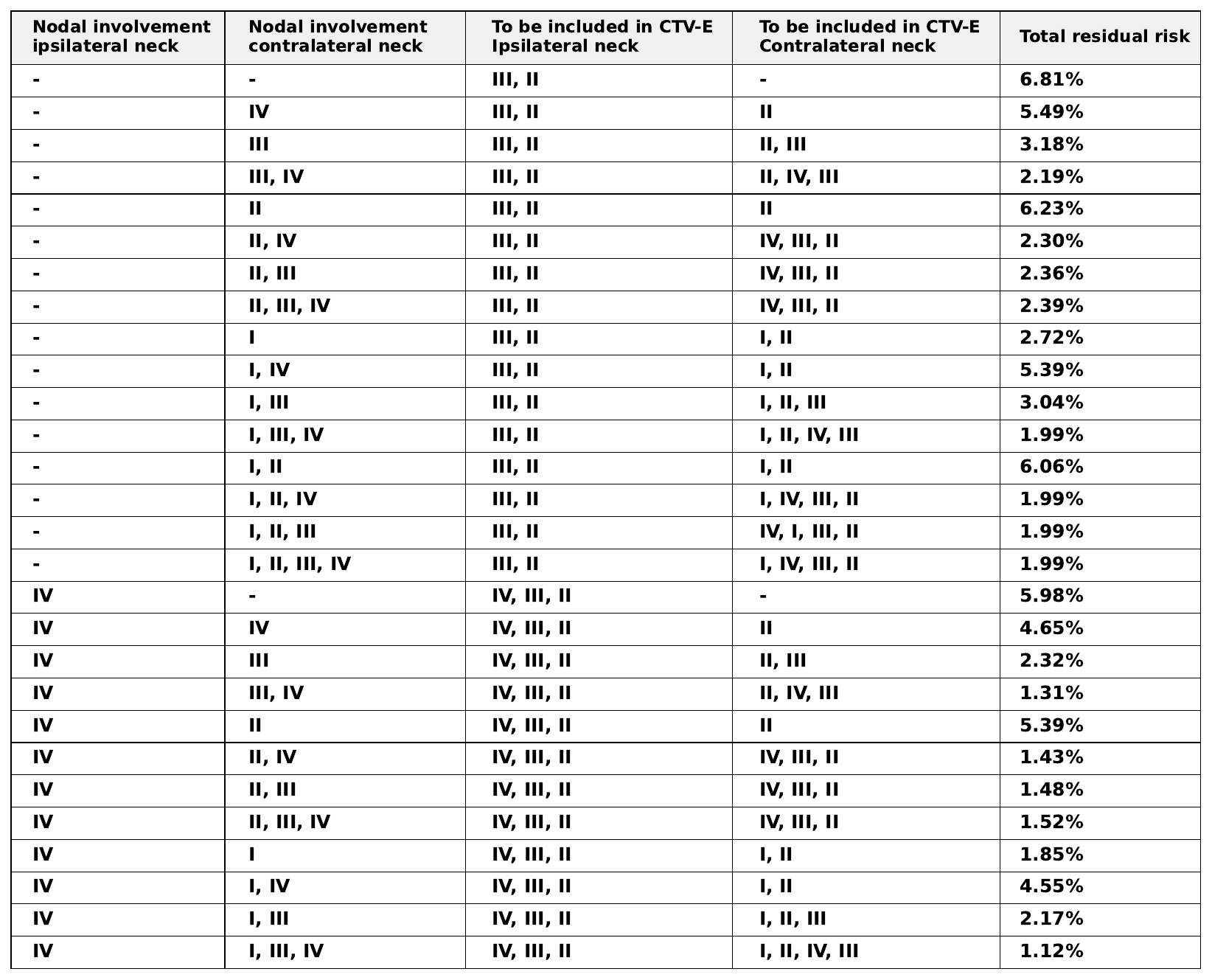}
\includepdf[
  pages=2-,
  pagecommand={}
]{pdf_files/Proposed_Treatment_Protocols_tr_0.10_early}

\includepdf[
  pages=1,
  pagecommand={\section{Treatment Protocols for Advanced-stage Tumours with Risk Threshold of 10\%}}
]{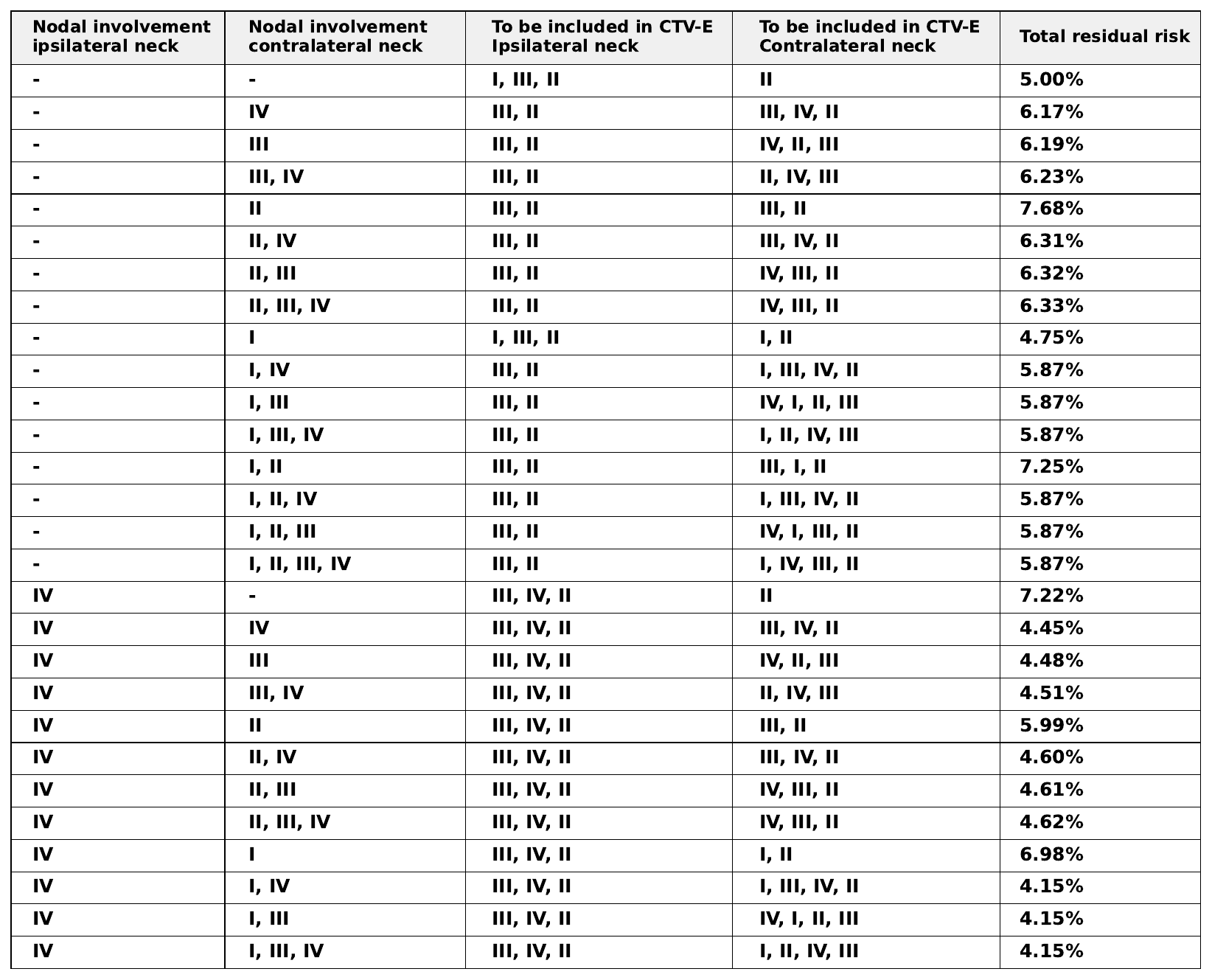}
\includepdf[
  pages=2-,
  pagecommand={}
]{pdf_files/Proposed_Treatment_Protocols_tr_0.10_advanced}

\includepdf[
  pages=1,
  pagecommand={\section{Treatment Protocols for Early-stage Tumours with Risk Threshold of 12\%}}
]{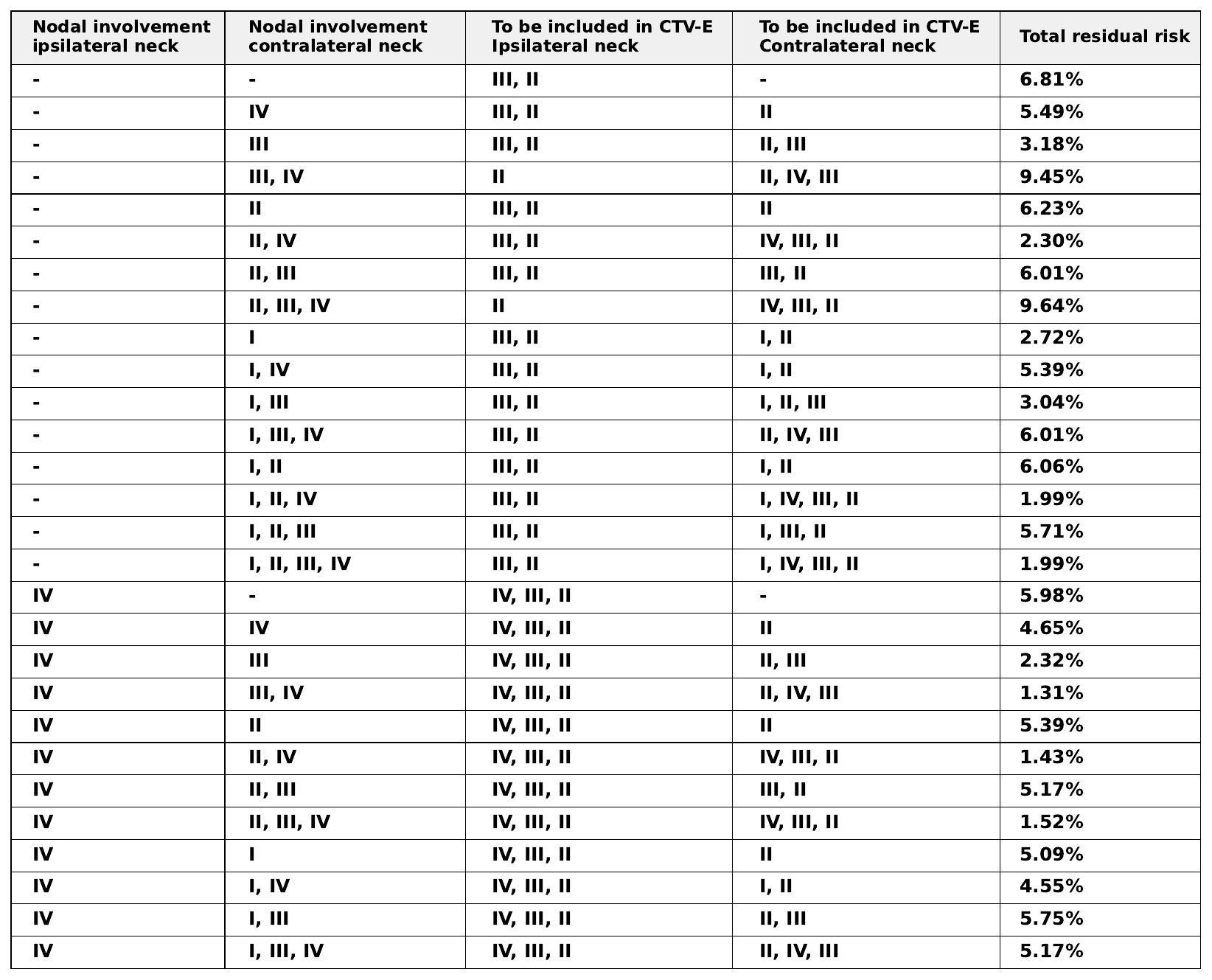}
\includepdf[
  pages=2-,
  pagecommand={}
]{pdf_files/Proposed_Treatment_Protocols_tr_0.12_early}

\includepdf[
  pages=1,
  pagecommand={\section{Treatment Protocols for Advanced-stage Tumours with Risk Threshold of 12\%}}
]{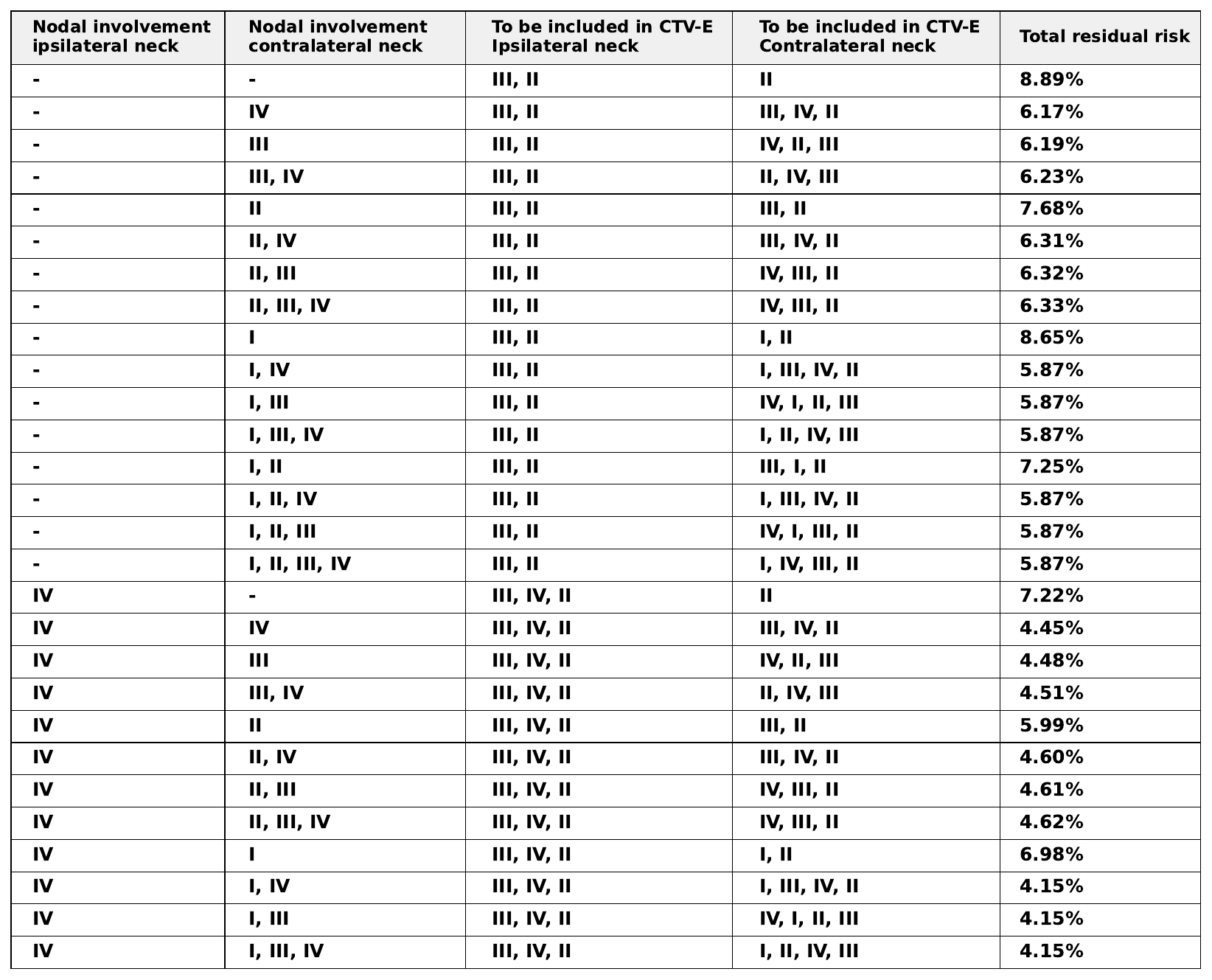}
\includepdf[
  pages=2-,
  pagecommand={}
]{pdf_files/Proposed_Treatment_Protocols_tr_0.12_advanced}

\includepdf[
  pages=1,
  pagecommand={\section{Treatment Protocols for Early-stage Tumours with Risk Threshold of 15\%}}
]{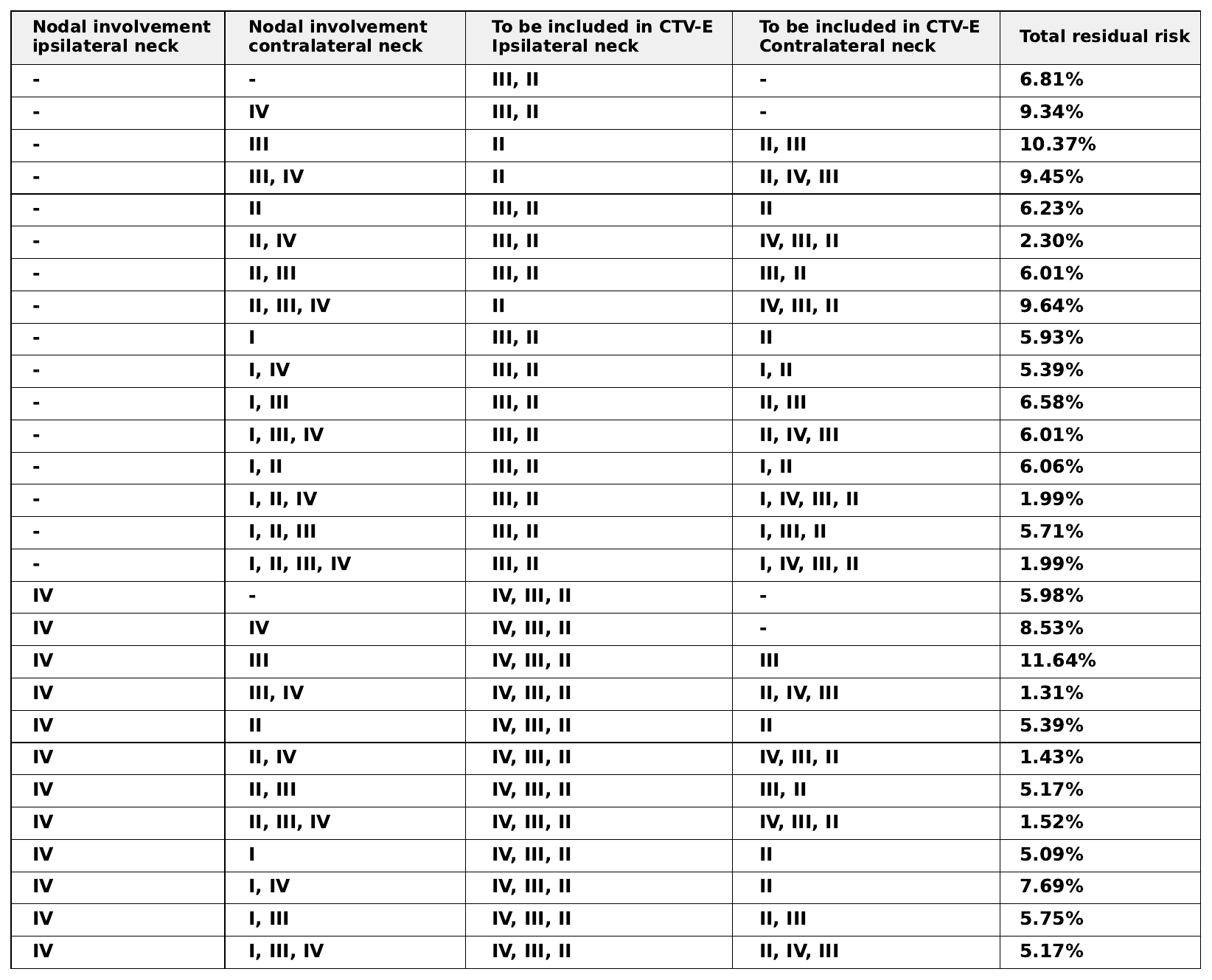}
\includepdf[
  pages=2-,
  pagecommand={}
]{pdf_files/Proposed_Treatment_Protocols_tr_0.15_early}

\includepdf[
  pages=1,
  pagecommand={\section{Treatment Protocols for Advanced-stage Tumours with Risk Threshold of 15\%}}
]{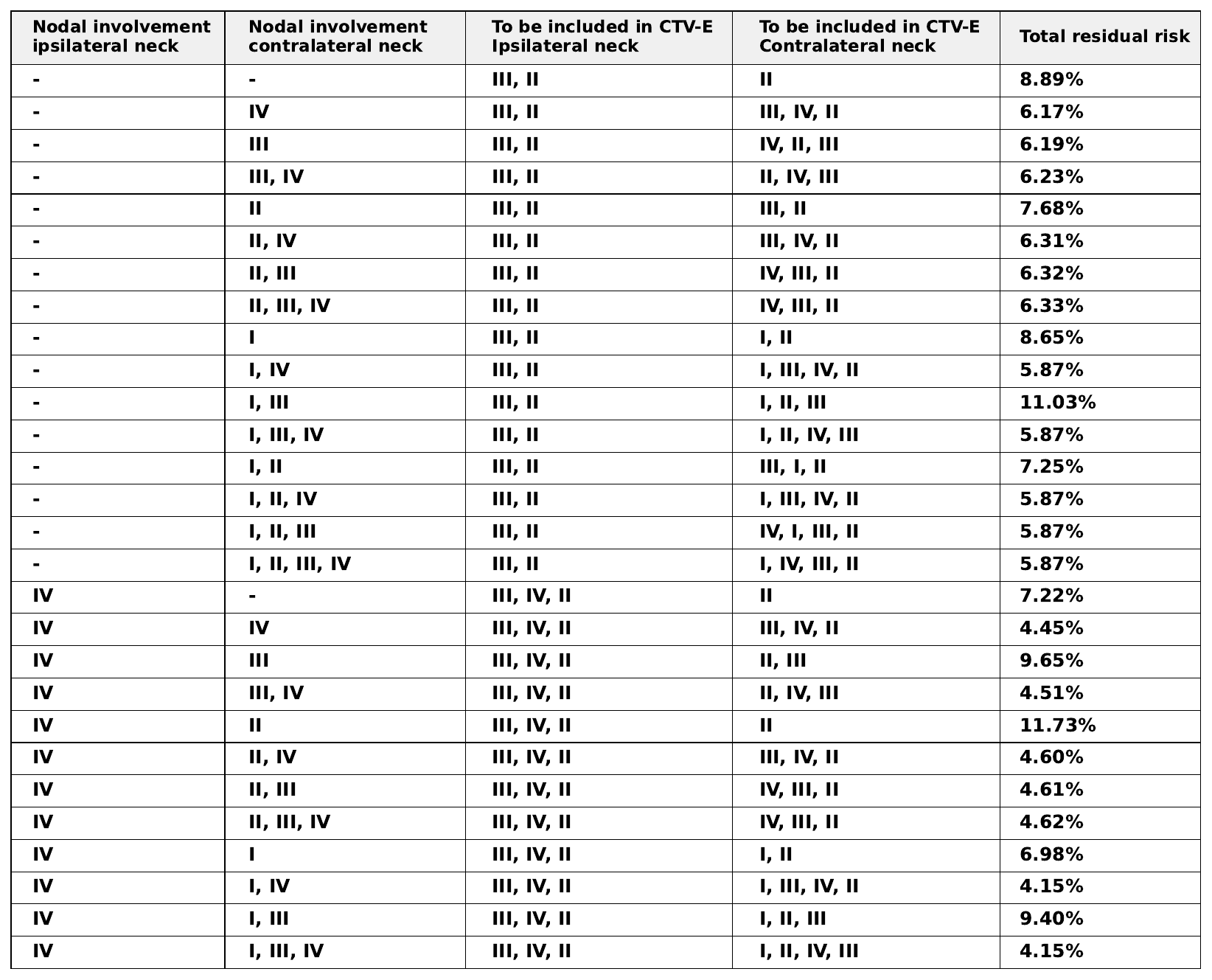}
\includepdf[
  pages=2-,
  pagecommand={}
]{pdf_files/Proposed_Treatment_Protocols_tr_0.15_advanced}

\includepdf[
  pages=1,
  pagecommand={\section{Treatment Protocols for Early-stage Tumours with Risk Threshold of 20\%}}
]{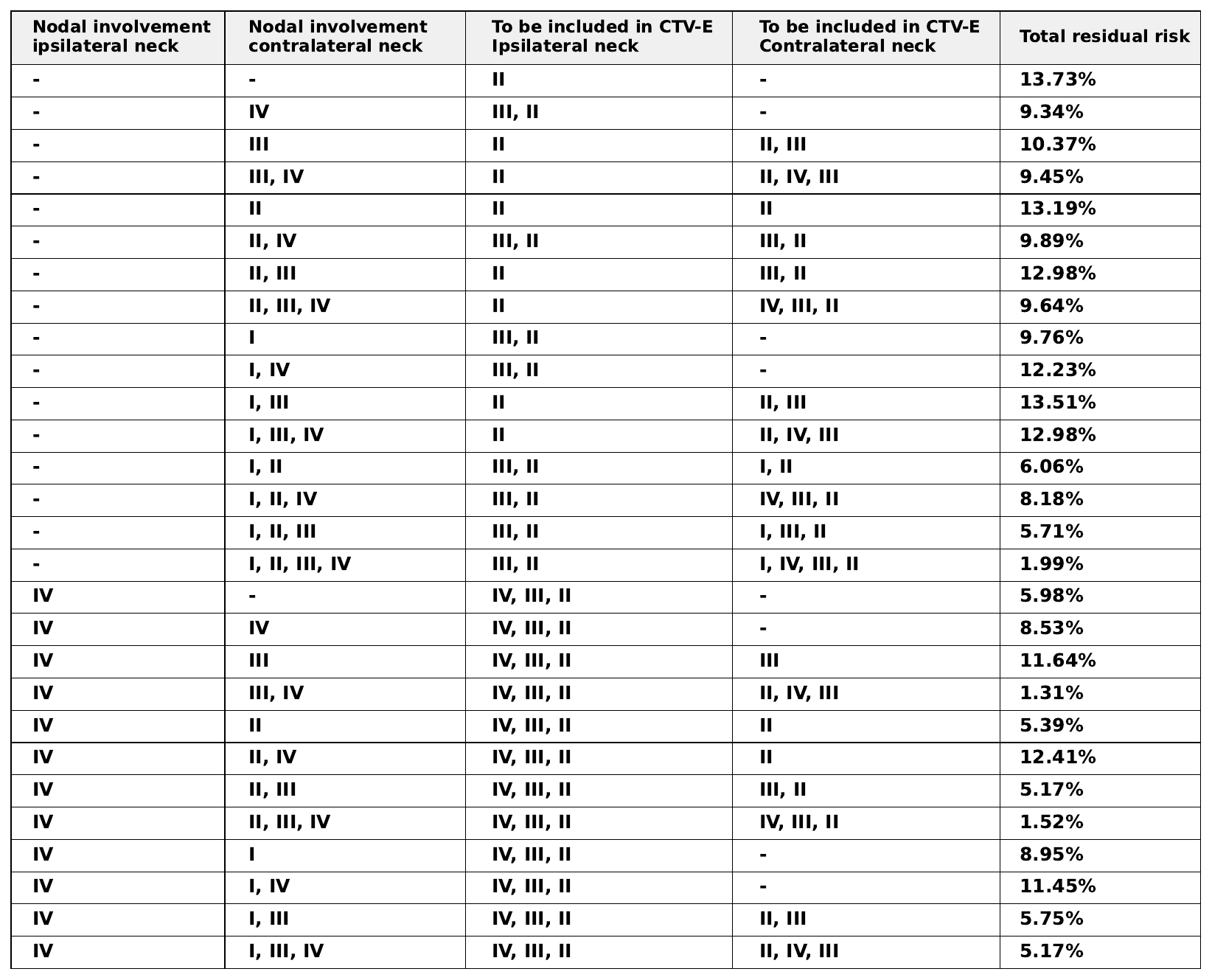}
\includepdf[
  pages=2-,
  pagecommand={}
]{pdf_files/Proposed_Treatment_Protocols_tr_0.20_early}

\includepdf[
  pages=1,
  pagecommand={\section{Treatment Protocols for Advanced-stage Tumours with Risk Threshold of 20\%}\label{app:proto_advanced_20}}
]{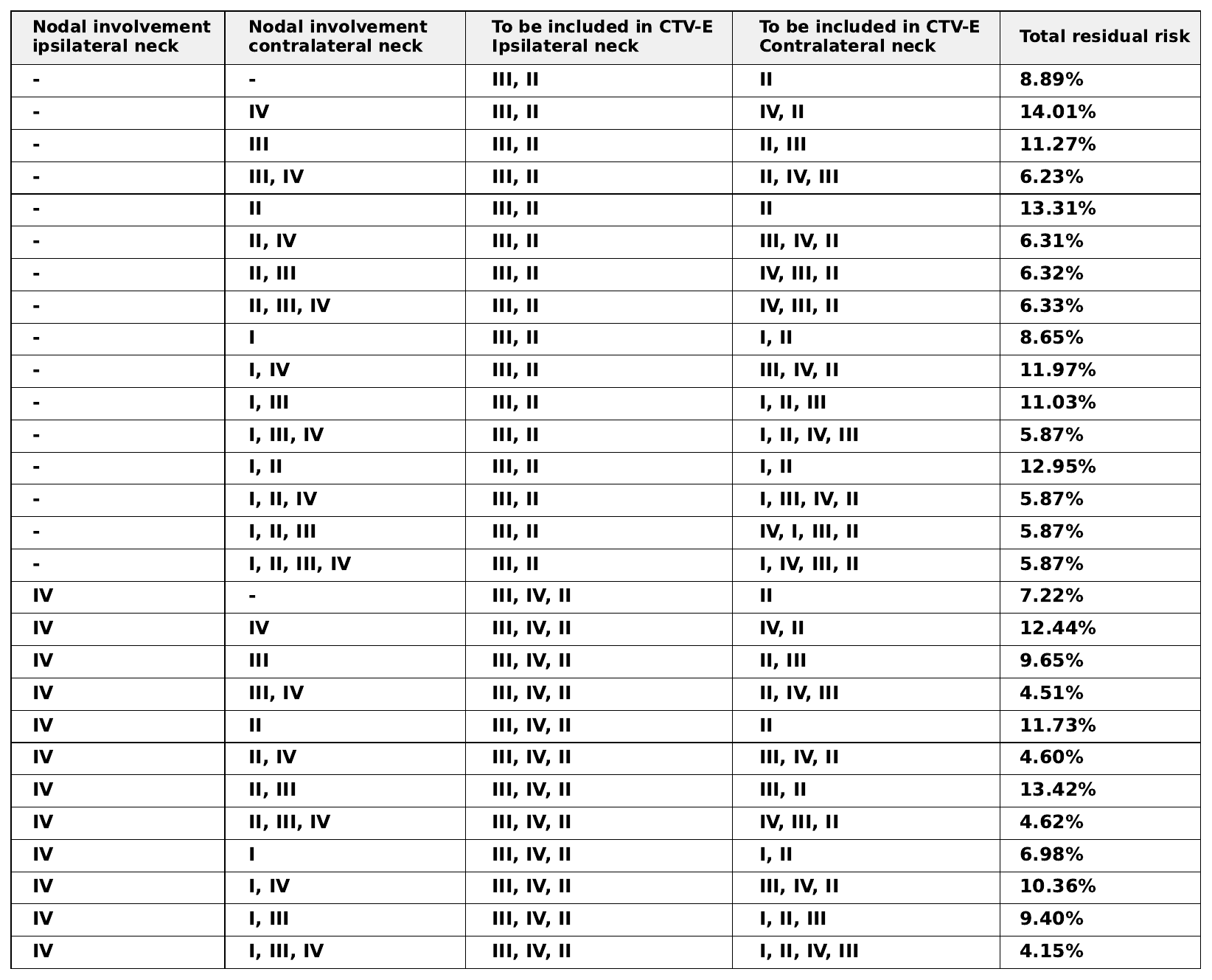}
\includepdf[
  pages=2-,
  pagecommand={}
]{pdf_files/Proposed_Treatment_Protocols_tr_0.20_advanced}

\end{document}